%% file: draft_paper_collision_preprint.tex
\documentclass[10pt]{ismpreprint}

\input{macros_preprint}

\begin{document}

\frontmatter
\date{6th March 2025}
\title{A collision model for very flexible Cosserat rods and immersed-boundary fluid--structure coupling}
\titlerunning{A collision model for very flexible Cosserat rods and fluid--structure coupling}   
\author{Bastian~Löhrer\textsuperscript{a}\footnotemark[1] \and Rolf~Krause\textsuperscript{b,c} \and Jochen~Fröhlich\textsuperscript{a}}
\affiliation{\textsuperscript{\normalfont a}\,Institute~of~Fluid~Mechanics, Technische~Universität~Dresden, Dresden, Germany\\
             \textsuperscript{\normalfont b}\,Euler Institute, Università della Svizzera italiana, Lugano, Switzerland\\
             \textsuperscript{\normalfont c}\,Division Computer, Electrical and Mathematical Sciences and Engineering, King Abdullah University of Science and Technology, Thuwal, Saudi Arabia
             }
\footnotetext[1]{Corresponding author: \href{mailto:bastian.loehrer@tu-dresden.de}{bastian.loehrer@tu-dresden.de}}
\maketitle
\renewcommand*{\thefootnote}{\fnsymbol{footnote}}
\begin{abstract}
\input{./chapters/abstract}
\end{abstract}

\tableofcontents

\mainmatter

\subimport{./chapters/}{main}

\begin{acknowledgements}[Funding]
\input{./chapters/funding}
\end{acknowledgements}

\begin{acknowledgements}
\input{./chapters/acknowledgements}
\end{acknowledgements}

\clearpage
\appendix
\subimport{./chapters/}{appendix}

\clearpage
\printbibliography[heading=bibintoc]

\typeout{get arXiv to do 4 passes: Label(s) may have changed. Rerun} 

\end{document}

%% file: macros_preprint.tex
\usepackage[utf8]{inputenc}

\usepackage[ngerman,british]{babel}

\usepackage{import}
\usepackage{ifdraft}
\usepackage{graphicx}

    \usepackage[T1]{fontenc}
    
    \usepackage{amsmath}
    \usepackage{amssymb}
    \usepackage{mathtools} 
    \usepackage{siunitx}
    \DeclareSIUnit{\coreh}{core\,h}
    
    \usepackage{accents}   
    \usepackage{relsize}
    \usepackage{microtype}
    \usepackage{pifont}
    \usepackage{textalpha}  
    \usepackage{textcomp}  
    
    \usepackage[OMLmathsfit]{isomath} 
    \newcommand{\matrixsymconst}[1]{\mathbf{#1}}
    
    \usepackage{xfrac}  

    \usepackage{rotating}          
    \usepackage{placeins}          
    \usepackage[export]{adjustbox} 
    \usepackage[style=american]{csquotes}
    \usepackage[inline]{enumitem}
    \usepackage{coseoul}
    \usepackage{hyperref}
    \hypersetup{
        colorlinks   = true,
        citecolor    = black,
        linkcolor    = black,
    }
    \usepackage{booktabs}
    \usepackage{multirow}
    \usepackage[inkscapearea=page,inkscapepath=svgsubdir,inkscapedpi=600]{svg}
    \usepackage[vlined,algoruled,linesnumbered]{algorithm2e} \SetAlCapNameFnt{\small}\SetAlCapFnt{\small}
    \usepackage{setspace}  
    \usepackage[nospace]{varioref}
    \usepackage[noabbrev]{cleveref} 
    \crefname{lstlisting}{listing}{listings}
    \Crefname{lstlisting}{Listing}{Listings}
    \crefname{algocf}{algorithm}{algorithms}  
    \Crefname{algocf}{Algorithm}{Algorithms}  
    \AtBeginEnvironment{appendices}{\crefalias{section}{appendix}}  

    \newcommand{\usuk}[2]{#2}

\input{macros/math_misc}

    \input{macros/math_diffs}
\input{macros/math_moreaccents}
\input{macros/misc_subcaptioning}
    \input{macros/misc_algorithmelements}

\input{macros/commands/includestandalonewithpath}
\subimport{macros/}{bib_biblatex.tex}

\usepackage{threeparttable}
\subimport{nomenclature/}{newcommands.tex}
\subimport{nomenclature/}{acronyms.tex}
\usepackage[mode=image|tex]{standalone}

\let\cites\undefined \subimport{macros/commands/}{cites}

\let\oldcite\cite  
\let\oldtextcite\textcite
\let\oldTextcite\Textcite
\renewcommand{\cite}[2][]{\oldcite{#2}}
\renewcommand{\textcite}[2][]{\oldtextcite{#2}}
\renewcommand{\Textcite}[2][]{\oldTextcite{#2}}

\newcommand{\codesnippet}[4]{#1}
\newcommand{\withLES}[1]{}
\newcommand{\othermasterelement}{\ensuremath{{\tilde{\masterelement}}}}

\renewcommand{\zero}[1]{#1}
\newcommand{\primecoderef}[1]{}
\newcommand{\sidecaption}{}

%% file: macros/math_misc.tex
\ifdefined\Real   \else\newcommand{\Real}{{\mathrm{I\!R}}}\fi
\ifdefined\Nat    \else\newcommand{\Nat}{{ \mathrm{I\!N}}}\fi
\ifdefined\Complex\else\newcommand{\Complex}{\mathbb{C}}\fi

\newcommand*{\ind}[1]{\ensuremath{_\mathrm{#1}}}
\newcommand*{\Ind}[1]{\ensuremath{^\mathrm{#1}}}

\usepackage{mleftright}
\newcommand{\of}[1]{\mleft({#1}\mright)}
\newcommand{\braceof}[1]{\mleft\lbrace{#1}\mright\rbrace}

\newcommand{\snorm}[1]{\left\lvert#1\right\rvert}
\newcommand{\vnorm}[1]{\left\lVert#1\right\rVert}

\newcommand{\eqdot}{\ \text{.}}
\newcommand{\eqcomma}{\ \text{,}}

\DeclareMathOperator{\diag}{diag}

%% file: macros/math_diffs.tex
\newcommand{\pdiff}[1]{\partial #1}

\newcommand{\tdiff}[1]{\mathrm{d} #1 \,}

\newcommand{\Tdiff}[1]{\nTdiff{}{#1}}

\newcommand{\nTdiff}[2]{\mathrm{D^{#1}} #2 \,}

%% file: macros/math_moreaccents.tex
\begingroup\makeatletter\nfss@catcodes
\gdef\ImportFromMnSymbolaux#1#2#3{%
	\DeclareFontFamily{U}{MnSymbol#2#3}{}
	\DeclareFontShape{U}{MnSymbol#2#3}{m}{n}
	{ <-6>    s*[#1] MnSymbol#25
		<6-7>   s*[#1] MnSymbol#26
		<7-8>   s*[#1] MnSymbol#27
		<8-9>   s*[#1] MnSymbol#28
		<9-10>  s*[#1] MnSymbol#29
		<10-12> s*[#1] MnSymbol#210
		<12->   s*[#1] MnSymbol#212
	}{}
	\DeclareFontShape{U}{MnSymbol#2#3}{b}{n}
	{ <-6>    s*[#1] MnSymbol#2-Bold5
		<6-7>   s*[#1] MnSymbol#2-Bold6
		<7-8>   s*[#1] MnSymbol#2-Bold7
		<8-9>   s*[#1] MnSymbol#2-Bold8
		<9-10>  s*[#1] MnSymbol#2-Bold9
		<10-12> s*[#1] MnSymbol#2-Bold10
		<12->   s*[#1] MnSymbol#2-Bold12
	}{}
	\DeclareSymbolFont{MnSy#2#3}{U}{MnSymbol#2#3}{m}{n}
	\SetSymbolFont{MnSy#2#3}{bold}{U}{MnSymbol#2#3}{b}{n}
}
\endgroup
\newcommand{\ImportFromMnSymbol}[3][1]{\ImportFromMnSymbolaux{#1}{#2}{#3}}

\newcommand\DeclareMnSymbol[4]{\DeclareMathSymbol{#1}{#2}{MnSy#3}{#4}}

\ImportFromMnSymbol{C}{100}
\DeclareMnSymbol{\smalltriangleup}{\mathord}{C100}{73}
\DeclareMnSymbol{\smalltriangledown}{\mathord}{C100}{75}
\DeclareMnSymbol{\hcrossing}{\mathord}{C100}{144}

\ImportFromMnSymbol{D}{100}
\DeclareMnSymbol{\hateq}{\mathrel}{D100}{61}

\newcommand{\accsmalltriangleup}[1]{\accentset{\smalltriangleup}{#1}}
\newcommand{\accsmalltriangledown}[1]{\accentset{\smalltriangledown}{#1}}

%% file: macros/misc_subcaptioning.tex
\usepackage{subcaption}
\newcommand{\subcaptionmarker}[1]{(\textit{#1})}

%% file: macros/misc_algorithmelements.tex
\newcommand{\assign}{\mathrel{\longleftarrow}}

\newcommand{\plusassign}{\mathrel{\overset{\mathsmaller{+}}{\assign}}}
\newcommand{\minusassign}{\mathrel{\overset{\mathsmaller{-}}{\assign}}}

%% file: macros/bib_biblatex.tex
%


\usepackage[style=ext-numeric,
            backend=biber,
            maxbibnames=10,
            maxcitenames=2,  
            giveninits=true,
            isbn=false,
            articlein=false,
            defernumbers=true, 
            url=true]{biblatex}

\DeclareFieldFormat
  [article,inbook,incollection,inproceedings,patent,thesis,unpublished,report,book,online]
  {titlecase:title}{\MakeSentenceCase*{#1}}

\DeclareSourcemap{
  \maps[datatype=bibtex]{
    \map{\step[fieldsource=langid, matchi=\regexp{^de|de-DE|german}, final]\step[fieldset=language, fieldvalue=German]}
    \map{\step[fieldsource=langid, matchi=\regexp{^en|en-.+|english}, final]\step[fieldset=language, null]}
  }
}


\AtEveryBibitem{%
  \ifthenelse{\ifentrytype{article}\OR%
              \ifentrytype{book}\OR%
              \ifentrytype{incollection}\OR%
              \ifentrytype{report}}%
  {%
   \clearfield{month}%
   \clearfield{day}%
   \clearfield{urlyear}%
   \clearfield{urlmonth}%
   \clearfield{urlday}%
  }{}%
  \clearfield{note}%
  \clearfield{pagetotal}%
}

\DefineBibliographyStrings{english}{%
  page             = {p.},
  pages            = {pp.},
}

\usepackage{xurl}

\newcommand{\citep}[1]{\cite{#1}}


\DefineBibliographyExtras{english}{%
  \DeclareBibstringSet{latin}{andothers,ibidem}%
  \DeclareBibstringSetFormat{latin}{\mkbibemph{#1}}%
}
\UndefineBibliographyExtras{english}{%
  \UndeclareBibstringSet{latin}%
}
\DefineBibliographyExtras{british}{%
  \DeclareBibstringSet{latin}{andothers,ibidem}%
  \DeclareBibstringSetFormat{latin}{\mkbibemph{#1}}%
}
\UndefineBibliographyExtras{british}{%
  \UndeclareBibstringSet{latin}%
}

%% file: nomenclature/acronyms.tex
\usepackage{acro}
\acsetup{
    short-plural-ending={},
}

\DeclareAcronym{rms}{short={RMS}, long={root mean square}, short-indefinite={an}}
\DeclareAcronym{pdf}{short={PDF}, long={probability density function}}
\DeclareAcronym{jpdf}{short={JPDF}, long={joint probability density function}}
\DeclareAcronym{les}{short={LES}, long={large-eddy simulation}, short-indefinite={an}}
\DeclareAcronym{DNS}{short={DNS}, long={direct numerical simulation}}
\DeclareAcronym{rans}{short={RANS}, long={Reynolds-averaged Navier--Stokes}}
\DeclareAcronym{ibm}{short={IBM}, long={immersed-boundary method}, short-indefinite={an}, long-indefinite={an}}
\DeclareAcronym{KH}{short={KH}, long={Kelvin--Helmholtz}}
\DeclareAcronym{HU}{short={HU}, long={head-up}, short-indefinite={an}}
\DeclareAcronym{SO}{short={SO}, long={streamwise-oriented}, short-indefinite={an}}
\DeclareAcronym{LS}{short={LS}, long={low-speed}, short-indefinite={an}}
\DeclareAcronym{HS}{short={HS}, long={high-speed}, short-indefinite={an}}
\DeclareAcronym{LDPE}{short={LDPE}, long={low-density polyethylene}, short-indefinite={an}}
\DeclareAcronym{CFL}{short={CFL}, long={Courant--Friedrichs--Lewy}}
\DeclareAcronym{NSE}{short={NSE}, long={Navier--Stokes equations}, short-indefinite={an}}
\DeclareAcronym{RK}{short={RK}, long={Runge--Kutta}}
\DeclareAcronym{FSI}{short={FSI}, long={fluid--structure interaction}, short-indefinite={an}, long-indefinite={an}}
\DeclareAcronym{EOM}{short={EOM}, long={equation of motion}, short-indefinite={an}, long-indefinite={an}}
\DeclareAcronym{LCP}{short={LCP}, long={linear complementary problem}, short-indefinite={an}}
\DeclareAcronym{PGSM}{short={PGSM}, long={projected Gauß--Seidel method}}
\DeclareAcronym{PVC}{short={PVC}, long={Polyvinylchlorid}}
\DeclareAcronym{ADV}{short={ADV}, long={Acoustic Doppler velocimetry}}
\DeclareAcronym{TKE}{short={TKE}, long={turbulence kinetic energy}}
\DeclareAcronym{ODE}{short={ODE}, long={ordinary differential equation}}

%% file: macros/commands/cites.tex

\makeatletter
\NewDocumentCommand{\cites}{oom}{\gdef\mykeys{#3}\next@cites}
\NewDocumentCommand{\next@cites}{oog}{%
  \IfNoValueTF{#3}{\cite{\mykeys}}{\g@addto@macro\mykeys{,#3}\next@cites}%
}
\makeatother

%% file: chapters/abstract.tex
%
The paper presents a constraint-based collision model for Cosserat rods,
able to handle dynamic or static contact between a large number of highly flexible structures.
The model provides the required collision impulses prior to updating the solution of the rods,
with the impulses accounted for as external loads.
The procedure avoids the need to modify the structure solver itself 
and circumvents any iteration between the collision model and the solver for the Cosserat rods,
maintaining the efficiency of any chosen Cosserat solver.
The collision model is adopted from Tschisgale \textit{et al.} (Arch. Appl. Mech., 2019)
and extended towards higher stability,
which is found necessary in the case of very flexible rods.
Furthermore, the model is supplemented with additional terms
that arise when the colliding rods are immersed in a fluid.
The latter is accounted for by an immersed-boundary method.
A large number of tests are conducted to demonstrate the functionality of the final model.

%% file: chapters/keywords.tex
Cosserat rod \and Collision model \and Fluid--structure interaction \and Immersed-boundary method \and Multiple simultaneous collisions

%% file: chapters/main.tex

\section{Introduction}
\label{sec:introduction}

    \subimport{./collision-model/chapters/}{motivation}

    \subimport{./collision-model/chapters/}{literature}
    \label{sec:previous-method}
    \subimport{./collision-model/chapters/}{context_previous_method}
    
    The paper is structured as follows.
    \Cref{sec:numerical-method} recalls the numerical method,
    which the new collision model will be embedded in.
    This involves the fluid solver,
    the solver for the motion of the Cosserat rods,
    and the coupling of the two.
    Both analytical and discretized views are presented, 
    since properties of the discretization have implications for 
    the design of the collision model.
    The basic concept of the collision model itself is briefly summarized in \cref{sec:collision-model-concept}.
    The detailed aspects of the updated implementation are presented in \cref{sec:collision-model-new},
    forming the core of this paper.
    Finally, the new collision model formulation 
    is applied in \cref{sec:collision-model_configurations}
    where a range of dry and wet test cases are presented.
    
\section{Numerical Method}
\label{sec:numerical-method}
    
    \subsection{Model for the structure}
        \label{sec:eom-continuous}
        \subimport{./structure-solver/chapters/}{eom-continuous}%
        \subimport{./structure-solver/chapters/}{eom-continuous-quaternionbased}%
        \subimport{./structure-solver/chapters/}{remark-viscoelastic-material}%
        \subimport{./structure-solver/chapters/}{shear-correction}%
        \subimport{./structure-solver/chapters/}{remark-damping}
    \subsection{Equations for the fluid}
        \subimport{./fluid-solver/}{brief-collisionpaper}
        
    \subsection{Fluid-structure coupling}
        \subimport{./fluid-structure-coupling/chapters/}{introduction}\par
        \subimport{./fluid-structure-coupling/chapters/}{analytical}
    
    \subsection{Time integration of the coupled problem}
        \label{sec:time-integration}
        \subimport{./fluid-structure-coupling/chapters/}{semi-implicit-coupling-scheme}

    \subsection{Discretization of the Cosserat equations}
        \label{sec:eom-discretized}
        \subimport{./structure-solver/chapters/}{eom-discretized}
        \subimport{./fluid-structure-coupling/chapters/}{figure_structure_elements_markers}
        \subimport{./fluid-structure-coupling/chapters/}{discretized}
        \Cref{eq:str_EOM_numerical,eq:str_EOM_numerical_withfluid2} correspond to the algorithm in \cite{TschisgaleNumerical2018,TschisgaleImmersed2020}.

\setcounter{currentlevel}{6}
\subimport{./collision-model/}{main}

\FloatBarrier\clearpage
\section{Conclusions}
\subimport{./collision-model/}{conclusions}

%% file: chapters/collision-model/chapters/motivation.tex
%
\Ac{FSI} involving slender rods can be found in many contexts 
ranging from biological flows to engineering applications \cite{PramanikComputational2024}.
Examples include
    the locomotion of aquatic animals and of microorganisms \cite{GrayStudies1933,WuHydromechanics1971,LiChemomechanical2023}, 
    soft robots \cite{RuckerStatics2011,ArmaniniSoft2023}, 
    technical applications employing deformable \usuk{fibers}{fibres} \cite{AndoNumerical2022},
    and aquatic canopy flows \cite{RaupachTurbulence1981,FinniganTurbulence2000,NepfFlow2012,BrunetTurbulent2020},
    where collision and contact between the individual \usuk{fibers}{fibres} add to the complexity of the numerical method.

\begin{figure}[b!]
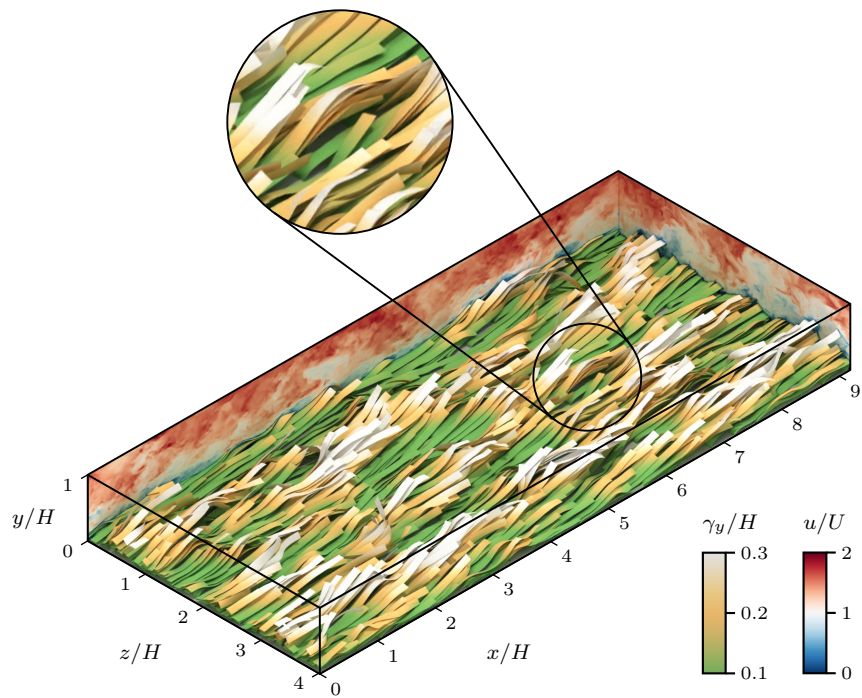

  \begingroup%
  \newcommand{\datapath}{figures/}%
  \includestandalone[]{\datapath/SampleFlowHighCauchy}%
  \endgroup
    \caption{Snapshot from a canopy flow simulation involving flexible blades,
        characterized by a high Cauchy number flow requiring the updated collision handling presented in the present contribution.
        The snapshot was extracted from a simulation which was set up identically to the one in \cite{LohrerFirst2020},
        but with horizontal domain extents increased and periodic boundaries imposed also in $z$-direction.
        The visualization shows the streamwise velocity in vertical slices,
        and the blades are \usuk{colored}{coloured} according to the vertical coordinate of each point on the surface.}
    \label{fig:SampleFlowHighCauchy}
\end{figure}

\begin{figure}[htb]
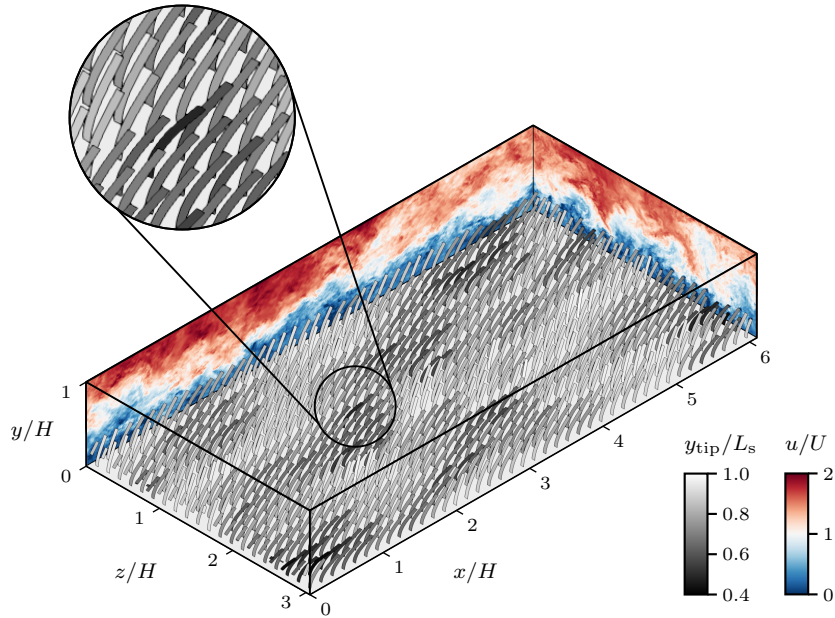
%
  \begingroup%
  \newcommand{\datapath}{figures/}%
  \includestandalone[]{\datapath/SampleFlowMediumCauchy}%
  \endgroup
    \caption{Snapshot from a canopy flow simulation involving flexible blades,
        characterized by a mdoerate Cauchy number,
        as investigated in \cite{TschisgaleLarge2021}.
        The visualization shows the streamwise velocity in vertical slices,
        and the blades are uniformly \usuk{colored}{coloured} according to the vertical coordinate of their tips.}
    \label{fig:SampleFlowMediumCauchy}
\end{figure}

Here, we present a numerical method
to treat aquatic model canopy flows as depicted in \cref{fig:SampleFlowHighCauchy},
designed for long and flexible rods, also termed \textit{blades} or \textit{structures} below.
However, our method is not limited to this use case.
It can as well be employed for stiffer rods,
cf.~\cref{fig:SampleFlowMediumCauchy},
or freely flowing \usuk{fibers}{fibres}.

Fluid--structure coupling can be accomplished by a number of different methods,
ranging from body-fitted grids to immersed-boundary methods 
\cite{BungartzFluidStructure2006,PeskinNumerical1977,MittalImmersed2005,SotiropoulosImmersed2014}.
Here, the \ac{ibm} of \cite{TschisgaleImmersed2020} is applied since it avoids 
the need to adapt the mesh of the fluid solver to the rather complex, time-variable geometry of the immersed rods.
Instead, a static, Cartesian background grid is employed,
and the fluid--structure coupling is established through force terms which are applied 
on the surfaces of the rods and distributed to the \usuk{neighboring}{neighbouring} fluid grid cells.

%% file: chapters/collision-model/chapters/literature.tex
%
Methods for the representation of collisions of rod-like structures
can be classified into 
    penalty-methods and
    constraint-based methods,
        the latter of which are also referred to as \enquote{non-penetration methods}.
The terminology may be misleading,
    as Spillmann noted \cite[19]{SpillmannCoRdE2008},
    since both approaches serve to handle constraints.
Early works presenting these methods include
    \cite{LenoirSurgical2002,PhillipsSimulated2002,ChoeSimulating2005,BertailsLinear2009,VetterFinite2013},
    with application to the collision between filaments,
    and \cite{BridsonRobust2002,ThomaszewskiConsistent2006,StumppGeometric2008},
    addressing the contact between cloths, or membrane-like structures.

Penalty methods determine the contact force based on a measure of interpenetration,
    such as the depth of penetration between colliding objects.
    While conceptually simple,
        the difficulty of applying these models arises from the need 
        to prescribe an appropriate stiffness and possibly also damping parameter
        to translate the penetration into a force.
    Applications to rod-like structures include
    \cite{LenoirSurgical2002,BertailsSuperhelices2006,VetterFinite2013,BertailsLinear2009}.

Constraint-based methods, instead, incorporate position constraints or velocity constraints into the motion of the colliding objects,
    hence avoiding mutual interpenetration by design.
    Applications were presented in \cite{PhillipsSimulated2002,ChoeSimulating2005}.
    Spillmann \textit{et al.} \cite{SpillmannCoRdE2007,SpillmannNoniterative2007}
    proposed a specialized collision model for Cosserat rods 
    that combines the accuracy of a constraint-based method 
    with the efficiency of a penalty method.

Rod models, such as the Cosserat rod of interest in the present paper,
assume rigid cross-sections.
Hence, the collision force instantaneously
changes the momentum of colliding rod segments,
and is also transferred immediately through any chains of collisions,
as encountered, \usuk{e.g.,}{e.g.} in a stack of colliding blades.
The collision of multiple objects then becomes a global problem
which requires special numerical techniques.
In the context of penalty methods,
artificial elasticity is commonly introduced,
so that the nature of a local problem is maintained also in case of simultaneous collisions.
This has the disadvantage of requiring disproportionately small time steps 
to achieve numerical stability and physical correctness \cite{SpillmannCoRdE2008,BergouDiscrete2008}.

In the context of constraint-based methods,
the renowned work by Guendelman \cite{GuendelmanNonconvex2003}
proposed an impulse-based collision model
with collision impulses 
computed globally by means of an iterative procedure, suitable for rigid body dynamics.
Comparable strategies in this problem group were proposed by 
Bender and Schmitt \cite{BenderConstraintbased2006}
and Tonge \textit{et al.} \cite{TongeMass2012}.

The situation is very different for soft, deformable structures.
Unlike collisions of rigid bodies, 
the non-penetration constraints can be 
accounted for locally where two surfaces are in contact,
if compression and decompression during the collision process are resolved in time.
Hence, a situation involving multiple collisions can be approximated in a non-iterative manner, employing
constraint-based methods or penalty methods
\cite{SpillmannCoRdE2008,SpillmannNoniterative2007}.%

%% file: chapters/collision-model/chapters/context_previous_method.tex
%
The presented method extends the collision model previously developed by Tschisgale \textit{et al.} \cite{TschisgaleConstraintbased2019}.
The basic underlying concept is summarized in \cref{sec:collision-model-concept}.
This model is a constraint-based method which considers collisions between individual elements of the discretized elastic beams.
The non-penetration constraint is employed to determine corrective collision impulses in the spirit of the 
model by Guendelman \textit{et al.} \cite{GuendelmanNonconvex2003}.
The flexible structures are \usuk{modeled}{modelled} as geometrically exact Cosserat rods.
This formulation is among the most complex models for 1D beams, capable of representing 
rigid-body motion, extension, bending, shearing, and torsion \cite{AntmanNonlinear1995,SimoFinite1985,AuricchioGeometrically2008,LangMultibody2011}.
The collision model of \cite{TschisgaleConstraintbased2019} was tested with a number of dry test cases
and finally employed in the study depicted in \cref{fig:SampleFlowMediumCauchy} and reported in \cite{TschisgaleLarge2021}.
In all tests and applications the method worked very well 
and can be recommended for cases with comparatively large stiffness and comparatively simple situations
as obtained, \usuk{e.g.,}{e.g.} with the top of a structure hitting the down stream blade at one point 
(zoom in \cref{fig:SampleFlowMediumCauchy}).

While in general properties associated with the beam cross sections can be varied along the length of the rods,
enabling, \usuk{e.g.,}{e.g.} tapered geometries \cite{SchoppmannEfficient2021},
slender ribbons of constant width are considered in the present work.

\label{sec:scope-new-model}

Straightforward application of the collision model of \cite{TschisgaleConstraintbased2019} 
in simulations of configurations with high flexibility of the rods, 
similar to the one of \cref{fig:SampleFlowHighCauchy},
leads to instabilities.
Efforts to identify the source of this issue eventually led to a revision of the algorithm.
While the overall previous model, 
as described in \cite{TschisgaleConstraintbased2019},
could be taken over,
modifications were needed at more detailed levels.
This includes using
one collision element per two structure elements instead of the 1:1 relationship previously employed, and
discriminating between bending and twisting deformation due to collision impulses,
with a number of implications.

A substantial extension of the work of Tschisgale \textit{et al.} \cite{TschisgaleConstraintbased2019}
concerns the application to structures that are coupled to a fluid.
Tschisgale \textit{et al.} only considered the dry situation, and 
fluid--structure coupling loads need to be 
accounted for as external loads acting on the colliding structure elements.
These coupling terms are developed in the present paper,
compatible with a semi-implicit \ac{ibm}-based coupling scheme.
The main difficulty in doing so is connected from the fact that only a part of the coupling loads are known when solving the collision problem.
The other part results from the implicit update of the structure dynamics,
which is determined only after the collision handling,
while also requiring the collision impulses.
This second part of the coupling loads takes on the form of an inertia,
which is incorporated into our revised collision model.

%% file: chapters/structure-solver/chapters/eom-continuous.tex
%
The colliding beam-like objects, submerged in a fluid are \usuk{modeled}{modelled} 
in a one-dimensional way, \usuk{i.e.,}{i.e.} all material and geometric properties as well as position and deformation are a function of a single coordinate.
The geometrically exact Cosserat rod model
is one such model and among the most versatile as it accounts for many more effects than,
\usuk{e.g.,}{e.g.} 
the Euler model of a beam.
The motion of such a Cosserat rod can be expressed by the equations for the linear and angular momentum
\cite{SimoFinite1985,AntmanNonlinear1995,LangMultibody2011}
\begin{subequations}\label{eq:str_EOM}
    \begin{gather}
        \label{eq:str_EOM_lin}
        \bdensity\bladeA\ttderiv{\vectorsym{\centerlinepos}} = \frac{\partial\vectorsym{\internalforce}}{\partial\arclength} + \externalforcedensity
        \eqqcolon \rhsvecforce
        \\
        \label{eq:str_EOM_rot}
        \bdensity \left( \crosssectiontensorofinertia\bcdot\tderiv{\vectorsym{\angvel}} + \vectorsym{\angvel}\times\crosssectiontensorofinertia\bcdot\vectorsym{\angvel} \right)
        = \frac{\partial\vectorsym{\internaltorque}}{\partial\arclength}
        + \frac{\partial\vectorsym{\centerlinepos}}{\partial\arclength} \times \vectorsym{\internalforce}
        + \externaltorquedensity
        \eqqcolon \rhsvecomega
        \eqcomma
    \end{gather}
\end{subequations}
where $\vectorsym{\centerlinepos} = \vectorsym{\centerlinepos}\of{\arclength, t}$ 
is the position of the skeleton line in the three-dimensional laboratory coordinate system,
while $\arclength$ is 
the coordinate along this \usuk{centerline}{centreline} of the structure,
such that $\partial\vectorsym{\centerlinepos}/\partial\arclength$ gives the orientation of the structure skeleton.
The velocity of rotation of cross-sections of the beam is denoted by $\vectorsym{\angvel}\of{\arclength, t}$.
Internal forces and torques are denoted $\vectorsym{\internalforce}$ and $\vectorsym{\internaltorque}$, respectively,
while the arc-length specific external forces and torques imposed on the structure are given by $\externalforcedensity$ and $\externaltorquedensity$.
The translational inertia and rotational inertia of cross-sectional segments are given through the density $\bdensity$ of the structure material,
the cross-sectional area $\bladeA$ of the blades, and the 
tensor of inertia $\crosssectiontensorofinertia$.

The numerical procedure for solving \eqref{eq:str_EOM} was
implemented according to \cite[]{LangMultibody2011}
(see also \cite[]{LangLagrangian2009,LinnDerivation2013}).

%% file: chapters/structure-solver/chapters/eom-continuous-quaternionbased.tex
%
It uses quaternions $\quaternion$ instead of angular velocity $\vectorsym{\angvel}$,
therefore solving an equation for $\quaternion$, which is derived from \eqref{eq:str_EOM_rot}.
In addition, interior loads are expressed in the reference configuration, denoted with the subscript $0$ in the following.
The resulting set of equations reads

\begin{subequations} \label{eq:str_EOM_quaternion}
\begin{gather}
    \label{eq:str_EOM_quaternion_lin}
    \bdensity\bladeA\ttderiv{\vectorsym{\centerlinepos}} 
    = \rhsvecforce
    \\
    \label{eq:str_EOM_quaternion_rot}
    \ttderiv{\quaternion}
    = \frac{2}{\bdensity}\inversequaternionmatrixofinertia\of{\quaternion}\bcdot
    \Big(
              4\bdensity\tderiv{\quaternion}\qcdot\quaternionmatrixofinertia\bcdot(\tderiv{\qconj{\quaternion}}\qcdot\quaternion)
        + 
        \rhsvecquaternion
    \Big)
    - \vnorm{\tderiv{\quaternion}}^2 \quaternion
    \eqcomma
\end{gather}
subject to the constraint
\begin{gather}
     \label{eq:str_EOM_quaternion_additional}
    0 = \vnorm{\quaternion}^2 - 1
    \eqdot
\end{gather}
The loads in \eqref{eq:str_EOM_quaternion} are
\begin{gather}
    \rhsvecforce
    =
    \frac{\partial\quaternion\qcdot\vectorsym{\relref{\internalforce}}\qcdot\qconj{\quaternion}}{\partial\arclength} + \externalforcedensity
    \\
    \rhsvecquaternion
    \coloneqq
      \frac{\partial\quaternion}{\partial\arclength}\qcdot\vectorsym{\relref{\internaltorque}}
    + \frac{\partial\quaternion\qcdot\vectorsym{\relref{\internaltorque}}}{\partial\arclength}
    + \frac{\partial\vectorsym{\centerlinepos}}{\partial\arclength}\qcdot\quaternion\qcdot\vectorsym{\relref{\internalforce}}
    + \externaltorquedensity\qcdot\quaternion
   \eqdot
\end{gather}
The asterisk $\qcdot$ denotes the multiplication of two quaternions, which is non-com\-mu\-ta\-tive.
The quaternion matrix of inertia is
\begin{equation}
    \quaternionmatrixofinertia \coloneqq \left(\begin{array}{c|ccc} 0&0&0&0 \\\hline 0 & \bladeI_1 & 0 & 0 \\ 0 & 0 & \bladeI_2 & 0 \\ 0 & 0 & 0 & \bladeJ \end{array}\right),
    \quad
    \begin{aligned}
        &\bladeI_1 = \bladeW\bladeT^3 / 12, \\
        &\bladeI_2 = \bladeT\bladeW^3 / 12, \\
        &\bladeJ   = \bladeI_1+\bladeI_2\eqcomma
    \end{aligned}
\end{equation}
where $\bladeT$ is the thickness of the rod and $\bladeW$ its width.
The \enquote{tangential inverse} \cite[9]{LangLagrangian2009} quaternion matrix of inertia is
\begin{equation}
    \inversequaternionmatrixofinertia\of{\quaternion} \coloneqq \frac{1}{4} \quaternionmatrix\of{\quaternion} \bcdot \quaternionmatrixofinertia^{-1} \bcdot \transposed{\quaternionmatrix}\!\of{\quaternion}
\end{equation}
with
\begin{equation}
    \quaternionmatrix\of{\quaternion}
    = \left(
        \begin{array}{c|c} 
            \qreal{\quaternion} & -\transposed{\qimag{\quaternion}} \\\hline 
            \qimag{\quaternion} & \qreal{\quaternion}\identity + \skewmatrix{\qimag{\quaternion}} 
        \end{array}
    \right)
    = \left(\begin{array}{r|rrr} q_0 & -q_1 & -q_2 & -q_3 \\\hline q_1 & q_0 & -q_3 & q_2 \\ q_2 & q_3 & q_0 & -q_1 \\ q_3 & -q_2 & q_1 & q_0 \end{array}\right)
    \eqdot
\end{equation}
\end{subequations}

The accelerating terms of \cref{eq:str_EOM},
needed for the collision model below, 
are maintained in $\rhsvecforce$ 
and those of $\rhsvecomega$ can be inferred from $\rhsvecquaternion$, 
since (c.f.~\cref{sec:deriving-quaternion-based-equations})
\begin{equation}
    \rhsvecquaternion
    = \quaternion\qcdot\relref{{\rhsvecomega}} - \quaternion\qcdot\vectorsym{\eta}\ind{r}
   \eqcomma
\end{equation}
where $\vectorsym{\eta}\ind{r}$ is a purely real quaternion, such that
\begin{equation}
    \label{eq:rhsvecomega_rhsvecquaternion}
    \rhsvecomega
    = \quaternion\qcdot\relref{{\rhsvecomega}}\qcdot\qconj{\quaternion}
    = \rhsvecquaternion \qcdot \qconj{\quaternion}
    \eqdot
\end{equation}

%% file: chapters/structure-solver/chapters/remark-viscoelastic-material.tex
%
The internal forces $\vectorsym{\internalforce}$ and moments $\vectorsym{\internaltorque}$
obey viscoelastic material \usuk{behavior}{behaviour} \cite{LangMultibody2011,LangLagrangian2009}.
\begin{equation}
    \vectorsym{\relref{\internalforce}}
    =    \Cstrain \bcdot \vectorsym{\relref{\strain}}
    + 2\,\Cstrainrate \bcdot \vectorsym{\relref{\dot{\strain}}}
    ,\qquad
    \vectorsym{\relref{\internaltorque}}
    =    \Ccurvature \bcdot \vectorsym{\relref{\curvature }}
    + 2\,\Ccurvaturerate \bcdot \vectorsym{\relref{\dot{\curvature}}}
    \eqcomma
\end{equation}
with the Hookean-like matrices
\begin{equation}\label{eq:Hookean_matrices}
    \Cstrain = \diag\of{\bladekOne\bladeG\bladeA, \bladekTwo\bladeG\bladeA, \bladeE\bladeA}
    ,\qquad
    \Ccurvature = \diag\of{\bladeE\bladeI_1, \bladeE\bladeI_2, \bladekt\bladeG\bladeJ}
    \eqcomma
\end{equation} 
and the strain vector $\vectorsym{\strain}$ and curvature vector $\vectorsym{\curvature}$, both in the material reference frame. 
The matrices $\Cstrainrate$ and $\Ccurvaturerate$
provide dissipative internal force and torque contributions~\cite{LangMultibody2011,LangLagrangian2009}.
\begin{equation}\label{eq:damping_matrices_}
    \Cstrainrate
      = \diag\of{\bladeCg{1}, \bladeCg{2}, \bladeCk{3}}
    ,\qquad
    \Ccurvaturerate
     = \diag\of{\bladeCk{1}, \bladeCk{2}, \bladeCg{3}}
    \eqdot
\end{equation}

%% file: chapters/structure-solver/chapters/shear-correction.tex
%
The factors $\bladeks{1}, \bladeks{2} \in [0,1]$ in \cref{eq:Hookean_matrices} are Timoshenko shear correction factors
which depend on the geometry of the cross-section
and serve to account for the non-uniformity of stresses and strain within cross-sections~\cite{CowperShear1966}.
The factor $\bladekt \in [0,1]$ approximates the effect of torsional out-of-plane warping~\cite{LinnDerivation2013}.
In fact, the Cosserat model, given its one-dimensional formulation, is inherently unable to represent this effect,
since it would require deformable cross-sections.
The shear-correction factors are a standard approach to cope with this issue,
with numerous expressions for obtaining these values proposed in the literature 
\cite{CowperShear1966,KanekoTimoshenko1975,HutchinsonShear2000,GruttmannShear2001}. 
The effect of warping was addressed, 
for example, by \textcite{DongMuch2010,FreundWarping2016},
extending Saint--Venant's theory of uniform torsion~\cite[385]{SimoGeometricallyexact1991}.
%

%% file: chapters/structure-solver/chapters/remark-damping.tex

For the simulations carried out below,
the matrices associated with damping in \cref{eq:damping_matrices_} were
estimated as \cite[4]{LinnDerivation2013}
\begin{subequations}
\begin{align}\label{eq:damping_matrices}
    \Cstrainrate
       &= \diag\of{\bladekOne \bladeG \bladeA \retardationtimes,
                  \bladekTwo \bladeG \bladeA \retardationtimes,
                  \bladeE \bladeA            \retardationtimee}
    \\
    \Ccurvaturerate
      &= \diag\of{\bladeE \bladeI_1        \retardationtimee, 
                 \bladeE \bladeI_2        \retardationtimee,
                 \bladekt \bladeG \bladeJ \retardationtimes}
    \eqcomma
\end{align}
\end{subequations}
the validity of which is subject to certain conditions,
including moderate curvature of the rod in its reference configuration and small strains \cite[3]{LinnDerivation2013}.
The damping parameters in \cref{eq:damping_matrices} are determined 
by the 
retardation time $\retardationtimes$ for shearing and 
$\retardationtimee$ for extension,
which relate the bulk and shear viscosities of a viscoelastic Kelvin--Voigt solid 
to the shear modulus $\bladeG$ and 
Young's modulus $\bladeE$~\cite[3--4]{LinnDerivation2013}.
They were computed according to \cite[4,31]{LinnDerivation2013}
for the special case of incompressible material,
\usuk{i.e.,}{i.e.}~$\bladePoissonsratio = 0.5$ and 
$\retardationtimee = \retardationtimes$.
The retardation time $\retardationtimee$
was estimated as
\begin{equation}\label{eq:retardation_time_dampingratio}
    \retardationtimee \approx \dampingratio T \pi
    \eqcomma
\end{equation}
by choosing the desired damping ratio $\dampingratio$
with $\dampingratio=1$ corresponding to a \emph{critical} damping of the vibrating system, 
and $0<\dampingratio\ll1$ indicating \emph{weak} damping.
Here, the variable $T$ is the oscillation period of the fundamental transverse vibration of a cantilevered beam \cite[31]{LinnDerivation2013},
estimated with the value obtained from Euler--Bernoulli theory
\begin{equation}
	T \approx \frac{2\pi}{\num{1.875}^2}\sqrt{\frac{\bdensity \bladeA \bladeL^4}{\bladeE \bladeI_1}}
	\eqdot
\end{equation}

%% file: chapters/fluid-solver/brief-collisionpaper.tex
%
The considered fluid is incompressible, Newtonian, with constant viscosity,
and described by the Navier-Stokes equations (NSE),
with the usual nomenclature,~\cite{TruesdellNonLinear2004}
\begin{subequations}
\begin{gather}
    \label{eq:NSE}
    \fdensity\frac{\partial\vectorsym{\vel}}{\partial t} + \fdensity\nabla\bcdot\left(\vectorsym{\vel}\dyaddot\vectorsym{\vel}\right)
    = 
    \nabla\bcdot\tensorsym{\hydrodynamicstress} 
    + \fdensity\massspecificforce
    \\
    \nabla\bcdot\vectorsym{\vel} = 0
    \eqcomma
\end{gather}
\end{subequations}
where $\vectorsym{u} = \transposed{(u, v, w)}$ is the velocity vector
with its components along the Cartesian laboratory coordinates $x$, $y$, $z$,
while $t$ represents the time, and
$\fdensity$ the fluid density.
The hydrodynamic stress tensor is
\begin{equation}\label{eq:stresstensor}
    \tensorsym{\hydrodynamicstress} = -\pressure\,\identity + \fdensity\,\kinvis2\tensorsym{\Sh}
    ,\qquad
    \tensorsym{\Sh} = \tfrac{1}{2} (\nabla\vectorsym{\vel} + \transposed{\nabla\vectorsym{\vel}})
    \eqcomma
\end{equation}
where $\pressure$ is the pressure,
$\kinvis$ is the kinematic viscosity, and 
$\tensorsym{\Sh}$ the strain rate tensor.
The term $\vectorsym{\massspecificforce}$ is a specific force, \usuk{e.g.,}{e.g.} due to gravity.

%% file: chapters/fluid-structure-coupling/chapters/introduction.tex
%
The kinematic and dynamic coupling between fluid and structure is enforced with a
continuous-forcing Immersed Boundary Method (IBM) 
involving an own semi-implicit temporal coupling scheme \cite{TschisgaleImmersed2020}.
%
%
In the following, a brief overview is provided 
to make the paper self-contained and to provide 
the terms arising from the fluid-structure coupling in the \usuk{modeling}{modelling} of collisions.

%% file: chapters/fluid-structure-coupling/chapters/analytical.tex
%
Fluid and structure motion are physically coupled at their common interfaces
as illustrated in \cref{fig:controlplane}.
\begin{figure}[b]%
    {\phantomsubcaption\label{fig:controlplane}}%
    {\phantomsubcaption\label{fig:controlplanezerowidth}}%
  \begingroup%
  \newcommand{\datapath}{./figures/}%
  \includestandalone[]{\datapath/coupling_surface}%
  \endgroup
    \caption{Control plane cutting a Cosserat rod structure along one of its rigid cross-sections $\structurecrosssectionspace$.
        \subref{fig:controlplane}~Blade with finite thickness $\bladeT>0$,
        \subref{fig:controlplanezerowidth}~infinitely thin blade with $\bladeT\to0$.
    }
    \label{fig:situationwithcontrolplane}
\end{figure}
Considering a control plane $\omega$ aligned with a cross-section of the structure,
the kinematic coupling condition reads
\begin{equation}
    \label{eq:kinematiccoupling}
    \vectorsym{\vel}\of{\xvec = \vectorsym{\centerlinepos} + \vectorsym{\relpos}} = \tderiv{\vectorsym{\centerlinepos}} + \vectorsym{\angvel}\times\vectorsym{\relpos}
    \eqdot
\end{equation}
Dynamic coupling is established through additional external load densities,
$\externalforcedensityrelfsinterface$ and $\externaltorquedensityrelfsinterface$, 
which are added to \cref{eq:str_EOM}, resulting in
\begin{subequations}\label{eq:str_EOM_withfluidloads}
    \begin{gather}
        \bdensity\bladeA\ttderiv{\vectorsym{\centerlinepos}}
            = \rhsvecforce + \externalforcedensityrelfsinterface \\
        \bdensity\crosssectiontensorofinertia\bcdot\tderiv{\vectorsym{\angvel}} + \vectorsym{\angvel}\times\bdensity\crosssectiontensorofinertia\bcdot\vectorsym{\angvel}
            = \rhsvecomega + \externaltorquedensityrelfsinterface
        \eqdot
    \end{gather}
\end{subequations}
These loads are formally obtained by 
integrating the fluid shear stress 
along the closed curve $\partial\Upsilon$ representing the boundary of the blade cross-section,
\begin{equation}
    \label{eq:dynamiccoupling}
    \externalforcedensityrelfsinterface  = -\int\limits_{\partial\Upsilon}\!\tensorsym{\hydrodynamicstress}\bcdot\normalvec\,\tdiff{C}
    ~,\qquad
    \externaltorquedensityrelfsinterface = -\int\limits_{\partial\Upsilon}\!
    \vectorsym{\relpos}
    \times(\tensorsym{\hydrodynamicstress}\bcdot\normalvec)\,\tdiff{C}
    \eqdot
\end{equation}

Since the thickness of the rods is very small,
the limit of vanishing thickness is taken,
but with non-vanishing mass,
the situation depicted in \cref{fig:controlplanezerowidth}.
The distance between the one-sided limits 
    of the hydrodynamic stress $\tensorsym{\relpside{\hydrodynamicstress}}$ and $\tensorsym{\relmside{\hydrodynamicstress}}$
    at the planar interface $\fsinterface$ 
collapses, 
generating a jump of stress across the interface representing the structure~(c.f.~\cref{sec:dynamiccoupling_surfacespecific}).

In the framework of the IBM,
this surface-specific force is smoothed in the proximity of the interface, $\forcingvolumespace$,
as illustrated in \cref{fig:controlplanezerowidth},
to obtain a volume-specific representation $\vectorsym{\volumecouplingforce}$ that is imposed in the Navier Stokes equations, 
then reading
\begin{subequations}\label{eq:NSE_}
\begin{gather}
    \fdensity\frac{\Tdiff{\vectorsym{\vel}}}{\tdiff{t}}
    = \nabla\bcdot\tensorsym{\hydrodynamicstress}
    + \fdensity\massspecificforce
    + \fdensity\vectorsym{\volumecouplingforce}
    \\
    \nabla\bcdot\vectorsym{\vel} = 0
    \eqdot
\end{gather}
\end{subequations}
In the framework of the present method the value of $\vectorsym{\volumecouplingforce}$ is determined for discrete time spans,
\usuk{e.g.,}{e.g.} when computing the velocity field $\vectorsym{\vel}\of{t^{\smash{(\rkstep)}}}$
from $\vectorsym{\vel}\of{t^{\smash{(\rkstep-1)}}}$ (c.f.~\mbox{\cref{sec:kinematicoupling}}).
The required effective value of $\vectorsym{\volumecouplingforce}$ is determined by direct forcing according to
\begin{equation}
    \label{eq:FSI_analytical_volumecouplingforce}
    \vectorsym{\volumecouplingforce}^{(\rkstep)}
    \coloneqq
    \Spr\braceof{
        \frac{\vectorsym{\vel}\relfsinterface^{(\rkstep)} - \uprelimvec\relfsinterface}{t^{(\rkstep)}-t^{(\rkstep-1)}}
    }
    ,\quad
    \uprelimvec\relfsinterface
    \coloneqq
    \IntOp\braceof{\uprelimvec}
    \eqcomma
\end{equation}
where
$\uprelimvec\relfsinterface$ 
    is the preliminary fluid velocity at the interface for $t=t^{(\rkstep)}$ but without the fluid structure coupling,
$\vectorsym{\vel}\relfsinterface^{(\rkstep)}$
    is the updated velocity of the interface, 
$\Spr$ is a spreading operator,
$\IntOp$ an interpolation operator,
and $\rkstep$ the time step counter.
The coupling load densities are then determined from
\begin{equation}\label{eq:coupled3_dynamiccoupling}
    \externalforcedensityrelfsinterface 
        = -\int\limits_\forcingcrosssectionspace\!\fdensity\,
            \vectorsym{\volumecouplingforce}
          \,\dSurface
    ,\quad
    \externaltorquedensityrelfsinterface
        = -\int\limits_\forcingcrosssectionspace\!\vectorsym{\relpos}\times\fdensity\,
             \vectorsym{\volumecouplingforce}
          \,\dSurface
    \eqcomma
\end{equation}
where 
    $\lambda\subset\Lambda$ is the cross-sectional area affected by the IBM force
    and $\Lambda$ the volume around the blade constituting the support of $\vectorsym{\volumecouplingforce}$.

%% file: chapters/fluid-structure-coupling/chapters/semi-implicit-coupling-scheme.tex
%
The \ac{NSE} \eqref{eq:NSE_} are integrated as in \cite{UhlmannImmersed2005,KempeNumerical2013},
employing a low-storage three-step \ac{RK} scheme of second order which 
can be derived from the one of Le and Moin \cite[§2]{LeImprovement1991}.
Each (outer) time step is divided into three \ac{RK} stages, numbered $\rkstep=1,2,3$ below,
and each one of the three \ac{RK} stages involves a series of further steps given in \vref{eq:NSE_IBM}.
At the beginning of the first \ac{RK} stage, 
$\vectorsym{\vel}^{(0)}, \pre^{(0)}$
is set to the solution of the previous time step,
or to the initial condition in case of the first time step, respectively.
A particularity of this scheme, and a prerequisite for the \ac{ibm},
is the first step at the beginning of each \ac{RK} stage
resembling a physically sound, explicit forward step. 
This step in \cref{eq:NSE_IBM_explicit} includes all terms, except for fluid-structure coupling forces.

The fluid-structure coupling introduced in the previous section 
is integrated into \cref{eq:NSE_IBM_FSI_volumeforce}.
Different from \cref{eq:FSI_analytical_volumecouplingforce},
$\rkstep$ in \cref{eq:NSE_IBM_FSI_volumeforce} now refers to \ac{RK} stages instead of outer time steps.
The required intermediate velocity field without fluid--structure coupling is provided by
\begin{subequations}
    \label{eq:NSE_IBM}
    \begin{gather}
        \label{eq:NSE_IBM_explicit}
        \begin{aligned}
            \uprelimvec 
            = \vectorsym{\vel}^{(\rkstep-1)} + \dt \Big[
            &- 2 \alpha_{\rkstep} \codesnippet{ \nabla \pre^{(\rkstep-1)}                             }{he1,he2,he3              }{he1,he2,he3   }{he1,he2,he3   }
             + 2 \alpha_{\rkstep} \codesnippet{ \left(\nabla\bcdot  \kinvis 2\tensorsym{\Sh} \right)^{(\rkstep-1)} }{rud,rvd,rwd              }{rud,rvd,rwd   }{rud,rvd,rwd   }
             \withLES{+   \gamma_{\rkstep} \codesnippet{ \left(\nabla\bcdot  \sgsvis 2\tensorsym{\Sh} \right)^{(\rkstep-1)} }{rud1,rvd1,rwd1           }{rud2,rvd2,rwd2}{rud1,rvd1,rwd1}}
             \withLES{+    \zeta_{\rkstep} \codesnippet{ \left(\nabla\bcdot  \sgsvis 2\tensorsym{\Sh} \right)^{(\rkstep-2)} }{\text{---}\vphantom{rud1}}{rud1,rvd1,rwd1}{rud2,rvd2,rwd2}}\\
            &+ 2 \alpha_{\rkstep} \codesnippet{ \vectorsym{\force}^{(\rkstep)}                              }{fx,fy,fz                 }{fx,fy,fz      }{fx,fy,fz      }
             -   \gamma_{\rkstep} \codesnippet{ \nabla\bcdot(\vectorsym{\vel}\dyaddot\vectorsym{\vel})^{(\rkstep-1)}}{ru1,rv1,rw1              }{ru2,rv2,rw2   }{ru1,rv1,rw1   }
             -    \zeta_{\rkstep} \codesnippet{ \nabla\bcdot(\vectorsym{\vel}\dyaddot\vectorsym{\vel})^{(\rkstep-2)}}{\text{---}\vphantom{rud1}}{ru1,rv1,rw1   }{ru2,rv2,rw2   }
            \Big]
        \end{aligned}
        \\
        t^{(\rkstep)} = t^{(\rkstep-1)} + 2\alpha_{\rkstep}\dt
        \eqdot
    \end{gather}
    The coefficients are
    $\alpha_{\rkstep} = \sfrac{4}{15}, \sfrac{1}{15}, \sfrac{1}{6}$,
    $\gamma_{\rkstep} = \sfrac{8}{15}, \sfrac{5}{12}, \sfrac{3}{4}$,
    $ \zeta_{\rkstep} = 0, \sfrac{-17}{60}, \sfrac{-5}{12}$
    \mbox{\cite{LeImprovement1991,RaiDirect1991}}.
    The resulting preliminary field $\uprelimvec$ is interpolated to the old geometry of the fluid-structure interface,
    \begin{equation}\label{eq:RKscheme_uprelimvec}
        \uprelimvec\relfsinterface\of{\xvec\relfsinterface^{(\rkstep-1)}} = \IntOp\braceof{\uprelimvec\of{\xvec}}
        \eqdot
    \end{equation}
    Next, the motion of the structure is updated,
    updating 
    $\big[\vectorsym{\centerlinepos}, \tderiv{\vectorsym{\centerlinepos}}, \quaternion, \tderiv{\quaternion} \big]^{(\rkstep-1)}$
    $\mapsto$\linebreak
    $\big[\vectorsym{\centerlinepos}, \tderiv{\vectorsym{\centerlinepos}}, \quaternion, \tderiv{\quaternion} \big]^{(\rkstep)}$
    according to
    \begin{gather}
        \label{eq:NSE_IBM_structureseolver_first}
        \bdensity\bladeA\ttderiv{\vectorsym{\centerlinepos}} 
        = \rhsvecforce + \externalforcedensityrelfsinterface
        \\
        \ttderiv{\quaternion}
        = \frac{2}{\bdensity}\inversequaternionmatrixofinertia\of{\quaternion}\bcdot
        \left( 4\bdensity\tderiv{\quaternion}\qcdot\quaternionmatrixofinertia\bcdot(\tderiv{\qconj{\quaternion}}\qcdot\quaternion)
              + \rhsvecquaternion + \externaltorquedensityrelfsinterface\qcdot\quaternion
        \right)
        - \vnorm{\tderiv{\quaternion}}^2 \quaternion
        \\
        0 = \vnorm{\quaternion}^2 - 1
        \eqdot
    \end{gather}
    The coupling loads are treated implicitly due to their dependency on the updated interface motion.
    They are defined as
    \begin{equation}
        \label{eq:NSE_IBM_structurecouplingforcedensity}
        \externalforcedensityrelfsinterface 
            = -\int\limits_\forcingcrosssectionspace\!\fdensity\,
                \vectorsym{\volumecouplingforce}^{(\rkstep)}
              \,\dSurface
        ,\quad
        \externaltorquedensityrelfsinterface
            = -\int\limits_\forcingcrosssectionspace\!\vectorsym{\relpos}\times\fdensity\,
                 \vectorsym{\volumecouplingforce}^{(\rkstep)}
              \,\dSurface
        \eqcomma
    \end{equation}
    where $\vectorsym{\volumecouplingforce}$ is the volumetric representation
    of the fluid-structure coupling force,
    \begin{gather}\label{eq:NSE_IBM_FSI_volumeforce}
        \vectorsym{\volumecouplingforce}^{(\rkstep)}\of{\xvec}
        = \Spr\braceof{
            \frac{ \vectorsym{\vel}\relfsinterface^{(\rkstep)}\of{\xvec\relfsinterface^{(\rkstep-1)}} 
                 - \uprelimvec\relfsinterface\of{\xvec\relfsinterface^{(\rkstep-1)}} }{2\alpha_{\rkstep}\dt}
          }
        \eqcomma
    \end{gather}
    which relates the updated velocity of the fluid-structure interface
    \begin{equation}
        \vectorsym{\vel}\relfsinterface^{(\rkstep)}\of{\xvec\relfsinterface^{(\rkstep-1)}} 
        = \tderiv{\vectorsym{\centerlinepos}}^{(\rkstep)} + \vectorsym{\angvel}^{(\rkstep)}\times\vectorsym{\relpos}^{(\rkstep-1)}
        \eqdot
    \end{equation}
    to the preliminary fluid velocity at the interface,
    with $\Spr$ the spreading operator realized by means of the regularized delta function of Roma~\cite{RomaAdaptive1999}.
    In \eqref{eq:NSE_IBM_structurecouplingforcedensity},
    $\externalforcedensityrelfsinterface$ is a length-specific force,
    $\externaltorquedensityrelfsinterface$ a length-specific moment,
    and $\forcingcrosssectionspace$ a two-dimensional plane,
    so that spreading and integration do not cancel out.
    \Cref{eq:NSE_IBM_FSI_volumeforce}
    is used in the update of the preliminary velocity field and pressure,
    \begin{gather}
        \nabla^2\vectorsym{\vel}^\ast - \frac{\vectorsym{\vel}^\ast}{\kinvis\alpha_\rkstep\dt} = 
           \frac{1}{\kinvis}\left(
              \left(\nabla\bcdot\kinvis2\tensorsym{\Sh}\right)^{(\rkstep-1)} 
            - \frac{\uprelimvec + 2\alpha_{\rkstep}\dt\vectorsym{\volumecouplingforce}^{(\rkstep)}}{\alpha_{\rkstep}\dt}
            \right) \\
        \nabla^2 \phi = \nabla \bcdot \vectorsym{\vel}^\ast  \\
        \vectorsym{\vel}^{(\rkstep)} = \vectorsym{\vel}^\ast - \nabla\phi      \\
        \pre^{(\rkstep)}  = \pre^{(\rkstep-1)} + \frac{\phi}{2 \alpha_\rkstep \dt} - \tfrac{1}{2}\nu \nabla\bcdot\vectorsym{\vel}^\ast
        \eqcomma
    \end{gather}
    to complete the Runge-Kutta step.
    The discretization of the derivatives in space is performed by a staggered Finite-Volume Method,
    as described in \cite{KempeImproved2012}.
\end{subequations}

%% file: chapters/structure-solver/chapters/eom-discretized.tex
%
The motion of a Cosserat rod is computed on a staggered, one-dimensional grid
which represents its \usuk{centerline}{centreline}.
To this end, the \usuk{centerline}{centreline} is split into discrete elements of equal size
(\cref{fig:CosseratDiscretization}).
Discrete values for linear motion are stored at the nodes between adjacent structure elements,
while values for angular motion are stored at the element \usuk{centers}{centres}
as proposed in \cite{LangMultibody2011}.
This is reflected by the choice of indices.
Indices $\structureelement\in\Nat$ refer to element \usuk{centers}{centres} while 
indices offset by $\sfrac{1}{2}$, refer to structure nodes,
\begin{subequations}\label{eq:str_EOM_numerical}
    \begin{equation}
        \label{eq:str_EOM_numerical_lin}
        \ttderiv{\vectorsym{\centerlinepos}}_{\structureelement+\sfrac{1}{2}}
            = \frac{1}{\bdensity\bladeA}\left(
                  {\rhsvecforce}_{\structureelement+\sfrac{1}{2}}
                + {\externalforcedensityrelfsinterface}_{\structureelement+\sfrac{1}{2}}
            \right)
            ,\quad
            \structureelement = 0, 1, \dots, \bladeNe
    \end{equation}
    \begin{multline}
        \label{eq:str_EOM_numerical_ang}
        \relstructureelement{\ttderiv{\quaternion}}{\structureelement}
            = 
            \frac{2}{\bdensity}\inversequaternionmatrixofinertia\of{\relstructureelement{\quaternion}{\structureelement}}\bcdot
            \Big( 4\bdensity\relstructureelement{\tderiv{\quaternion}}{\structureelement}\qcdot\quaternionmatrixofinertia\bcdot(\relstructureelement{\tderiv{\qconj{\quaternion}}}{\structureelement}\qcdot\relstructureelement{\quaternion}{\structureelement})
                  + \relstructureelement{{\rhsvecquaternion}}{\structureelement}
                  + \relstructureelement{{\externaltorquedensityrelfsinterface}}{\structureelement}\qcdot\relstructureelement{\quaternion}{\structureelement}
            \Big)
            - \vnorm{\relstructureelement{\tderiv{\quaternion}}{\structureelement}}^2 \relstructureelement{\quaternion}{\structureelement}
            ,\\
            \structureelement = 1, 2, \dots, \bladeNe
    \end{multline}
    \begin{equation}
        0 = \vnorm{\relstructureelement{\quaternion}{\structureelement}}^2 - 1
        \eqcomma
    \end{equation}
    with
    \begin{equation}
        \label{eq:str_EOM_numerical_rhsvecforce}
        {\rhsvecforce}_{\structureelement+\sfrac{1}{2}}
           = \frac{  \quaternion_{\structureelement+1}\qcdot{\vectorsym{\relref{\internalforce}}}_{\structureelement+1}\qcdot\qconj{\quaternion}_{\structureelement+1}
                    - \quaternion_{\structureelement  }\qcdot{\vectorsym{\relref{\internalforce}}}_{\structureelement  }\qcdot\qconj{\quaternion}_{\structureelement  } }
                   {\darclength_{\structureelement+\sfrac{1}{2}}}
            + \externalforcedensity_{\structureelement+\sfrac{1}{2}}
    \end{equation}
    \begin{multline}
        \label{eq:str_EOM_numerical_rhsvectorque}
        \relstructureelement{{\rhsvecquaternion}}{\structureelement}
           = \frac{  \quaternion_{\structureelement+1}\qcdot{\vectorsym{\relref{\internaltorque}}}_{\structureelement+\sfrac{1}{2}}
                    - \quaternion_{\structureelement-1}\qcdot{\vectorsym{\relref{\internaltorque}}}_{\structureelement-\sfrac{1}{2}}
                   }{\relstructureelement{\darclength}{\structureelement}}
            \\
            + \frac{\vectorsym{\centerlinepos}_{\structureelement+\sfrac{1}{2}} - \vectorsym{\centerlinepos}_{\structureelement-\sfrac{1}{2}}}{\relstructureelement{\darclength}{\structureelement}}
              \qcdot\relstructureelement{\quaternion}{\structureelement}\qcdot{\vectorsym{\relref{\internalforce}}}_{\structureelement}
            + \relstructureelement{\externaltorquedensity}{\structureelement}\qcdot\relstructureelement{\quaternion}{\structureelement}
            \eqdot
    \end{multline}
\end{subequations}

%% file: chapters/fluid-structure-coupling/chapters/figure_structure_elements_markers.tex
%
\begin{figure}[t]
    \centerline{%
  \begingroup%
  \newcommand{\datapath}{figures/}%
  \includestandalone[]{\datapath/structure_elements_markers}%
  \endgroup}
    {\phantomsubcaption\label{fig:BladeGeometryZeroT}}
    {\phantomsubcaption\label{fig:CosseratDiscretization}}
    {\phantomsubcaption\label{fig:ElementWithMarkers}}
    \caption[]{Representations of the flexible blades in the numerical method with 
        an arbitrary element being highlighted.
        \subref{fig:BladeGeometryZeroT}~Geometric representation of the rod elements accounting for their vanishing thickness;
        \subref{fig:CosseratDiscretization}~one-dimensional representation of the corresponding discretized Cosserat rod 
            with the staggered locations of solution variables indicated:
            \raisebox{1pt}{\includegraphics{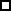}}~orientation and angular velocity,
            \raisebox{1pt}{\includegraphics{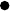}}~position and linear velocity.
        \subref{fig:ElementWithMarkers}~Marker points employed in the \ac{ibm} (\raisebox{1pt}{\includegraphics{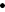}}),
        with the surface patch associated with a single marker point at an arbitrary position $\relmarker[\markerelement]{\xvec}$ being highlighted.
        The sketches are not to scale, with 
        the thickness $\bladeT$ and the length $\bladeLe$ exaggerated.
    }
    \label{fig:structure_elements_markers}
\end{figure}

%% file: chapters/fluid-structure-coupling/chapters/discretized.tex
%
The coupling loads associated with a structure segment of length $\darclength$,
between $\vectorsym{\centerlinepos}_{\structureelement-\sfrac{1}{2}}$ and $\vectorsym{\centerlinepos}_{\structureelement+\sfrac{1}{2}}$,
read
\begin{subequations}\label{eq:str_EOM_numerical_withfluid2}
    \begin{alignat}{3}
        \label{eq:str_EOM_numerical_withfluid2_fsiloads_lin}
        &{\externalforcedensityrelfsinterface}_{\structureelement+1/2}
            &&= - \bigg[ \frac{ \vectorsym{\linmom}\of{t} - \vectorsym{\linmom}^{(\rkstep-1)} }{\darclength(t-t^{(\rkstep-1)})} \bigg]_{\structureelement+1/2}
              &&+ {\modexternalforcedensityrelfsinterface}_{\structureelement+1/2}
        \\
        \label{eq:str_EOM_numerical_withfluid2_fsiloads_ang}
        &{\externaltorquedensityrelfsinterface}_{\structureelement}
            &&= - \relstructureelement{\bigg[ \frac{ \vectorsym{\angmom}\of{t} - \vectorsym{\angmom}^{(\rkstep-1)} }{\darclength(t-t^{(\rkstep-1)})}\bigg]}{\structureelement}
              &&+ {\modexternaltorquedensityrelfsinterface}_{\structureelement}
        \eqcomma
    \end{alignat}
    where
    \begin{equation}
        {\modexternalforcedensityrelfsinterface}_{\structureelement+1/2}
            \coloneqq \bigg[\frac{ \vectorsym{\linmom}^{(\rkstep-1)}   - \vectorsym{{\tilde{\linmom}}}}{\darclength(t^{(\rkstep)}-t^{(\rkstep-1)})} \bigg]_{\structureelement+1/2}
        ,\qquad
        {\modexternaltorquedensityrelfsinterface}_{\structureelement}
            \coloneqq \relstructureelement{\bigg[\frac{ \vectorsym{\angmom}^{(\rkstep-1)} - \vectorsym{{\tilde{\angmom}}} }{\darclength(t^{(\rkstep)}-t^{(\rkstep-1)})} \bigg]}{\structureelement}
        \eqdot
    \end{equation}
The momentum vectors are evaluated as
    \begin{gather}
    \begin{alignat}{3}
        \relstructureelement{\vectorsym{\linmom}}{{\structureelement+1/2}}
            &\coloneqq \tfrac{1}{2}( \relstructureelement{\vectorsym{\linmom}}{{\structureelement}} 
                                   + \relstructureelement{\vectorsym{\linmom}}{{\structureelement+1}} )
        ,\qquad
        &&\relstructureelement{\vectorsym{\linmom}}{\structureelement}
            &&\coloneqq \relstructureelement{\tilde{m}}{\structureelement}
                        \relstructureelement{\tderiv{\vectorsym{\centerlinepos}}}{\structureelement}
              \zero{\ +       \relstructureelement{\vectorsym{\forcinginertialmoment}}{\structureelement}
                        \qcdot\relstructureelement{\tderiv{\quaternion}}{\structureelement}
                        \qcdot\relstructureelement{\qconj{\quaternion}}{\structureelement} 
                      +       \relstructureelement{\quaternion}{\structureelement}
                        \qcdot\relstructureelement{\qconj{\tderiv{\quaternion}}}{\structureelement}
                        \qcdot\relstructureelement{\vectorsym{\forcinginertialmoment}}{\structureelement} }
        \\
        \vectorsym{{\tilde{\linmom}}}_{\structureelement+1/2}
            &\coloneqq \tfrac{1}{2}(\vectorsym{{\tilde{\linmom}}}_{\masterelement} + \vectorsym{{\tilde{\linmom}}}_{\masterelement+1})
        ,\qquad
        &&\relstructureelement{\vectorsym{{\tilde{\linmom}}}}{\structureelement} 
            &&\coloneqq \sum_\markerelement \uprelimvec\of{\relmarker[\markerelement]{\xvec}^{(\rkstep-1)}} \relmarker[\markerelement]{\mass}
    \end{alignat}\\
    \begin{align}
        \relstructureelement{\vectorsym{\angmom}}{\structureelement}
            &= \zero{ \relstructureelement{\vectorsym{\forcinginertialmoment}}{\structureelement}\times\tderiv{\vectorsym{\centerlinepos}}_{\structureelement} \ +\ }
               \relstructureelement{\forcingtensorofinertia}{\structureelement}\bcdot\relstructureelement{\vectorsym{\angvel}}{\structureelement}
        \\
        \relstructureelement{\vectorsym{{\tilde{\angmom}}}}{\structureelement}
            &= \sum_\markerelement \relmarker[\markerelement]{\vectorsym{\relpos}}^{(\rkstep-1)}\times\uprelimvec\of{\relmarker[\markerelement]{\xvec}^{(\rkstep-1)}}\relmarker[\markerelement]{\mass}
    \end{align}
    \end{gather}
with
    \begin{gather}
        \relstructureelement{\tderiv{\vectorsym{\centerlinepos}}}{\structureelement}
            \coloneqq \tfrac{1}{2}(\tderiv{\vectorsym{\centerlinepos}}_{\structureelement-1/2}+\tderiv{\vectorsym{\centerlinepos}}_{\structureelement+1/2})
        ,\qquad
        \relstructureelement{\vectorsym{\angvel}}{\structureelement} = \qimag{2\relstructureelement{\tderiv{\quaternion}}{\structureelement}\qcdot\relstructureelement{\qconj{\quaternion}}{\structureelement}}
        \\
        \label{eq:forcinginertias}
        \relstructureelement{\forcingmass}{\structureelement} \coloneqq \sum_\markerelement \relmarker[\markerelement]{\mass}
        ,\qquad
        \relstructureelement{\forcingtensorofinertia}{\structureelement} \coloneqq \sum_\markerelement \transposed{\skewmatrix{\relmarker[\markerelement]{\vectorsym{\relpos}}^{(\rkstep-1)}}}\bcdot\skewmatrix{\relmarker[\markerelement]{\vectorsym{\relpos}}^{(\rkstep-1)}}\relmarker[\markerelement]{\mass}
        \\
        \zero{\relstructureelement{\vectorsym{\forcinginertialmoment}}{\structureelement} \coloneqq \sum_\markerelement \relmarker[\markerelement]{\vectorsym{\relpos}}^{(\rkstep-1)} \relmarker[\markerelement]{\mass}
        }\eqdot
    \end{gather}
\end{subequations}%
The operator $\skewmatrix{\dots}$ produces a skew matrix from its vector-valued argument with the property
$\skewmatrix{\vectorsym{a}}\bcdot\vectorsym{b} = \vectorsym{a}\times\vectorsym{b}$.
Note: due to the symmetry of the present problem
certain terms in \cref{eq:str_EOM_numerical_withfluid2} vanish,
\begin{equation}
    \relstructureelement{\vectorsym{\forcinginertialmoment}}{\structureelement} = 0
    ,\qquad
    \relstructureelement{\vectorsym{\forcinginertialmoment}}{\structureelement}
    \qcdot\relstructureelement{\tderiv{\quaternion}}{\structureelement}
    \qcdot\relstructureelement{\qconj{\quaternion}}{\structureelement} 
   +\relstructureelement{\quaternion}{\structureelement}
    \qcdot\relstructureelement{\qconj{\tderiv{\quaternion}}}{\structureelement}
    \qcdot\relstructureelement{\vectorsym{\forcinginertialmoment}}{\structureelement} = 0
    ,\qquad
    \relstructureelement{\vectorsym{\forcinginertialmoment}}{\structureelement}\times\tderiv{\vectorsym{\centerlinepos}}_{\structureelement} = 0
    \eqdot
\end{equation}

%% file: chapters/collision-model/main.tex

\newcommand{\modelvariant}[1]{\textit{#1}}

\leveldown{The collision model of Tschisgale \emph{et al.}}
\label{sec:collision-model-concept}

    \leveldown{Basic approach}
    \subimport{./chapters/}{basic-apporach}

    \levelstay{Single collision}
    \label{sec:singlecollision}
    \subimport{./chapters/}{collisiondistance}
    \subimport{./chapters/}{singlecollision-dry}

    \levelstay{Multiple collisions}
    \label{sec:multiplecollision}
    \subimport{./chapters/}{multiplecollisions}

\levelup{Enhanced collision model}
\label{sec:collision-model-new}

    \leveldown{Broadening the concept}
    \subimport{./chapters/}{basic-concept-revisions}

    \levelstay{Identifying collision partners}
    \label{sec:identifyingcollisionpartners}
    \subimport{./chapters/}{pairdetection}

    \levelstay{Collision element length and distance between \usuk{neighboring}{neighbouring} collisions}
    \label{sec:fix_removal_competing_collisions}
    \subimport{./chapters/}{introducelementlength}\par
    \subimport{./chapters/}{removecompeting}    An increased length $\bladeLc>\bladeLe$ associated with the collision elements
    also requires corresponding loads and inertial terms to be evaluated for this length.
    This is addressed in the following sections.
    
    \FloatBarrier
    \levelstay{Collisions in the presence of fluid--structure coupling}
    \label{sec:collisionelementfsi}
    \subimport{./chapters/}{singlecollision-wet}
    
    \levelstay{Inertia of collision elements and system matrices}
    \label{sec:collisionelementinertia}
    \subimport{./chapters/}{revision-collisionelementinertia}

    \FloatBarrier
    \levelstay{Velocity due to accelerating loads}
    \label{sec:vrelvecw}
    \subimport{./chapters/}{revision-vrelvecw}
    
    \levelstay{Imposing the collision impulses on the structures}
    \label{sec:imposecollision}
    \label{sec:fix_Mcv2}
    \subimport{./chapters/}{imposingcollisionimpulses}

\clearpage
\levelup{Assessment of the new collision model}
\label{sec:collision-model_configurations}

    \leveldown{Objective and strategy}
    \subimport{./chapters/}{tests-introduction}

    \FloatBarrier
    \levelstay{Simple prototypical test cases for evaluating the different model variants and verification}
    \label{sec:tests-simple}
    \subimport{./chapters/}{tests-simple}

    \FloatBarrier
    \levelup{Complex test cases}
    \label{sec:tests-complex}
    \subimport{./chapters/}{tests-complex}

%% file: chapters/collision-model/chapters/basic-apporach.tex
\begin{figure}[b!]
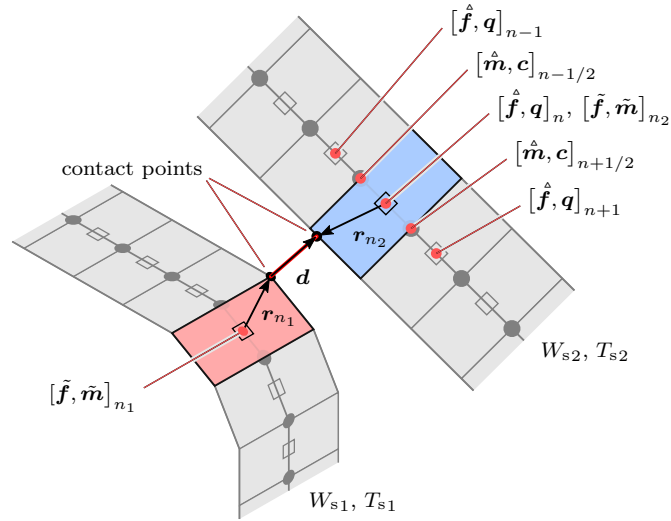

  \begingroup%
  \newcommand{\datapath}{figures/}%
  \includestandalone[]{\datapath/colliding_elements_oldmodel}%
  \endgroup
    \caption{Collision between two structures, 1 on the left, 2 on the right.
        The collision takes place between two structure elements, 
            $n_1$ on structure 1, and 
            $n_2$ on structure 2.
        Relevant quantities of the discretized \ac{EOM} are marked 
        on one of the structure segments 
        in the \usuk{neighborhood}{neighbourhood} of element $n_2$.}
    \label{fig:singlecollision_oldmodel}
\end{figure}

%
This section recalls the model proposed in \cite{TschisgaleConstraintbased2019} for later reference
since it serves as the basis for the proposed improvements.
This model is a constraint-based model 
employing collision impulses in the equation of motion of the structure.
These impulses are determined iteratively 
so as to \usuk{fulfill}{fulfil} the kinematic and dynamic relations imposed between colliding segments of the structures, 
identified with collisions between individual structure elements,
 as illustrated in \cref{fig:singlecollision_oldmodel} for two structures of vanishing thickness.
This approach allows to also account for the collision of a structure with itself 
by finding the corresponding associated colliding structure element on the same structure.
The situation of a single collision, \usuk{i.e.,}{i.e.} the collision between two structure elements, is discussed first.
The situation when more than two structure elements of the same or different structures collide, is termed multiple collision and is discussed afterwards.

Here, a collision event is defined as the situation when the shortest distance between structure elements,
$\vectorsym{\distance}$ in \cref{fig:singlecollision_oldmodel}, is less than a fixed critical distance $\dcrit$,
and when a collision impulse needs to be imposed on both elements to prevent penetration.
This comprises the case of static contact between the structure segments.
The critical distance could, ideally, be set to the mean thickness of the colliding rods
of thickness ${\bladeT}_1$ and ${\bladeT}_2$,
and of width ${\bladeW}_1$ and ${\bladeW}_2$, respectively, 
\begin{equation}\label{eq:dcrit}
    \dcrit = {\bladeT}_1/2 + {\bladeT}_2/2 \ll {\bladeW}_1, {\bladeW}_2
    \eqdot
\end{equation}
In the illustrations below a relatively large value $\dcrit\gg0$ is used for better visibility.

%% file: chapters/collision-model/chapters/collisiondistance.tex
%
The first step of the numerical model is to determine the distance between potentially colliding elements.
The vector $\vectorsym{\distance}$
in \cref{fig:singlecollision_oldmodel}
indicates the distance vector between two structure elements,
identified with the shortest distance between them.
This distance is the distance between two points, named contact points,
one on each structure element.
Their coordinates are geometrically obtained with an orthogonal projection method summarized in \cite[§3.2.2]{TschisgaleConstraintbased2019}.
The two structure elements are considered to be colliding when their shortest distance 
is smaller than a safety distance $\vnorm{\vectorsym{\distance}} < \distance\ind{crit}$.

%% file: chapters/collision-model/chapters/singlecollision-dry.tex
%
The current relative velocity of the contact points,
\usuk{i.e.,}{i.e.} before considering any collision impulse, 
is
\begin{equation}
    \vrelvecv \coloneqq
    {\tderiv{\vectorsym{\distance}}} 
        = \relcollisionelement{\Big[ \tderiv{\vectorsym{\centerlinepos}} + \vectorsym{\angvel} \times \contactpointvec \Big]}{\collisionelement_2} 
        - \relcollisionelement{\Big[ \tderiv{\vectorsym{\centerlinepos}} + \vectorsym{\angvel} \times \contactpointvec \Big]}{\collisionelement_1}
    \eqcomma
\end{equation}
where the notation $[\dots]_{n}$ denotes that the contained expression is evaluated for the $n$th element.
The relative velocity, here,
is decomposed into components co-linear with $\vectorsym{\distance}$ 
    and normal to the direction of $\vectorsym{\distance}$,
\begin{equation}
    \vrelvecv = {\vrelvecv}\ind{n} + {\vrelvecv}\ind{t}
        ,\qquad
        {\vrelvecv}\ind{n} \coloneqq (\vrelvecv\bcdot\distancedir)\distancedir
        ,\qquad
        {\vrelvecv}\ind{t} \coloneqq \vrelvecv - {\vrelvecv}\ind{n}
        ,\qquad
        \distancedir \coloneqq \vectorsym{\distance} / \vnorm{\vectorsym{\distance}}
    \eqdot
\end{equation}
Based on the restitution coefficient and the coefficients of friction,
the desired post-collision velocity of the contact points can be determined,
which gives the desired change of velocity $\dvtot$.
This is accomplished with the strategy of~\cite{GuendelmanNonconvex2003}.
In a first step it is assumed that the tangential component of the relative velocity, perpendicular to $\distancedir$,
gets reduced to zero by the impact of the collision,
whereas the normal component of the post-collision relative velocity is dictated by the Poisson hypothesis,
\begin{equation}\label{eq:PoissonsHypothesis}
    \postcollision{\vrelvecv} \overset{!}{=} {\postcollision{\vrelvecv}}\ind{n}
        ,\qquad
        {\postcollision{\vrelvecv}}\ind{n} \coloneqq -\restitutioncoefficient {\vrelvecv}\ind{n}
    \qquad\Rightarrow\qquad
    \dvtot 
        \coloneqq \postcollision{\vrelvecv} - \vrelvecv
        = - \vrelvecv - \restitutioncoefficient(\vrelvecv\bcdot\distancedir)\distancedir
    \eqdot
\end{equation}

To derive the corresponding collision impulse $\dptot$,
an invertible, so-called system matrix $\systemmatrix$ is needed, satisfying
\begin{equation}
    \dvtot = \systemmatrix\bcdot\dptot
    \eqdot
\end{equation}
This system matrix encapsulates how the relative velocity of the contact points 
would react to any collision impulse acting on the two involved contact points,
but with opposite sign.
The system matrix is derived from the equations of motion for the Cosserat rod
by adding the collision impulse $\dptot$ to the right hand sides.
The model of \cite{TschisgaleConstraintbased2019} 
    does not consider fluid-structure coupling so that
the modified equations~\eqref{eq:str_EOM} give
\begin{subequations}\label{eq:EOM_with_collisionloads}
\begin{gather}
    \bdensity\bladeA\ttderiv{\vectorsym{\centerlinepos}} = \rhsvecforce + \diraccollision\dptot
    \\
    \bdensity \left( \crosssectiontensorofinertia\bcdot\tderiv{\vectorsym{\angvel}} + \vectorsym{\angvel}\times\crosssectiontensorofinertia\bcdot\vectorsym{\angvel} \right)
    = \rhsvecomega + \diraccollision(\contactpointvec\times\dptot)
    \eqdot
\end{gather}
\end{subequations}
Equations~\eqref{eq:EOM_with_collisionloads}
are integrated over the two blade elements involved in the collision,
$\collisionelement_1$ and $\collisionelement_2$
in \cref{fig:singlecollision_oldmodel},
and over the time step during which the collision occurs,
giving
\begin{subequations}\label{eq:CMEOMIntegrated_dry}
\begin{gather}
    \label{eq:collisionelementacceleration_dry}
    \begin{multlined}
    {\mathsmaller\Delta}\relcollisionelement{\tderiv{\vectorsym{\centerlinepos}}}{\collisionelement}
        =
        \relcollisionelement{{\collisionelementmass}}{\collisionelement}^{-1} \left( \dt \relcollisionelement{\forceoncollisionelement}{\collisionelement} + \dptot \right)
    ,\qquad
    \relcollisionelement{\Delta\vectorsym{\angvel}}{\collisionelement}
        =
        \relcollisionelement{\tensorofinertia}{\collisionelement}^{-1} \bcdot
        \left( \dt\relcollisionelement{\torqueoncollisionelement}{\collisionelement} + \contactpointvec\times\dptot \right)
    ,\\
    \collisionelement=\collisionelement_1, \collisionelement_2
    \eqdot
    \end{multlined}
\shortintertext{Here}
    {\mathsmaller\Delta}\relcollisionelement{\tderiv{\vectorsym{\centerlinepos}}}{\collisionelement} 
        \coloneqq \int\limits_{\dt}\!\relcollisionelement{\ttderiv{\vectorsym{\centerlinepos}}}{\collisionelement}\,\tdiff{t}
    ,\qquad
    \relcollisionelement{\Delta\vectorsym{\angvel}}{\collisionelement}
        \coloneqq \int\limits_{\dt}\!\relcollisionelement{\tderiv{\vectorsym{\angvel}}}{\collisionelement}\,\tdiff{t}
    \\
    \label{eq:collisionelementinertia_dry}
    \relcollisionelement{{\collisionelementmass}}{\collisionelement}
        \coloneqq \int\limits_{\hspace{-1ex}\mathrlap{\arclength_n - \bladeLe/2}}^{\hspace{-1ex}\mathrlap{\arclength_n + \bladeLe/2}}\!\bdensity\bladeA\,\dCurve
        = \mass\eind
        = \bdensity\bladeA\bladeLe
    ,\qquad
    \relcollisionelement{\tensorofinertia}{\collisionelement} 
        \coloneqq \int\limits_{\hspace{-1ex}\mathrlap{\arclength_n - \bladeLe/2}}^{\hspace{-1ex}\mathrlap{\arclength_n + \bladeLe/2}}\!\bdensity\crosssectiontensorofinertia\,\dCurve
        = \relstructureelement{\rotationmatrix}{\structureelement} \bcdot \relstructureelement{{\relref{\tensorofinertia}}}{\structureelement}
\shortintertext{with}
        \label{eq:tensorofinertia_original}
        \relstructureelement{{\relref{\tensorofinertia}}}{\structureelement}
        = \frac{\bdensity\bladeW\bladeT\bladeLe}{12} \diag\of{ \bladeLe^2+\bladeT^2, \bladeLe^2+\bladeW^2, \bladeW^2 +\bladeT^2}
\intertext{and $\relstructureelement{\rotationmatrix}{\structureelement}$ the rotation matrix which rotates the material reference frame into the current orientation of the element.
Force and torque acting on the respective element in \eqref{eq:collisionelementacceleration_dry} are indicated by a tilde and obtained from}
    \relcollisionelement{\forceoncollisionelement}{\collisionelement} 
        \coloneqq 
        \int\limits_{\hspace{-1ex}\mathrlap{\arclength_n - \bladeLe/2}}^{\hspace{-1ex}\mathrlap{\arclength_n + \bladeLe/2}}\! \rhsvecforce \,\dCurve
    ,\qquad
    \relcollisionelement{\torqueoncollisionelement}{\collisionelement} 
        \coloneqq
        - \dt^{-1}\int\limits_{\dt}\!\relcollisionelement{\vectorsym{\angvel}}{\collisionelement} \times \relcollisionelement{\tensorofinertia}{\collisionelement} \bcdot\relcollisionelement{\vectorsym{\angvel}}{\collisionelement}\,\tdiff{t}
        + \int\limits_{\hspace{-1ex}\mathrlap{\arclength_n - \bladeLe/2}}^{\hspace{-1ex}\mathrlap{\arclength_n + \bladeLe/2}}\!\rhsvecomega\,\dCurve
    \eqdot
\end{gather}
\end{subequations}
From these expressions,
an equation for the change of the velocity of a contact point is obtained,
\begin{subequations}
\begin{align}
\relcollisionelement{\dvtot}{\collisionelement} 
    =  \relcollisionelement{\Big[ {\mathsmaller\Delta}\tderiv{\vectorsym{\centerlinepos}} + {\mathsmaller\Delta}\vectorsym{\angvel} \times \contactpointvec \Big]}{\collisionelement}
    &= \relcollisionelement{\Big[ \mass^{-1}\left(\dt\,\forceoncollisionelement + \dptot \right) 
            + \left(\tensorofinertia^{-1}\bcdot\left(
              \dt\,\torqueoncollisionelement + \contactpointvec\times\dptot
            \right)\right)\times\contactpointvec
      \Big]}{\collisionelement}
   \\
   \label{eq:CMDeltavn}
   &=  \relcollisionelement{{\vrelvecw}}{\collisionelement}
   + \relcollisionelement{\systemmatrix}{\collisionelement}\bcdot\dptot
   \eqdot
\end{align}
\end{subequations}
The second equality introduces
the velocity change due to accelerating loads, $\relcollisionelement{{\vrelvecw}}{\collisionelement}$,
and the system matrix for the individual collision element, $\relcollisionelement{\systemmatrix}{\collisionelement}$,
reading 
\begin{gather}
    \label{eq:CMvwn}
    \relcollisionelement{{\vrelvecw}}{\collisionelement}
        \coloneqq \relcollisionelement{\Big[ \dt\left( \mass^{-1}\forceoncollisionelement + \left(\tensorofinertia^{-1}\bcdot\torqueoncollisionelement\right)\times\contactpointvec \right) \Big]}{\collisionelement}
    \\
    \label{eq:CMK}
    \relcollisionelement{\systemmatrix}{\collisionelement}
        \coloneqq \relcollisionelement{\Big[\transposed{ \skewmatrix{\contactpointvec} } \bcdot \left(\tensorofinertia^{-1}\bcdot\skewmatrix{\contactpointvec}\right) + \identity\,\mass^{-1}\Big]}{\collisionelement}
    \eqdot
\end{gather}
The relative velocity of the two contact points in collision, 
is then obtained from the difference in $\dvtot$,
\begin{subequations}
\begin{gather}
    \dvtot 
    = \relcollisionelement{\dvtot}{\collisionelement_2}
        - \relcollisionelement{\dvtot}{\collisionelement_1}
    = \vrelvecw + \systemmatrix\bcdot\dptot
    \eqcomma
    \label{eq:collisioneq}
    \\
    \vrelvecw \coloneqq \relcollisionelement{{\vrelvecw}}{\collisionelement_2} - \relcollisionelement{{\vrelvecw}}{\collisionelement_1}
    \\
    \label{eq:systemmatrix}
    \systemmatrix \coloneqq \relcollisionelement{\systemmatrix}{\collisionelement_1}+\relcollisionelement{\systemmatrix}{\collisionelement_2}
    \eqdot
\end{gather}
\end{subequations}
Inserting $\dvtot$ from \eqref{eq:PoissonsHypothesis} into \eqref{eq:collisioneq}
yields the corresponding collision impulse
as the sum of a contribution from 
the change of the current relative velocity to satisfy the Poisson hypothesis,
and 
an impulse compensating the relative acceleration of the contact points due to loads acting on the colliding elements,
\usuk{i.e.,}{i.e.}
\begin{equation}\label{eq:dptot_dpv_dpw}
    \dptot = \dpv + \dpw
    ,\qquad
    \dpv \coloneqq -\systemmatrix^{-1}\bcdot(\restitutioncoefficient(\vrelvecv\bcdot\distancedir)\distancedir + \vrelvecv)
    ,\qquad
    \dpw \coloneqq -\systemmatrix^{-1}\bcdot\vrelvecw                                                                       
    \eqdot
\end{equation}
This impulse 
achieves the correct reaction 
in the direction of $\vectorsym{\distance}$, 
but its tangential component
$\dptott$
is such that the 
tangential relative velocity gets reduced to zero, \usuk{i.e.,}{i.e.}
\begin{gather}
    \dptot = \dptotn + \dptott
    ,\qquad
    \dptotn \coloneqq (\dptot\bcdot\distancedir)\distancedir
    ,\qquad
    \dptott \coloneqq \dptot - \dptotn
    \\
    \dvtot\ind{t} \coloneqq \dvtot - (\dvtot\bcdot\distancedir)\distancedir = \vectorsym{0}
    \eqdot
\end{gather}
This corresponds to a sticking contact, 
which, according to the Coulomb friction model
employed here,
motivated by~\cite{GuendelmanNonconvex2003},
is physically valid only if 
    $\vnorm{\dptott} \le \frictioncoeffstatic \vnorm{\dptotn}$.
Otherwise, a sliding contact is encountered and $\vnorm{\dptott}$ must be reduced to 
$\frictioncoeffdynamic\vnorm{\dptotn}$, giving
\begin{equation}\label{eq:CoulombFrictionModel}
\left\lbrace
\begin{alignedat}[c]{3}
    \vnorm{ \dptott } &\le \frictioncoeffstatic  \vnorm{ \dptotn } \quad&&\Leftrightarrow\quad \dvtot\ind{t} &= 0, \\
    \vnorm{ \dptott } &=   \frictioncoeffdynamic \vnorm{ \dptotn } \quad&&\Leftrightarrow\quad \dvtot\ind{t} &\ne 0
\end{alignedat}\right\rbrace
,\qquad
\frictioncoeffdynamic\le\frictioncoeffstatic
\eqdot
\end{equation}
In the latter case, the tangential velocity does not vanish,
implying a relative motion.
The entire procedure is presented in algorithm~\ref{alg:singlecollision} of the appendix.  

The total collision impulse $\dptot = \dptotn + \dptott$ acts on the two contact points.
With respect to the associated structure element \usuk{center}{centre} this
impulse constitutes a linear part $\dptot$ and an angular momentum
$\vectorsym{\angmom}\ind{c} = \contactpointvec\times\dptot$,
as written in \cref{eq:EOM_with_collisionloads}.
To update the structure motion,
 the discrete \ac{EOM} \eqref{eq:str_EOM_numerical} are solved, 
with the equivalent force and torque densities
$\dptot/(\dt\darclength)$ and $\contactpointvec\times\dptot/(\dt\darclength)$, respectively,
added to the discrete vectors of external force density and external torque density,
$\externalforcedensityrelfsinterface$ and $\externaltorquedensityrelfsinterface$, respectively.

%% file: chapters/collision-model/chapters/multiplecollisions.tex
%
As soon as 
a single collision element 
is in contact with more than one other element, 
collisions can no longer be treated independently from one another.
Instead, the entire system of collisions must be solved, 
creating a so-called \acf{LCP}.
This is handled in an iterative manner 
employing a \acf{PGSM}~\cite{TschisgaleConstraintbased2019}.

For each colliding structure element,
the final, converged linear and angular impulses
resulting from all collisions,
are summed up in $\vectorsym{\linmom}_\masterelement$ and $\vectorsym{\angmom}_\masterelement$, respectively.
The corresponding procedure is shown in algorithm~\ref{alg:multiplecollisions} of the appendix
for convenience.
As for the single-collision case, 
the summarized collision-element \usuk{centered}{centred}
force density $\vectorsym{\linmom}_\masterelement/(\dt\darclength)$ and
torque density $\vectorsym{\angmom}_\masterelement/(\dt\darclength)$
is added to the discrete vectors of external force and torque density,
$\externalforcedensityrelfsinterface$ and $\externaltorquedensityrelfsinterface$, respectively,
in the discrete \ac{EOM} \eqref{eq:str_EOM_numerical}.

%% file: chapters/collision-model/chapters/basic-concept-revisions.tex
%
In order to enhance the collision model for applications involving highly flexible
and frequently colliding structures embedded in a fluid,
we revisit the concept outlined above and check for explicit or implicit choices being made, 
which possibly could be adapted to the case of \ac{FSI}.

In particular,
in \cite{TschisgaleConstraintbased2019}
    each segment of length $\bladeLe$ of the discretized structure is identified with a
    solid colliding body.
    This is done to closely relate the model to a model for individual solid particles.
    This is now generalized by
    defining \enquote{collision elements} of length $\bladeLc$, 
    possibly different from the length of structure elements.
The same applies to the lengths over which internal and external loads are integrated,
    although in either case restrictions occur due to the staggered discretization scheme of the structure solver.

%% file: chapters/collision-model/chapters/pairdetection.tex
%
Instead of evaluating the shortest distance between all possible 
pairs of structure elements, 
a subset of potential pairs is identified first.
Not all of these pairs satisfy $\vnorm{\distance}<\dcrit$,
but any pair excluded in this preliminary selection 
has a distance larger than the critical distance $\dcrit$.
In a first step, for each element
$\masterelement\in\setofelements$,
pairs involving elements from a structure $\structure$ are considered.
\begin{equation}\label{eq:collisionpairs1}%
    \relelement{\CollPairs}{\masterelement,\structure} 
    = \Big\lbrace
    (\masterelement, \othermasterelement)
    \Bigm|
    \left.\othermasterelement \in {\setofelements}_\structure,\right.
    \left.\dsph\of{\masterelement,\othermasterelement} < \dcrit,\right.
    \left.\masterelement\ne\othermasterelement,\right.
    \left.\text{$(\masterelement, \othermasterelement)$
                 not \usuk{neighboring}{neighbouring}}\right.
    \Big\rbrace
    \eqdot
\end{equation}
Here, 
    $\setofelements$ is the set of all structure elements and
    ${\setofelements}_\structure$ contains the elements belonging to the $\structure$th structure.
The function $\dsph\of{\masterelement,\othermasterelement}$ provides the distance between spheres enveloping the two elements $\masterelement$, $\othermasterelement$,
    with corresponding radii $\Rsph_\masterelement$ and $\Rsph_\othermasterelement$, respectively, \usuk{i.e.,}{i.e.}
\begin{equation}
    \dsph\of{\masterelement,\othermasterelement} 
    = \vnorm{ \vectorsym{\centerlinepos}_\masterelement - \vectorsym{\centerlinepos}_\othermasterelement }
    - {\Rsph}_\masterelement
    - {\Rsph}_\othermasterelement
    \eqdot
\end{equation}
This function serves as a preliminary criterion
that is evaluated instead of the actual shortest distance 
    $\vnorm{\vectorsym{\distance}\of{\masterelement,\othermasterelement}}$ 
for speed up.
It is suitable since $\dsph\of{\masterelement,\othermasterelement} < \vnorm{\vectorsym{\distance}\of{\masterelement,\othermasterelement}}$.
\usuk{Neighboring}{Neighbouring} elements on the same structure never collide,
although satisfying $\dsph\of{\masterelement,\othermasterelement} < \dcrit$.
For this reason an element $\othermasterelement$ is skipped in \cref{eq:collisionpairs1},
if it is in the \usuk{neighborhood}{neighbourhood} of $\masterelement$.

Next, the sets $\relelement{\CollPairs}{\masterelement,\structure}$ are reduced by keeping at most $\npartners$ pairs in each,
for which $\dsph$ are smallest, \usuk{i.e.,}{i.e.}
\begin{equation}\label{eq:collisionpairs2}
    \newcommand{\pairone}{\masterelement,\othermasterelement}
    \newcommand{\pairtwo}{\masterelement,m}
    \newcommand{\bpairone}{(\pairone)}
    \newcommand{\bpairtwo}{(\pairtwo)}
    \begin{split}
    \relelement{\CollPairs}{\masterelement,\structure}^{\npartners}
    = \Big\lbrace
    \bpairone \in \relelement{\CollPairs}{\masterelement,\structure}
    \Bigm|
    \left.\snorm{\big\lbrace
        \bpairtwo \in \relelement{\CollPairs}{\masterelement,\structure}
        \mid
        \left.\dsph\of{\pairtwo} < \dsph\of{\pairone}\right.
        \big\rbrace} < \npartners\right.
    \Big\rbrace
    ,\\
    \snorm{\relelement{\CollPairs}{\masterelement,\structure}^{\npartners}} \le \npartners\eqdot
    \end{split}
\end{equation}
A value of $\npartners>1$ is generally required to include all pairs 
for which not only $\dsph$ is small, 
but also the actual distance $\vnorm{\vectorsym{\distance}\of{\masterelement,\othermasterelement}} < \dcrit$.
In the present case with structures of vanishing thickness, this value is 
\begin{equation}\label{eq:Npartners}
    \npartners = \left\lceil \bladeW / \bladeLe \right\rceil
    \eqcomma
\end{equation}
as illustrated in the examples shown in \cref{fig:situation_with_distance}.
Here, using $\dsph\rightarrow\min$ as a criterion is unable to identify the collision pair 
with $\dsph\of{\masterelement,\othermasterelement}\to\min$.
In \cref{fig:S03_96b_610_analysis}, for instance,
    this criterion would only find the elements marked 1
    while the desired elements with the shortest actual distance are the ones marked 5,
    \usuk{i.e.,}{i.e.} the element on the other structure with only the fifth shortest $\dsph$.
    According to \eqref{eq:Npartners}, $\npartners = 7$ elements need to be considered 
    in this example to actually guarantee that the 
    closest other element is considered.
\begin{figure}[b]
  \begingroup%
  \newcommand{\datapath}{figures/}%
  \includestandalone[]{\datapath/situation_with_distance}%
  \endgroup
    {\phantomsubcaption\label{fig:situation_with_distance_01}}
    {\phantomsubcaption\label{fig:S03_96b_610_analysis}}
    \caption{Collision situations where $\dsph\rightarrow\min$ fails to identify the element pair 
        with the shortest distance between colliding elements.
        \subref{fig:situation_with_distance_01}~A prototypical, particularly dangerous case.
        \subref{fig:S03_96b_610_analysis}~A situation actually observed in a simulation where the 
            elements \usuk{colored}{coloured} in red are colliding.
            For either element, elements on the other structures are ranked by their \usuk{center-to-center}{centre-to-centre} distance, with 1 indicating the closest element.
    }
    \label{fig:situation_with_distance}
\end{figure}%
By contrast, Tschisgale \textit{et al.} \cite{TschisgaleConstraintbased2019}
implicitly imposed $\npartners=1$
which was suitable for the applications considered in that study.
With highly flexible structures a larger number is needed.

Another exemplary situation of colliding structures is sketched in \cref{fig:competingcollisionsa}.
    Here, structures $s_1$ and $s_2$ collide in the surrounding of elements $\masterelement_1\dots\masterelement_4$,
    with the collision pairs
    $\lbrace\masterelement_1,\masterelement_3\rbrace$,
    $\lbrace\masterelement_2,\masterelement_3\rbrace$, and
    $\lbrace\masterelement_2,\masterelement_4\rbrace$ defined.
    In other words, for two structures physically colliding at a single spot along their lengths, 
    the numerical procedure determines three \usuk{modeled}{modelled} collisions among four structure elements.

After the pre-selection, the final set of pairs is obtained from the union of 
$\relelement{\CollPairs}{\masterelement,\structure}^{\npartners}$
according to
\begin{equation}
    \CollPairs 
    = \Big\lbrace
    \left\lbrace\masterelement, \othermasterelement\right\rbrace 
    \in 
    \bigcup_{\masterelement,\structure}
    \relelement{\CollPairs}{\masterelement,\structure}^{\npartners}
    \Bigm|
    \left.\vnorm{\vectorsym{\distance}\of{\masterelement,\othermasterelement}} < \dcrit\right.
    \Big\rbrace
    \eqcomma
\end{equation}
where 
    $\vectorsym{\distance} = \vectorsym{\distance}\of{\masterelement,\othermasterelement}$ 
is the actual minimal distance vector 
between two elements.
This distance is computed via a projection method suitable for rectangular geometries~\cite{TschisgaleConstraintbased2019}.

%% file: chapters/collision-model/chapters/introducelementlength.tex
%
At this point the following generalization is introduced:
instead of collisions between structure elements of length $\bladeLe$, 
structure segments of length
\begin{equation}
    \bladeLc = \dnLc\bladeLe
    ,\qquad
    \dnLc\in\Nat_+
    \eqcomma
\end{equation}
\usuk{co-centered}{co-centred} with the structure elements are considered. 
Such elements, termed \emph{collision elements} henceforth,
are illustrated in \crefrange{fig:competingcollisionsb}{fig:competingcollisionsd}.
\begin{figure}[b!]
  \begingroup%
  \newcommand{\datapath}{figures/}%
  \includestandalone[]{\datapath/competingcollisions}%
  \endgroup
    {\phantomsubcaption\label{fig:competingcollisionsa}}
    {\phantomsubcaption\label{fig:competingcollisionsb}}
    {\phantomsubcaption\label{fig:competingcollisionsc}}
    {\phantomsubcaption\label{fig:competingcollisionsd}}
    \caption{A collision of two structures $s_1$ and $s_2$,
        resulting in three collisions 
        which involve four structure elements $n_1$, $n_2$, $n_3$, $n_4$, 
        two on each structure, as indicated.
    \subref{fig:competingcollisionsa}~The affected structure elements are marked using different \usuk{colors}{colours} for \usuk{neighboring}{neighbouring} elements.
        The three shortest distances for the three collision events are shown in red.
    \subref{fig:competingcollisionsb}--\subref{fig:competingcollisionsd}~Separate 
        images for each collision event with the 
        collision elements drawn for $\bladeLc=2\bladeLe$.
    }
    \label{fig:competingcollisions}
\end{figure}
The inertia of such a collision element and the loads acting on it then also 
need to be extracted from the correspondingly larger structure segment of length $\bladeLc$
(with $\bladeLc$ adapted for the first and last element of a structure).
In \crefrange{fig:competingcollisionsb}{fig:competingcollisionsd}, for instance,
$\bladeLc = 2 \bladeLe > \bladeLe$. 
No consecutive coverage of the structures with collision elements was imposed in the final production runs, as this turned out to be unnecessary.
In tests with small numbers of elements per structure such a coverage was imposed in academic test cases to avoid jitter.

The need for such an increased length of the collision element is supported by the tests carried out in \cref{sec:collision-model_configurations} below.
It is linked to the staggered discretization of the Cosserat rod, as illustrated in \cref{fig:singlecollision_oldmodel},
with internal forces stored at element \usuk{centers}{centres} 
such that the acceleration of the element-\usuk{centered}{centred} collision element due to the internal force gradient can only be evaluated between the \usuk{neighboring}{neighbouring} element \usuk{centers}{centres}
constituting a length of $2\bladeLe$.

%% file: chapters/collision-model/chapters/removecompeting.tex
%
However, 
collision elements with $\bladeLc > \bladeLe$ from collisions involving \usuk{neighboring}{neighbouring} structure elements would then overlap,
and the related collision impulses would overcompensate the loads acting on the associated structure segments.
This raises the need to remove competing collisions.
Two collision pairs are considered competing if
(a)~each pair has a contact point on different elements of the same structure, and
(b)~these elements are separated by less than $\dnNoncompeting-1$ structure elements. 
The situation is resolved by comparing collision pairs, sorted by their respective collision distance,
and removing the pair with the larger distance for any two competing collision pairs. 
Doing so reduces the set of collision pairs to $\CollPairsNC$,
a set of non-competing collision pairs,
\begin{equation}\label{eq:collisionpairs_noncompeting}
    \CollPairsNC 
    = \Big\lbrace
    \left\lbrace\masterelement_1, \masterelement_2\right\rbrace 
    \in \CollPairs
    \Bigm|
    \forall
    \left\lbrace\othermasterelement_1,\othermasterelement_2\right\rbrace\in\CollPairsNC
    :
    \begin{aligned}[t]
       &     (\text{$\masterelement_i$ and $\othermasterelement_j$ on different blades}) \\
       &\vee (\masterelement_i - \othermasterelement_j = 0)                              \\
       &\vee (\snorm{\masterelement_i - \othermasterelement_j} \ge \dnNoncompeting)
       ,\quad i, j \in \lbrace1, 2\rbrace
    \,\Big\rbrace
    \eqdot
    \end{aligned}
\end{equation}
As a result, any two collision elements from different collision pairs of $\CollPairsNC$,
are either identical, or non-overlapping.

%% file: chapters/collision-model/chapters/singlecollision-wet.tex
%

Here, the collision model is extended to account for additional terms due to the \ac{ibm}-based fluid--structure coupling.
Consulting the corresponding \ac{EOM} \eqref{eq:str_EOM_withfluidloads} 
where the load densities $\externalforcedensityrelfsinterface$ and $\externaltorquedensityrelfsinterface$
account for \ac{FSI} loads determined with \ac{ibm},
it is obvious that these 
terms would also need to be integrated along the length of a collision element 
to determine suitable collision impulses.
It is less obvious, however, that the semi-implicit coupling scheme of 
\cref{sec:time-integration}
requires additional inertial terms to be accounted for in the explicit computation of collision impulses.
This was found necessary where the inertia generated by the virtual mass of surrounding fluid
is dominating over the inertia of the structures or the elastic forces intervening in the collisions,
as encountered with, 
\usuk{e.g.,}{e.g.} highly flexible structures in water.

Starting point is the procedure for computing collision impulses of \cite{TschisgaleConstraintbased2019},
which is provided in \cref{sec:singlecollision,sec:multiplecollision} and in the algorithms~\ref{alg:singlecollision} and \ref{alg:multiplecollisions} of the appendix.
To obtain a collision model that is compatible with the IBM,
the steps leading to equations~\eqref{eq:CMEOMIntegrated_dry} 
need to be reconsidered.
As demonstrated in \cref{sec:fluid-structure-coupling_appendix},
\vref{eq:str_EOM_withfluidloads}
can be reformulated to read
\begin{subequations}
\begin{gather}
    (\bdensity\bladeA + \fdensity\forcingcrosssectionarea ) \ttderiv{\vectorsym{\centerlinepos}} 
        = \rhsvecforce + \modexternalforcedensityrelfsinterface 
    \\
    (\bdensity\crosssectiontensorofinertia + \fdensity\forcingcrosssectiontensorofinertia)\bcdot\tderiv{\vectorsym{\angvel}}
        + \vectorsym{\angvel}\times(\bdensity\crosssectiontensorofinertia+\fdensity\forcingcrosssectiontensorofinertia)\bcdot\vectorsym{\angvel}
        = \rhsvecomega + \modexternaltorquedensityrelfsinterface
    \eqdot
\end{gather}
\end{subequations}
Here, parts of the coupling load densities that have the effect of inertia
are added to the inertial terms on the left hand side,
obtained in \cref{eq:structure_eom_w_coupling_final_forcingcrosssectioninertia} of \cref{sec:fluid-structure-coupling_appendix},
\begin{equation}
    \forcingcrosssectionarea = \int\limits_{\forcingcrosssectionspace}\!\dSurface
    ,\qquad
    \forcingcrosssectiontensorofinertia = \int\limits_{\forcingcrosssectionspace}\!\transposed{\skewmatrix{\vectorsym{\relpos}^{(\rkstep-1)}}}\bcdot\skewmatrix{\vectorsym{\relpos}^{(\rkstep-1)}}\,\dSurface
    \eqdot
\end{equation}
The remaining parts of 
$\externalforcedensityrelfsinterface$ and
$\externaltorquedensityrelfsinterface$
are kept in 
$\modexternalforcedensityrelfsinterface$ and
$\modexternaltorquedensityrelfsinterface$,
both of which are known after solving \cref{eq:RKscheme_uprelimvec} and 
taken constant during the collision.
The definitions of these terms are given in equations~\eqref{eq:structure_eom_coupled2} of the appendix.

Adding the, yet unknown, collision impulse to the right-hand sides of the equations gives
\begin{subequations}\label{eq:EOM_with_fluid_mass_and_collisionloads}
\begin{gather}
    (\bdensity\bladeA + \fdensity\forcingcrosssectionarea ) \ttderiv{\vectorsym{\centerlinepos}} 
        = \rhsvecforce + \modexternalforcedensityrelfsinterface + \diraccollision\dptot
    \\
    (\bdensity\crosssectiontensorofinertia + \fdensity\forcingcrosssectiontensorofinertia)\bcdot\tderiv{\vectorsym{\angvel}}
        + \vectorsym{\angvel}\times(\bdensity\crosssectiontensorofinertia+\fdensity\forcingcrosssectiontensorofinertia)\bcdot\vectorsym{\angvel}
        = \rhsvecomega + \modexternaltorquedensityrelfsinterface + \diraccollision(\contactpointvec\times\dptot)
    \eqcomma
\end{gather}
\end{subequations}
which differs from \cref{eq:EOM_with_collisionloads}, 
    due to the presence of fluid-structure coupling terms.
Equations \eqref{eq:EOM_with_fluid_mass_and_collisionloads} are integrated along the length of the collision element
and over the time step during which the collision occurs.
This results in
\begin{subequations}\label{eq:CMEOMIntegrated_wet}
    \begin{equation}
        {\mathsmaller\Delta}\tderiv{\vectorsym{\centerlinepos}}_\masterelement
            =
            \relcollisionelement{{\collisionelementmass}}{\collisionelement}^{-1} \left( \dt \forceoncollisionelement_\masterelement + \dptot \right)
        ,\qquad
        {\mathsmaller\Delta}\vectorsym{\angvel}_\masterelement
            =
            \tensorofinertia_\masterelement^{-1} \bcdot
            \left( \dt\torqueoncollisionelement_\masterelement + \contactpointvec\times\dptot \right)
        \eqcomma
    \end{equation}
as in equations~\eqref{eq:CMEOMIntegrated_dry}, 
but with modified parameters
    \begin{gather}
        \label{eq:CMEOMIntegrated_wet_inertia}
        \relcollisionelement{{\collisionelementmass}}{\collisionelement} 
        	\coloneqq \int\limits_{\hspace{-1ex}\mathrlap{\arclength_\masterelement - \bladeLc/2}}^{\hspace{-1ex}\mathrlap{\arclength_\masterelement + \bladeLc/2}}\!(\bdensity\bladeA + \fdensity\forcingcrosssectionarea)\,\tdiff{\arclength}
        ,\qquad
        \tensorofinertia_\masterelement \coloneqq \int\limits_{\hspace{-1ex}\mathrlap{\arclength_\masterelement - \bladeLc/2}}^{\hspace{-1ex}\mathrlap{\arclength_\masterelement + \bladeLc/2}}\!(\bdensity\crosssectiontensorofinertia + \fdensity\forcingcrosssectiontensorofinertia)\,\tdiff{\arclength}
        \\
        \label{eq:CMEOMIntegrated_wet_loads}
        \forceoncollisionelement_\masterelement 
            \coloneqq 
            \int\limits_{\hspace{-1ex}\mathrlap{\arclength_\masterelement - \bladeLc/2}}^{\hspace{-1ex}\mathrlap{\arclength_\masterelement + \bladeLc/2}}\! ( \rhsvecforce + \modexternalforcedensityrelfsinterface ) \,\tdiff{\arclength}
        ,\qquad
        \torqueoncollisionelement_\masterelement 
            \coloneqq
            - \dt^{-1}\int\limits_{\dt}\!\vectorsym{\angvel}_\masterelement \times \tensorofinertia_\masterelement \bcdot\vectorsym{\angvel}_\masterelement\,\tdiff{t}
            + \int\limits_{\hspace{-1ex}\mathrlap{\arclength_\masterelement - \bladeLc/2}}^{\hspace{-1ex}\mathrlap{\arclength_\masterelement + \bladeLc/2}}\!( \rhsvecomega + \modexternaltorquedensityrelfsinterface )\,\tdiff{\arclength}
        \eqdot
    \end{gather}
\end{subequations}
The integrals above are approximated using the discrete values associated with the structure segments.
This is addressed in the following sections.

%% file: chapters/collision-model/chapters/revision-collisionelementinertia.tex
%
The inertial terms in \cref{eq:CMEOMIntegrated_wet_inertia} 
depend only on the geometry of the 
blade and can therefore be determined fairly easily.
For blades of constant width and constant node spacing, 
they can also be related to the corresponding inertial terms of a structure element
through
{\let\structureelement\masterelement
\begin{subequations}\label{eq:CMinertiasMod}
	\begin{gather}
		\relcollisionelement{{\collisionelementmass}}{\collisionelement} 
		= 
		\frac{\bladeLc}{\bladeLe}
		\Big(
		\mass\eind
		+ \relstructureelement{\forcingmass}{\structureelement}
		\Big)
		\\ 
		\tensorofinertia_\masterelement 
		=
		\frac{\bladeLc}{\bladeLe}
		\left(
		\left( \relstructureelement{\rotationmatrix}{\structureelement} \bcdot {\relref{\tensorofinertia}}\eind \right) \bcdot \transposed{\relstructureelement{\rotationmatrix}{\structureelement}}
		+
		\relstructureelement{\forcingtensorofinertia}{\structureelement}
		\right)
		,\qquad
		{\relref{\tensorofinertia}}_\masterelement
		= 
		\frac{\bladeLc}{\bladeLe}
		\Big( {\relref{\tensorofinertia}}\eind 
		+ \transposed{\relstructureelement{\rotationmatrix}{\structureelement}}\bcdot(\relstructureelement{\forcingtensorofinertia}{\structureelement}\bcdot\relstructureelement{\rotationmatrix}{\structureelement})
		\Big)
		\eqcomma
	\end{gather}
\end{subequations}
with $\mass\eind$ from \cref{eq:collisionelementinertia_dry},
$\relstructureelement{\forcingmass}{\structureelement}$ and
$\relstructureelement{\forcingtensorofinertia}{\structureelement}$ from \cref{eq:forcinginertias},
and
$\relstructureelement{{\relref{\tensorofinertia}}}{\structureelement}$ 
from \cref{eq:tensorofinertia_original}.
}
The terms in \cref{eq:CMinertiasMod} describe the inertia of a freely mobile structure segment,
unaffected by any constraints imposed by the remainder of the respective structure.
As outlined in \cref{sec:singlecollision} for a single collision,
the collision model determines collision impulses based on the expected relative motion of colliding elements
by incorporating this inertia.
This is done explicitly before updating the motion of the structures.
The structure solver, on the other hand,
does account for the internal loads between \usuk{neighboring}{neighbouring} structure elements,
which restrict the motion of the individual colliding element.
This discrepancy between the collision element conceived as an isolated, mobile body by the collision model,
and the colliding element integrated into a longer structure by the structure solver
can lead to poor predictions of the collision model.

\begin{figure}[bth]
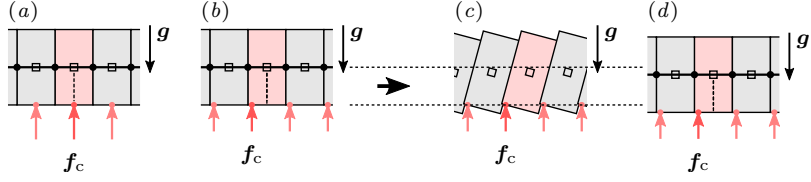
%
  \begingroup%
  \newcommand{\datapath}{figures/}%
  \includestandalone[]{\datapath/planar_issue_offcenter}%
  \endgroup
    \caption{Prototypical situation involving a straight structure segment 
        subject to gravity, 
        that collides laterally with a wall underneath.
        The wall is not shown, 
            and red arrows instead indicate the determined collision forces $\collisionforce$.
        \subcaptionmarker{a}~Ideal situation with contact points located underneath the element \usuk{centers}{centres}.
        \subcaptionmarker{b--d}~Realistic setting, in which detected contact points are \usuk{off-centered}{off-centred} 
            due to, \usuk{e.g.,}{e.g.} an ever so slight rotation of the structure to one side.
            \subcaptionmarker{b}~Initial situation;
            \subcaptionmarker{c}~hypothetical position of collision elements when allowed to rotate freely;
            \subcaptionmarker{d}~resulting displacement of the structure.}
    \label{fig:planar_issue_offcenter}
\end{figure}

A prototypical example where this plays a role is depicted in \cref{fig:planar_issue_offcenter}.
Here, a structure is in static contact with a flat surface.
Since interior force and torque terms are zero, 
    the expected collision force $\collisionforce$ is identical for all elements,
    compensating the gravitational force acting on each element.
When the contact points are underneath the element \usuk{centers}{centres}, as in \cref{fig:planar_issue_offcenter}a,
    this poses no problem to the collision model, and impulses (or forces) needed to maintain the equilibrium are correctly determined.
However, when contact points are \usuk{off-centered}{off-centred}, as in \cref{fig:planar_issue_offcenter}b, 
the collision model expects the individual element to rotate around their contact points (\cref{fig:planar_issue_offcenter}c).
Since this would result in the element \usuk{centers}{centres} translating towards the wall,
a smaller collision force than in \cref{fig:planar_issue_offcenter}a is determined this way.
As structure elements are not able to rotate individually like solid objects,
    the consequence is that the structure sinks in (\cref{fig:planar_issue_offcenter}d).
The needed collision impulses would actually be no different from those determined \cref{fig:planar_issue_offcenter}a,
    regardless of where the contact point is located along the length of the blade.

An improved \usuk{behavior}{behaviour} may be achieved by 
replacing $\tensorofinertia^{-1}$
with a different matrix 
${{\tensorofinertia\ind{t}}}^{-1}$
based on
$\relref{{\tensorofinertia\ind{t}}}^{-1}$,
in which the components
associated with rotations around any but the skeleton axis are set to zero,
\begin{equation}
	\label{eq:tensorofinertia_t}
	{\tensorofinertia\ind{t}}_\masterelement^{-1}
	\coloneqq
    \left[
	\left( \relstructureelement{\rotationmatrix}{\structureelement} \bcdot {\relref{{\tensorofinertia\ind{t}}}}^{-1} \right) \bcdot \transposed{\relstructureelement{\rotationmatrix}{\structureelement}}
    \right]_\masterelement
	,\qquad
	{\relref{{\tensorofinertia\ind{t}}}}_\masterelement^{-1}
	\coloneqq
	\diag(0,0,\left[{\relref{\tensorofinertia}}_\masterelement^{-1}\right]_{3,3})
	\eqdot
\end{equation}
This enforces infinite rotational inertia with respect to bending axes.
To enable a systematic investigation of this choice, 
the system matrix of a collision element, given in \eqref{eq:CMK},
is decomposed into a linear component, leading to translation, 
a second component that responds with torsional rotation around the spine,
and a third component producing the remaining rotational motion that would result in a bending, \usuk{i.e.,}{i.e.}
\begin{subequations}
	\begin{equation}\label{eq:specializedsystemmatrices}
		\systemmatrix_{\masterelement}
		= {\systemmatrix\ind{lin}}_{\masterelement} 
		+ {\systemmatrix\ind{t}}_{\masterelement}
		+ {\systemmatrix\ind{b}}_{\masterelement}
    \end{equation}
with
    \begin{align}
        {\systemmatrix\ind{lin}}_\masterelement 
		&\coloneqq \Bigl[\identity\,\mass^{-1}\Bigr]_\masterelement
		\\
		{\systemmatrix\ind{t}}_\masterelement
		&\coloneqq
		\Bigl[\transposed{ \skewmatrix{\contactpointvec} } \bcdot \left(\tensorofinertia\ind{t}^{-1}\bcdot\skewmatrix{\contactpointvec}\right)\Bigr]_\masterelement
		\\
		{\systemmatrix\ind{b}}_\masterelement 
		&\coloneqq \Bigl[\transposed{ \skewmatrix{\contactpointvec} } \bcdot \left(\tensorofinertia^{-1}\bcdot\skewmatrix{\contactpointvec}\right)\Bigr]_\masterelement
		- {\systemmatrix\ind{t}}_\masterelement
	\end{align}
\end{subequations}
This modification results in
\begin{equation}
    \systemmatrix_{\masterelement}
    \overset{\mathrm{mod.}}{=} {\systemmatrix\ind{lin}}_{\masterelement} 
    + {\systemmatrix\ind{t}}_{\masterelement}
\end{equation}
and is assessed in \cref{sec:collision-model_configurations} below.%

%% file: chapters/collision-model/chapters/revision-vrelvecw.tex
%
The staggered spatial discretization of the structures imposes restrictions
not reflected by \cref{eq:CMEOMIntegrated_dry,eq:CMEOMIntegrated_wet}.
Specifically, discrete values of $\rhsvecforce$,
    defined in \cref{eq:str_EOM_numerical_rhsvecforce},
    are only available at nodal positions $\arclength_{\structureelement\pm\sfrac{1}{2}}$
while discrete values of $\rhsvecomega$,
    defined in \cref{eq:rhsvecomega_rhsvecquaternion,eq:str_EOM_numerical_rhsvectorque},
    can only be determined at element \usuk{centers}{centres} $\arclength_{\structureelement}$.
Hence, the discrete accelerating force $\forceoncollisionelement$ in \cref{eq:CMEOMIntegrated_wet_loads}
can only be element-\usuk{centered}{centred} segment of length $\bladeLcf$
\begin{equation}\label{eq:bladeLcf}
    \bladeLcf = 2 k \bladeLe
    ,\qquad
    k \in \Nat
    \eqdot
\end{equation}
The same holds for the torque $\torqueoncollisionelement$,
also in \cref{eq:CMEOMIntegrated_wet_loads},
\begin{equation}\label{eq:bladeLcm}
    \bladeLcm = (2 l - 1) \bladeLe
    ,\qquad l \in \Nat
    \eqdot
\end{equation}
\Cref{fig:colliding_element_evaluation_lengthsa,fig:colliding_element_evaluation_lengthsb}
provide two examples.
\begin{figure}[htb]
  \begingroup%
  \newcommand{\datapath}{figures/}%
  \includestandalone[]{\datapath/colliding_element_evaluation_lengths}%
  \endgroup
    {\phantomsubcaption\label{fig:colliding_element_evaluation_lengths0}}
    {\phantomsubcaption\label{fig:colliding_element_evaluation_lengthsa}}
    {\phantomsubcaption\label{fig:colliding_element_evaluation_lengthsb}}
    \caption{Illustration of a structure with an element highlighted 
        and possible segments for the evaluation of loads \usuk{centered}{centred} around this element.
        \subref{fig:colliding_element_evaluation_lengths0}~Structure with element marked;
        \subref{fig:colliding_element_evaluation_lengthsa}~$\bladeLcf$ with $k=2$ in \cref{eq:bladeLcf} and 
                                                           $\bladeLcm$ with $l=1$ in \cref{eq:bladeLcm};
        \subref{fig:colliding_element_evaluation_lengthsb}~same, but for a different evaluation length $\bladeLcm$ with $l=3$.
    }
\end{figure}
\newcommand{\spos}{\relstructureelement{\arclength}{\masterelement}}%
With separate integration lengths $\bladeLcf$ and $\bladeLcm$ defined,
the loads $\forceoncollisionelement$, $\torqueoncollisionelement$
in \cref{eq:CMEOMIntegrated_dry,eq:CMEOMIntegrated_wet}
are computed with adapted expressions
\begin{equation}
    \forceoncollisionelement_\masterelement
    \overset{!}{=}
    \frac{\bladeLc}{\bladeLcf}
    \int\limits_{\spos\mathrlap{-\bladeLcf/2}}^{\spos\mathrlap{+\bladeLcf/2}}\! ( \rhsvecforce + \modexternalforcedensityrelfsinterface ) \,\dCurve
    ,\qquad
    \torqueoncollisionelement_\masterelement
    \overset{!}{=}
    \frac{\bladeLc}{\bladeLcm}
    \int\limits_{\spos\mathrlap{-\bladeLcm/2}}^{\spos\mathrlap{+\bladeLcm/2}}\!( \rhsvecomega + \modexternaltorquedensityrelfsinterface )\,\dCurve
    \eqcomma
\end{equation}
where $\spos$ provides the arc-length coordinate of the structure element 
\usuk{centered}{centred} at the collision element $\masterelement$.

Moreover, specialized versions of the collision model proposed here 
employ reduced variants of $\vrelvecw$ in \cref{eq:dptot_dpv_dpw}, 
by replacing the definition in \cref{eq:CMvwn}.
They are derived by decomposing the torque into a torsional part and a bending part,
\begin{equation}
    \label{eq:torque_on_collision_element_decomposed}
    \relstructureelement{\torqueoncollisionelement}{\masterelement}
    = \relstructureelement{[\torqueoncollisionelement\ind{t} + \torqueoncollisionelement\ind{b}]}{\masterelement}
    ,\qquad
    \relstructureelement{{\torqueoncollisionelement\ind{t}}}{\masterelement}
    \coloneqq \relstructureelement{[\left(\skeletonvector\bcdot\torqueoncollisionelement\right)\skeletonvector]}{\masterelement}
    ,\quad
    \relstructureelement{{\torqueoncollisionelement\ind{b}}}{\masterelement} 
    \coloneqq \relstructureelement{[\torqueoncollisionelement - \torqueoncollisionelement\ind{t}]}{\masterelement}
    \eqcomma
\end{equation}
with $\relstructureelement{\skeletonvector}{\masterelement}$ the tangential vector of the skeleton line at the element \usuk{center}{centre}.
Similar to the elemental system matrix $\systemmatrix_{\masterelement}$ above,
the change of contact point velocity due to external loads in \cref{eq:CMvwn}
is decomposed into a translational part, a bending rotation, and a torsional rotation,
\begin{subequations}
    \begin{gather}
        {\vrelvecw}_\masterelement
        = {{\vrelvecw}\ind{lin}}_\masterelement
        + {{\vrelvecw}\ind{b}}_\masterelement
        + {{\vrelvecw}\ind{t}}_\masterelement
        \\
        \begin{align}
            {{\vrelvecw}\ind{lin}}_\masterelement &\coloneqq \dt \Big[ \mass^{-1}\forceoncollisionelement                                                             \Big]_\masterelement \\
            {{\vrelvecw}\ind{b}}_\masterelement   &\coloneqq \dt \Big[ \left(\tensorofinertia^{-1}\bcdot\torqueoncollisionelement\ind{b}\right)\times\contactpointvec \Big]_\masterelement \\
            {{\vrelvecw}\ind{t}}_\masterelement   &\coloneqq \dt \Big[ \left(\tensorofinertia^{-1}\bcdot\torqueoncollisionelement\ind{t}\right)\times\contactpointvec \Big]_\masterelement
            \eqcomma
        \end{align}
    \end{gather}
\end{subequations}
with the loads and the contact point vector $\contactpointvec$ as indicated in \cref{fig:singlecollision_oldmodel}.

\let\spos\undefined

%% file: chapters/collision-model/chapters/imposingcollisionimpulses.tex
%
With all collision impulses computed, 
    the last step involves the distribution of these collision impulses to structure elements,
    such that they are included in the right-hand sides of the discretized \ac{EOM}
    \eqref{eq:str_EOM_numerical_lin} and \eqref{eq:str_EOM_numerical_ang}.
    For a single collision, these terms then represent the analytical expressions
    $\diraccollision\dptot$ and $ \diraccollision(\contactpointvec\times\dptot)$
    in \cref{eq:EOM_with_fluid_mass_and_collisionloads}.
To this end, 
effective collision forces and torques are computed,
\begin{equation}
    {\collisionforce}_\masterelement  
    \coloneqq 
    \begin{cases}
        \vectorsym{\linmom}_\masterelement/\dt & \text{if $\masterelement\in\CollPairsNC$}\\
        0                                      & \text{otherwise}
    \end{cases}
    ,\qquad
    {\collisiontorque}_\masterelement 
    \coloneqq 
    \begin{cases}
        \vectorsym{\angmom}_\masterelement/\dt & \text{if $\masterelement\in\CollPairsNC$}\\
        0                                      & \text{otherwise}
    \end{cases}
    \eqcomma
\end{equation}
and distributed along the length of the collision element in \cref{eq:str_EOM_numerical}
in accordance with the staggered discretization of the corresponding structure.
The following three variants were considered and are assessed in \cref{sec:collision-model_configurations}.

\begin{enumerate}[label=(\alph*)]
\item In the simplest manner these effective loads
are distributed to 
$[\darclength\externalforcedensity]_{\structureelement\pm\sfrac{1}{2}}$
and
$[\darclength\externaltorquedensity]_{\structureelement}$,
\usuk{i.e.,}{i.e.} the nearest nodes and the nearest element \usuk{center}{centre}, respectively.
The corresponding equation reads
\begin{equation}\label{eq:redistributing_collision_forces_old}
   \relstructureelement{\Big[\darclength\,\externalforcedensity\Big]}{\structureelement+\sfrac{1}{2}}
   = \tfrac{1}{2}{\collisionforce}_{\masterelement}   
   + \tfrac{1}{2}{\collisionforce}_{\masterelement+1} 
   ,\qquad
   \relstructureelement{\Big[\darclength\,\externaltorquedensity\Big]}{\structureelement}
   = {\collisiontorque}_\masterelement  
   \eqdot
\end{equation}

\item Alternatively, and similar to the treatment of accelerating torque due to external loads in 
\cref{eq:torque_on_collision_element_decomposed},
the collision torque is further decomposed into a twisting component and one that would invoke a bending,
\begin{equation}
	{\collisiontorque} = {{\collisiontorque}\ind{t}} + {{\collisiontorque}\ind{t}}
	,\qquad
	{{\collisiontorque}\ind{t}} \coloneqq ( {\collisiontorque} \bcdot \skeletonvector ) \skeletonvector  
	,\qquad
	{{\collisiontorque}\ind{b}} \coloneqq {\collisiontorque} - {{\collisiontorque}\ind{t}}  
	\eqdot
\end{equation}
The bending torque may, then, be imposed through forces acting on the structure nodes adjacent to the element \usuk{center}{centre},
with opposite signs, such that the total force and the total torque are maintained,
\begin{subequations}\label{eq:redistributing_collision_forces_Mcv2}
    \begin{alignat}{2}
    &\relstructureelement{\Big[\darclength\externalforcedensity \Big]}{\structureelement+\sfrac{1}{2}}
    &&= \left( \frac{\relcollisionelement{{\collisionforce}}{\collisionelement  }}{2} + \frac{\relcollisionelement{{{\collisiontorque}\ind{b}}}{\collisionelement  }\times\relstructureelement{\skeletonvector}{\structureelement  }}{\vnorm{\relstructureelement{\vectorsym{\centerlinepos}}{\structureelement+\sfrac{1}{2}} - \relstructureelement{\vectorsym{\centerlinepos}}{\structureelement-\sfrac{1}{2}}}} \right)  
      + \left( \frac{\relcollisionelement{{\collisionforce}}{\collisionelement+1}}{2} - \frac{\relcollisionelement{{{\collisiontorque}\ind{b}}}{\collisionelement+1}\times\relstructureelement{\skeletonvector}{\structureelement+1}}{\vnorm{\relstructureelement{\vectorsym{\centerlinepos}}{\structureelement+\sfrac{3}{2}} - \relstructureelement{\vectorsym{\centerlinepos}}{\structureelement+\sfrac{1}{2}}}} \right)  
    \\
    &\relstructureelement{\Big[\darclength\externaltorquedensity\Big]}{\structureelement}
    &&= \relcollisionelement{{{\collisiontorque}\ind{t}}}{\collisionelement} 
    \eqdot
\end{alignat}
\end{subequations}

\item In a third variant of the distribution procedure,
the now separated bending torque, is simply ignored,
    \begin{subequations}\label{eq:redistributing_collision_forces_nobending}
    \begin{alignat}{2}
       &\relstructureelement{\Big[\darclength\externalforcedensity\Big]}{\structureelement+\sfrac{1}{2}}
       &&= \tfrac{1}{2}{\collisionforce}_{\masterelement  }  
         + \tfrac{1}{2}{\collisionforce}_{\masterelement+1}  
       \\
       &\relstructureelement{\Big[\darclength\externaltorquedensity\Big]}{\structureelement}
       &&= {{\collisiontorque}\ind{t}}_\masterelement  
       \eqdot
    \end{alignat}
    \end{subequations}

\end{enumerate}

%% file: chapters/collision-model/chapters/tests-introduction.tex

The objective of this section is to 
evaluate possible design choices in the formulation of the collision model,
to obtain a variant that can be reliably applied in simulations involving massively colliding flexible Cosserat rods.
To this end, a number of test cases are presented and simulated.
A first set covers single collisions in very simple edge cases (Ss1--Ss5).
Another set of cases (Sm1--Sm2) addresses situations with many parallel collisions
in equally simple set-ups.
Lastly, complex cases (CS1, CR1) and entire canopy simulations (CC1, CC2)
are assessed qualitatively to ensure that the obtained model variant,
indeed, is capable of handling the targeted applications.

Tested model variants are given in  
\cref{tab:subconfigurations_final},
starting with a base configuration \modelvariant{bs11}
which corresponds to the method of \cite{TschisgaleConstraintbased2019},
and which is summarized in \cref{sec:collision-model-concept}, above. 
Going through the tests
Ss1, Ss2, Sm2, Ss3, Ss4 and Ss5 in \cref{sec:tests-simple}, 
the model was eventually refined into variant \modelvariant{nb22d} which passes all these tests.
A further adjustment of the way competing collisions are sorted out
was required when simulating test case Sm1 in \cref{sec:Sm1},
resulting in the model variant termed \modelvariant{final}.
This version was then further tested in the more complex test cases
presented in \cref{sec:tests-complex},
some of which also involve a coupled fluid phase.

Unless indicated differently, collisions were treated as fully inelastic, \usuk{i.e.,}{i.e.} $\restitutioncoefficient = 0$ in \cref{eq:dptot_dpv_dpw},
and tangential friction was turned off,
\usuk{i.e.,}{i.e.} $\frictioncoeffstatic = \frictioncoeffdynamic = 0$ in \cref{eq:CoulombFrictionModel}.
Some test case definitions list correction factors for shearing 
    $\bladeks{} = \bladeks{1} = \bladeks{2} \in [0\dots1]$, 
and warping 
    $\bladekt \in [0\dots1]$, 
both in \cref{eq:Hookean_matrices}.
These factors were determined according to \cite[§\,4.2]{FreundWarping2016}.
If no factor is provided in the description of a test case, a value of $1$ was employed.
Likewise, damping was not employed, unless either 
    the damping parameters of \cref{eq:damping_matrices_} are listed,
    or a damping ratio $\dampingratio$ is provided.
    In the latter case, the retardation times for shearing and elongation were assumed equal,
    \usuk{i.e.,}{i.e.} $\tau\ind{s} = \tau\ind{e}$,
    their value determined with \cref{eq:retardation_time_dampingratio} and inserted in 
    \cref{eq:damping_matrices_}.
Some of the test cases are static cases in which the colliding blades do not translate (after collision),
and possibly also do not deform, because movement is inhibited by the collision model.
Static contact is a particular case of a collision presenting its own challenges
\cite{KempeRelevance2014}.  
The situation is addressed here to ensure that also this challenge is successfully met.

\begin{sidewaystable}%
    \newcommand{\ElementlengthLinLoadsText}{$\bladeLcf/\bladeLe$}%
    \newcommand{\ElementlengthRotLoadsText}{$\bladeLcm/\bladeLe$}%
    \newcommand{\ElementlengthLinInertiaText}{$\relcollisionelement{{\collisionelementmass}}{\collisionelement}^{-1}$ evaluated for $\bladeLc/\bladeLe=\dots$}%
    \newcommand{\ElementlengthRotInertiaText}{$\tensorofinertia_\masterelement^{-1}$ evaluated for $\bladeLc/\bladeLe=\dots$}%
    \newcommand{\OtherRotInertia}{$-(\vectorsym{\angvel}_\masterelement \times \tensorofinertia_\masterelement \cdot\vectorsym{\angvel}_\masterelement)$}%
    \newcommand{\OtherRotInertiaText}{\OtherRotInertia{} accounted for}%
    \newcommand{\dnNoncompetingText}{$\dnNoncompeting$}%
    \newcommand{\DistNoncompetingText}{$\distNoncompeting/\bladeLe$}%
    \newcommand{\vvWithBendingRotationText}{${\vrelvecv}_0$ with bending rotation}%
    \newcommand{\vvWithTorsionalRotationText}{${\vrelvecv}_0$ with torsional rotation}%
    \newcommand{\vwWithBendingRotationText}{${\vrelvecw}_0$ with bending rotation}%
    \newcommand{\vwWithTorsionalRotationText}{${\vrelvecw}_0$ with torsional rotation}%
    \newcommand{\WithBendingInertiaText}{coll. el. may rotate around bending axis} 
    \newcommand{\WithTorsionalInertiaText}{coll. el. may rotate around torsional axis} 
    \newcommand{\CollisionBendingTorqueText}{coll. imposes bending torque \dots; via \dots}%
    \newcommand{\SpreadForceText}{coll. force imposed on \dots nodes}%
    \newcommand{\SpreadTorqueText}{coll. torque imposed on \dots elements}%
    \newcommand{\adfCrossCompensation}{force cross compensation}%
    \newcommand{\admCrossCompensation}{torque cross compensation}%
    \newcommand{\columnTextnobrot}{%
        $\displaystyle%
        \text{$\vrelvecv$ with bending rotation} = \begin{cases}%
            \text{yes} & \Rightarrow\quad {\vrelvecv}_\masterelement = \tderiv{\centerlinepos}_\masterelement +  \vectorsym{\angvel}_\masterelement                     \times \contactpointvec \\%
            \text{no}  & \Rightarrow\quad {\vrelvecv}_\masterelement = \tderiv{\centerlinepos}_\masterelement + (\vectorsym{\angvel}_\masterelement\cdot\dThree)\dThree \times \contactpointvec%
        \end{cases}$%
    }%
    \newcommand{\columnTextDist    }[1][2.5]{                \multirow{#1}{*}{\rotatebox[origin=c]{90}{rem.\tnote{a}}}}%
    \newcommand{\columnTextCompvw  }[1][5.0]{                \multirow{#1}{*}{\rotatebox[origin=c]{90}{$\vrelvecw$\tnote{b}}}}%
    \newcommand{\columnTextCompvv  }[1][2.5]{                \multirow{#1}{*}{\rotatebox[origin=c]{90}{$\vrelvecv$}}}%
    \newcommand{\columnTextK       }[1][5.0]{                \multirow{#1}{*}{\rotatebox[origin=c]{90}{$\systemmatrix$\tnote{d}}}}%
    \newcommand{\columnTextSpread  }[1][3.5]{                \multirow{#1}{*}{\rotatebox[origin=c]{90}{distr.\tnote{e}}}}%
    \newcommand{\columnTextCC      }[1][2.5]{                \multirow{#1}{*}{\rotatebox[origin=c]{90}{CC\tnote{f}}}}%
    \newcommand{\no}{\textcolor{gray}{no}}%
    \newcommand{\yes}{yes}%
    \newcommand{\dnm}{\textcolor{gray}{---}}%
    \newcommand{\ViaForces}{$\collisionforce$}%
    \newcommand{\ViaTorque}{$\collisiontorque$}%
    \newcommand{\SysMatrixLinText}{${\systemmatrix\ind{lin}}_{\masterelement}$ evaluated for $\bladeLc/\bladeLe=\dots$}%
    \newcommand{\SysMatrixRotText}{${\systemmatrix\ind{rot}}_{\masterelement}$ evaluated for $\bladeLc/\bladeLe=\dots$}%
    \begin{threeparttable}[t]%
        \newcommand{\shortermidrule}{\cmidrule{2-10}}
        \begin{tabular}{ c | l | c | c | c | c | c | c | c | c }
            \toprule
            \multicolumn{2}{l|}{\textbf{Variable}}           & \textbf{bs11} & \textbf{bs12} & \textbf{bs21} & \textbf{bs22} & \textbf{nb22a} & \textbf{nb22b} & \textbf{nb22c} & \textbf{nb22d} \\ \midrule[\heavyrulewidth]
            \columnTextDist   & \dnNoncompetingText          & 1             & 1             & 2             & 2             & 2              & 2              & 2              & 2              \\
            &                              &               &               &               &               &                &                &                &                \\ \midrule
            \columnTextCompvw & \ElementlengthLinLoadsText   & 1             & 2             & 1             & 2             & 2              & 2              & 2              & 2              \\
            \shortermidrule   & \ElementlengthRotLoadsText   & 1             & 1             & 1             & 1             & 1              & 1              & 1              & 1              \\
            & \vwWithBendingRotationText   & \yes          & \yes          & \yes          & \yes          & \yes           & \no            & \no            & \no            \\
            & \vwWithTorsionalRotationText & \yes          & \yes          & \yes          & \yes          & \yes           & \yes           & \yes           & \yes           \\ \midrule
            \columnTextCompvv & \vvWithBendingRotationText   & \yes          & \yes          & \yes          & \yes          & \yes           & \no            & \no            & \no            \\
            & \vvWithTorsionalRotationText & \yes          & \yes          & \yes          & \yes          & \yes           & \yes           & \yes           & \yes           \\ \midrule
            \columnTextK      & \SysMatrixLinText            & 1             & 1             & 2             & 2             & 2              & 2              & 2              & 2              \\
            \shortermidrule   & \SysMatrixRotText            & 1             & 1             & 2             & 2             & 2              & 2              & 2              & 2              \\
            & \WithBendingInertiaText      & \yes          & \yes          & \yes          & \yes          & \no            & \no            & \no            & \no            \\
            & \WithTorsionalInertiaText    & \yes          & \yes          & \yes          & \yes          & \yes           & \yes           & \yes           & \yes           \\ \midrule
            \columnTextSpread & \CollisionBendingTorqueText  & \ViaTorque    & \ViaTorque    & \ViaTorque    & \ViaTorque    & \ViaTorque     & \ViaTorque     & \no            & \no            \\
            & \SpreadForceText             & 2             & 2             & 2             & 2             & 2              & 2              & 2              & 2              \\
            & \SpreadTorqueText            & 1             & 1             & 1             & 1             & 1              & 1              & 1              & 2              \\ \bottomrule
        \end{tabular}
        \begin{tablenotes}%
            \item[a] See \vref{sec:fix_removal_competing_collisions}.%
            \item[b] See \vref{sec:vrelvecw}.%
            \item[d] See \vref{sec:collisionelementinertia}.%
            \item[e] See \vref{sec:imposecollision}, with%
            \enquote{$\collisiontorque$}~\cref{eq:redistributing_collision_forces_old};%
            \enquote{$\collisionforce$}~\cref{eq:redistributing_collision_forces_Mcv2};%
            \enquote{no}~\cref{eq:redistributing_collision_forces_nobending}.%
        \end{tablenotes}%
    \end{threeparttable}%
    \caption{Variants of the collision model assessed in the tests.}
    \label{tab:subconfigurations_final}
\end{sidewaystable}

%% file: chapters/collision-model/chapters/tests-simple.tex

\FloatBarrier
\leveldown{Beam colliding laterally, single contact, bending}
\label{sec:Ss1+Ss2}

The intent of this first test is to verify the correct, stable \usuk{behavior}{behaviour} when a beam is subject to a single collision,
as if it was balanced on the tip of a needle, and without any fluid present.
The arrangement is sketched in \cref{fig:Ss1_Ss2}.
To keep things simple the configuration was designed such that only bending deformation occurs.
The two variants of this test, Ss1 and Ss2, 
are different only by the number of elements used to discretize the structure, 
such that in 
Ss1 the collision occurs underneath the \usuk{center}{centre} of a structure element, 
whereas it occurs underneath a node in Ss2.
The relevant parameters are given in \cref{tab:Ss1_Ss2}.
\begin{figure}[htb]
    \begin{minipage}[c]{45mm}
        \begin{table}[H]
            \begin{tabular}[t]{l r @{\,} l r @{\,} l}
                \toprule
                &    \multicolumn{2}{c}{\textbf{Ss1}}     & \multicolumn{2}{c}{\textbf{Ss2}} \\ \midrule
                $\grav_y$   & \multicolumn{2}{c}{\cref{eq:Ss1_gravy}} &    \multicolumn{2}{c}{(same)}    \\
                $\bladeL$   & $\num{250}$   & $\si{mm}$               &    \multicolumn{2}{c}{(same)}    \\
                $\bladeW$   & $\num{15}$    & $\si{mm}$               &    \multicolumn{2}{c}{(same)}    \\
                $\bladeT$   & $\num{0.063}$ & $\si{mm}$               &    \multicolumn{2}{c}{(same)}    \\
                $\bladeE$   & $\num{250}$   & $\si{MPa}$              &    \multicolumn{2}{c}{(same)}    \\
                $\bdensity$ & $\num{920}$   & $\si{kg.m^{-3}}$        &    \multicolumn{2}{c}{(same)}    \\ \midrule
                $\bladeNe$  & $\num{17}$    &                         & $\num{18}$ &                     \\
                $\dt$       & $\num{1e-4}$  & $\si{s}$                &    \multicolumn{2}{c}{(same)}    \\ \bottomrule
            \end{tabular}
            \caption{Parameters of the test cases Ss1 and Ss2.}
            \label{tab:Ss1_Ss2}
        \end{table}
    \end{minipage}
    \hfil
    \begin{minipage}[c]{68mm}
        \begin{figure}[H]\hspace*{0pt}%
  \begingroup%
  \newcommand{\datapath}{figures/}%
  \includestandalone[]{\datapath/Ss1_Ss2}%
  \endgroup
            {\phantomsubcaption\label{fig:Ss1}}
            {\phantomsubcaption\label{fig:Ss2}}
            \caption{Illustration of the situations in the test cases
                \subref{fig:Ss1}~Ss1; \subref{fig:Ss2}~Ss2.
                The single structure collides at a single point underneath its \usuk{center}{centre} with a static object (not shown).}
            \label{fig:Ss1_Ss2}
        \end{figure}
    \end{minipage}
\end{figure}
To avoid an initial vibration of the elastic beam, 
gravitation is smoothly increased from zero with the only non-zero component defined as
\begin{equation}\label{eq:Ss1_gravy}
    \grav_y\of{t}
    = \SI{-9.81}{m.s^{-2}}
    \begin{cases}
        2 (t / t\ind{0})^2                        & \mathrm{if~} 0 \le t < t\ind{0}/2 \\
        1 - 2 ( \frac{t - t\ind{0}}{t\ind{0}})^2  & \mathrm{if~} t\ind{0}/2 \le t < t\ind{0} \\
        1                                         & \mathrm{if~} t \ge t\ind{0} \coloneqq \SI{1.2}{s} \eqcomma
    \end{cases}
\end{equation}
The exact collision force then is equal to the resulting weight of the structure,
and the height of the \usuk{center}{centre} section of the structure should remain constant.
Below, collision forces and vertical position of the contact point are reported.
The height of the contact point is given relative to its initial height,
which was deliberately placed at some distance inside the collision zone.
As a result the structures remain colliding even as the upper structure moves up a little.

\begin{figure}[htb]
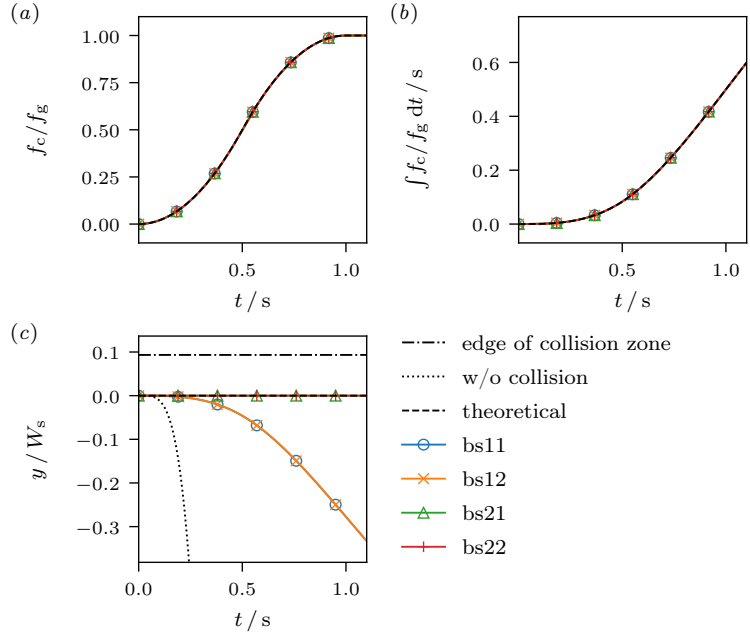

  \begingroup%
  \newcommand{\datapath}{figures/}%
  \includestandalone[]{\datapath/testresults_Ss1}%
  \endgroup
    {\phantomsubcaption\label{fig:Ss1_force}}
    {\phantomsubcaption\label{fig:Ss1_forceintegral}}
    {\phantomsubcaption\label{fig:Ss1_ycenter}}
    \caption{Results for case Ss1.
        \subref{fig:Ss1_force}        ~Instantaneous collision force;
        \subref{fig:Ss1_forceintegral}~time-integrated collision force;
        \subref{fig:Ss1_ycenter}      ~vertical position of the \usuk{center}{centre} of the structure.
        Only model variants 
        bs21, bs2 
        are able to keep the collision distance;
        bs, bs12 fail.
    }
    \label{fig:testresults_Ss1}
\end{figure}

\Cref{fig:testresults_Ss1} reports the results for test configuration Ss1.
Starting from the base variant of the model, \modelvariant{bs11},
the length of the collision element was varied between 
$\dnLc = \bladeLc/\bladeLe = 1$ (\modelvariant{bs11}, \modelvariant{bs12})
and $\dnLc = 2$ (\modelvariant{bs21}, \modelvariant{bs22}),
and
the length associated with forces was varied between
$\bladeLcf = 1 \bladeLe$ (\modelvariant{bs11}, \modelvariant{bs21})
and $2 \bladeLe$ (\modelvariant{bs12}, \modelvariant{bs22}).
All four variants reproduce the correct collision force (\cref{fig:Ss1_force,fig:Ss1_forceintegral}),
but only the variants \modelvariant{bs21} and \modelvariant{bs22}
are able to hold the structure in place.
This indicates the need for $\dnLc = \bladeLc/\bladeLe = 2$, which is also
the minimum length at which element-wise changes of internal forces can be evaluated (c.f.~\cref{sec:vrelvecw}),
and the length of the segment which is affected be the imposed collision force (c.f.~\cref{sec:imposecollision}).

\begin{figure}[htb]
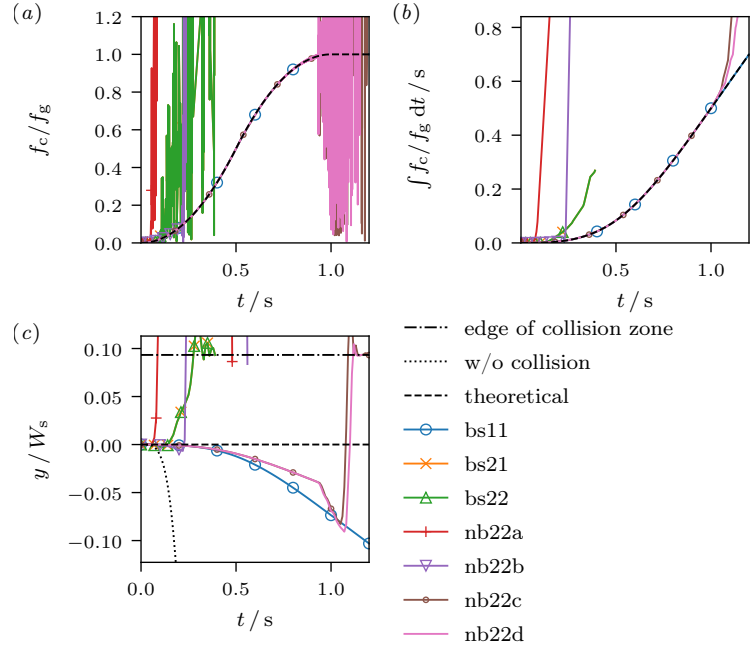

  \begingroup%
  \newcommand{\datapath}{figures/}%
  \includestandalone[]{\datapath/testresults_Ss2}%
  \endgroup
    \caption{Results for case Ss2.
        \subcaptionmarker{a}~Instantaneous collision force;
        \subcaptionmarker{b}~time-integrated collision force;
        \subcaptionmarker{c}~vertical position of the \usuk{center}{centre} of the structure.
    }
    \label{fig:testresults_Ss2}
\end{figure}

Results for the test case Ss2 are reported in \cref{fig:testresults_Ss2}.
Despite having failed Ss1, the base variant \modelvariant{bs11} is reported here again,
because it performs quite well in this case. 
This variant actually produces two collisions underneath the \usuk{center}{centre} node, 
each associated with one of the \usuk{neighboring}{neighbouring} elements.
This, apparently, compensates the deficiency of this variant in Ss1.
The base variants \modelvariant{bs21} and \modelvariant{bs22}, on the other hand,
after being successful in Ss1, fail this case, producing very large collision forces.
This instability was found to be related to the treatment of rotational inertia around the bending axes,
as discussed in \cref{sec:collisionelementinertia} above.
The additional variants \modelvariant{nb22a}, \modelvariant{nb22b}, \modelvariant{nb22c} and \modelvariant{nb22d}
address this issue. Each of them is derived from \modelvariant{bs22}, 
but with rotation of the collision element restricted to rotation around the skeleton axis,
\usuk{i.e.,}{i.e.} inhibiting any bending motion of the colliding segment due to a collision impulse in the surrounding of the contact point.
The four variants differ in the details concerning how rotation due to angular loads is treated,
how rotational velocity is accounted for, and how the torque due to the collision impulse is distributed to the element \usuk{centers}{centres} surrounding the contact point.
The model variants \modelvariant{nb22a} and \modelvariant{nb22b} produce excessive non-physical momentum, shooting the colliding structure upwards.
The variants \modelvariant{nb22c} and \modelvariant{nb22d}
produce satisfactory reactions with the correct collision force being reproduced for $t\le\SI{1}{s}$
and the structure sinking in only slightly.
During this phase one of the two elements surrounding the contact point is in steady contact,
resulting in the collision force being imposed slightly asymmetrically with regard to the \usuk{center}{centre} of the top structure.
At some point the element associated with the collision starts switching back and forth between the two \usuk{neighboring}{neighbouring} elements.
This results in a brief phase of accelerated decent. 
This decent comes to a halt when, again,
only one of the two elements is in steady contact, 
which occurs around $t=\SI{1.05}{s}$ here.
Since the dropping beam has picked up momentum, stopping the vertical motion at its \usuk{center}{centre} causes the beam to flex such that it translates upwards once its ends swing back up.
As a result, the beam is then lifted up to the edge of the collision zone.
Unlike the failing model variants, with \modelvariant{nb22c} and \modelvariant{nb22d}
this vertical displacement is controlled and does not result in the beam being shot upwards.

The variants \modelvariant{nb22c} and \modelvariant{nb22d} differ in how the torque due to collisions is distributed to the structure elements
which is not relevant for the present cases.
As outlined in \cref{sec:imposecollision} both variants do not produce a bending torque,
but they handle the remaining twisting torque differently.
With variant \modelvariant{nb22c} the torque is simply imposed on the collision-associated structure element.
With variant \modelvariant{nb22d}, only half the torque is imposed on this element, and the other half distributed to the \usuk{neighboring}{neighbouring} elements.
While this does not play a role for the cases considered so far,
only variant \modelvariant{nb22d} is able to accurately achieve the correct collision reaction, 
\usuk{e.g.,}{e.g.} when a beam collides parallel with a wall at an angle such that a rigid-body rotation around its skeleton line follows,
a case not reported here.
Hence, model variant \modelvariant{nb22d} is employed from here on.

\FloatBarrier
\levelstay{Beam colliding laterally with a wall} 

\begin{figure}[htb]
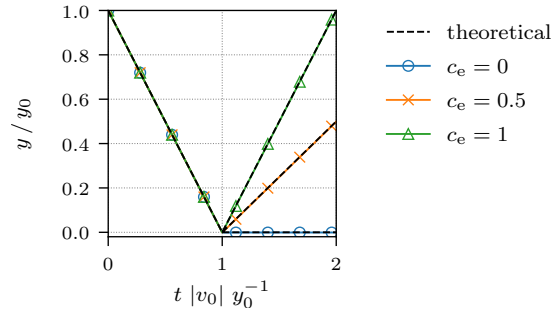

    \begin{minipage}[t]{84mm}
        \begin{figure}[H]
  \begingroup%
  \newcommand{\datapath}{figures/}%
  \includestandalone[]{\datapath/testresults_Sm2}%
  \endgroup
            \caption{Sketch of configurations involving a structure which collides laterally with a wall.}
            \label{fig:testresults_Sm2}
        \end{figure}
    \end{minipage}
\end{figure}

This setup was configured in \cite[§4.1]{TschisgaleConstraintbased2019} 
with a beam colliding flat with a wall as sketched in \cref{fig:Sm2}.
It was reconsidered here 
to verify that the correct \usuk{behavior}{behaviour} is still achieved 
with the present version of the collision model.
The beam was initialized with a downward vertical velocity $v_0<0$ at an arbitrary distance 
$y_0 = \SI{0.5}{\meter}$ above the wall.
Approaching this wall at constant velocity all elements collide simultaneously at $t = -y_0/v_0$,
resulting in a rigid-body collision problem.
The relevant parameters are listed in \cref{tab:Sm2}.
The velocity after collision solely depends on the restitution coefficient $\restitutioncoefficient$
and can be determined analytically.
\Cref{fig:testresults_Sm2} indicates an excellent reproduction of the expected \usuk{behavior}{behaviour}.
\begin{figure}[htb]%
    \begin{minipage}[c]{45mm}
        \begin{table}[H]
            \begin{tabular}[t]{l r @{\,} l}
                \toprule
                &        \multicolumn{2}{c}{\textbf{Sm2}}         \\ \midrule
                $\alpha$                  & $\ang{0}$    &                                  \\
                $\bladeL$                 & $\num{1}$    & $\si{m}$                         \\
                $\bladeW$                 & $\num{0.2}$  & $\si{m}$                         \\
                $\bladeT$                 & $\num{0.05}$ & $\si{m}$                         \\
                $\bdensity$               & $\num{1000}$ & $\si{kg.m^{-3}}$                 \\
                $v_0$                     & $\num{-5}$   & $\si{\meter\per\second}$         \\ \midrule
                $\bladeNe$                & $\num{5}$    &                                  \\
                $\dt$                     & $\num{1e-4}$ & $\si{s}$                         \\
                $\restitutioncoefficient$ & $0, 0.5, 1$  &                                  \\ \bottomrule
            \end{tabular}
            \caption{Parameters of the configuration Sm2 involving a beam which collides laterally with a wall.}
            \label{tab:Sm2}
        \end{table}
    \end{minipage}
    \hfil
    \begin{minipage}[c]{68mm}
        \begin{figure}[H]%
  \begingroup%
  \newcommand{\datapath}{figures/}%
  \includestandalone[]{\datapath/Sm2}%
  \endgroup
            \caption{Sketch of configuration Sm2 involving a beam which collides laterally with a wall.}
            \label{fig:Sm2}
        \end{figure}
    \end{minipage}
\end{figure}

\FloatBarrier
\levelstay{Beam colliding longitudinally, single contact, axial compression}  

{\newcommand{\tend}{T}
    
    This configuration considers the collision of a beam in the direction of its skeleton axis.
    Two cases were considered, Ss3 and Ss4.
    The cases Ss3 is sketched in \cref{fig:Ss3}.
    Here the beam initially touches the obstacle, and gravity oriented against it is gradually increased from zero, 
    \begin{equation}\label{eq:Ss3_gravy}
        \grav_y\of{t}
        = \SI{9.81}{m.s^{-2}}
        \begin{cases}
            2 (t / \tend)^2                        & \mathrm{if~} 0 \le t < \tend/2 \\
            1 - 2 ( \frac{t - \tend}{\tend})^2  & \mathrm{if~} \tend/2 \le t < \tend \\
            1                                         & \mathrm{if~} t \ge \tend \coloneqq \SI{2.5}{s} \eqcomma
        \end{cases}
    \end{equation}
    which is slow enough to neglect inertial effects.
    The exact collision force then should equal the gravitational force acting on the structure.
    The second configuration Ss4 in \cref{fig:Ss4}
    consists of a beam approaching the obstacle at a constant velocity $v_0$, 
    then being compressed while in contact, 
    and moving away from the obstacle when it has fully expanded again.
    This configuration had previously served as a validation case in \cite[§\,4.2]{TschisgaleConstraintbased2019}.
    Relevant parameters are given in \cref{tab:Ss3+Ss4}.
    \begin{figure}[htb]\centerline{%
            \begin{minipage}[c]{50mm}
                \begin{table}[H]
                    \begin{tabular}{l r @{\,} l r @{\,} l}
                        \toprule
                        &    \multicolumn{2}{c}{\textbf{Ss3}}     &   \multicolumn{2}{c}{\textbf{Ss4}}   \\ \midrule
                        $v_0$       & 0            &                          & $\num{5}$ & $\si{\meter\per\second}$ \\
                        $\grav_y$   & \multicolumn{2}{c}{\cref{eq:Ss3_gravy}} & 0         &                          \\
                        $\bladeL$   & $\num{1}$    & $\si{m}$                 &      \multicolumn{2}{c}{(same)}      \\
                        $\bladeA$   & $\num{0.1}$  & $\si{m^2}$               &      \multicolumn{2}{c}{(same)}      \\
                        $\bdensity$ & $\num{1000}$ & $\si{kg.m^{-3}}$         &      \multicolumn{2}{c}{(same)}      \\
                        $\bladeE$   & $\num{0.4}$  & $\si{MPa}$               &      \multicolumn{2}{c}{(same)}      \\ \midrule
                        $\bladeNe$  & $\num{25}$   &                          &      \multicolumn{2}{c}{(same)}      \\
                        $\dt$       & $\num{2e-5}$ & $\si{s}$                 &      \multicolumn{2}{c}{(same)}      \\ \bottomrule
                    \end{tabular}
                    \caption{Parameters of test cases Ss3 and Ss4.}
                    \label{tab:Ss3+Ss4}
                \end{table}
            \end{minipage}
            \hfil
            \begin{minipage}[c]{60mm}
                \begin{figure}[H]
                    \begin{adjustbox}{varwidth=\linewidth}
  \begingroup%
  \newcommand{\datapath}{figures/}%
  \includestandalone[]{\datapath/Ss3_Ss4}%
  \endgroup
                    \end{adjustbox}
                    {\phantomsubcaption\label{fig:Ss3}}
                    {\phantomsubcaption\label{fig:Ss4}}
                    \caption{Sketch of configurations involving a structure that collides longitudinally at one end.
                        \subref{fig:Ss3}~Static case Ss3; 
                        \subref{fig:Ss4}~Dynamic case Ss4, with $t_0 < t_1 < t_2 < t_3$.}
                \end{figure}
        \end{minipage}}
    \end{figure}
    
    In configuration Ss3, the contact force is well reproduced, as can be seen in \cref{fig:Ss3_force}.
    In both configurations, the end of the beam sinks in slightly (\cref{fig:Ss3_y0,fig:testresults_Ss4_y0_detail}),
    depending on the discretization of the problem,
    indicating a first-order accuracy of the model.
    
    \begin{figure}[htb]
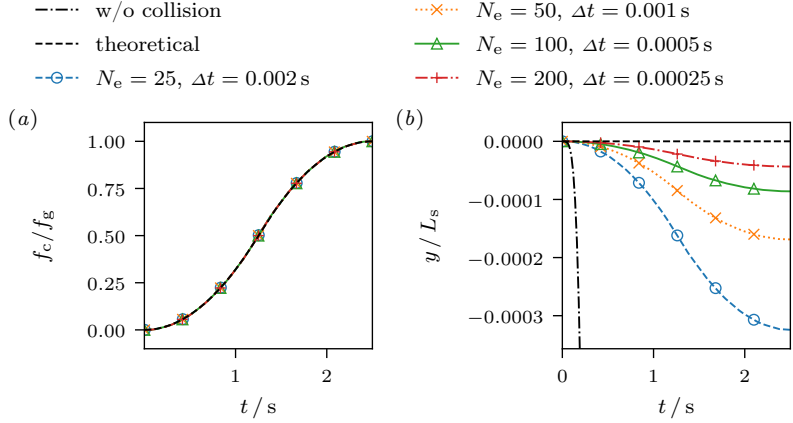

        \begin{adjustbox}{varwidth=\linewidth}
  \begingroup%
  \newcommand{\datapath}{figures/}%
  \includestandalone[]{\datapath/testresults_Ss3}%
  \endgroup
        \end{adjustbox}
        {\phantomsubcaption\label{fig:Ss3_force}}
        {\phantomsubcaption\label{fig:Ss3_y0}   }
        \caption{Results for case Ss3.
            \subref{fig:Ss3_force}~Collision force,
            \subref{fig:Ss3_y0}   ~position of colliding end.
        }
        \label{fig:testresults_Ss3}
    \end{figure}
    
    \begin{figure}[htb]
        \begin{adjustbox}{varwidth=\linewidth}
  \begingroup%
  \newcommand{\datapath}{figures/}%
  \includestandalone[]{\datapath/testresults_Ss4}%
  \endgroup
        \end{adjustbox}
        {\phantomsubcaption\label{fig:testresults_Ss4_y0}}
        {\phantomsubcaption\label{fig:testresults_Ss4_y0_detail}}
        \caption{Results for case Ss4.
            \subref{fig:testresults_Ss4_y0}~Position of colliding tip,
            \subref{fig:testresults_Ss4_y0_detail}~detail of the same plot.
        }
        \label{fig:testresults_Ss4}
    \end{figure}
}

\FloatBarrier
\vspace*{50mm}
\levelstay{Beam colliding laterally with single contact line, bending and torsion}

\begin{figure}[t!]
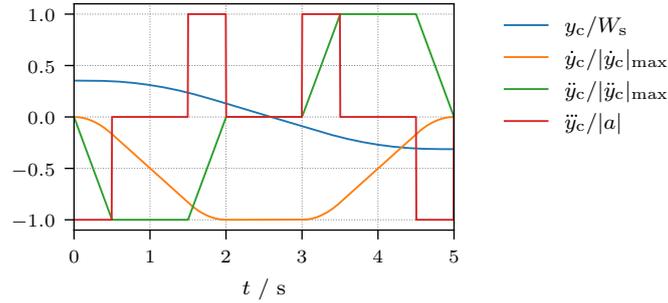

  \begingroup%
  \newcommand{\datapath}{figures/}%
  \includestandalone[]{\datapath/Ss5_ypos}%
  \endgroup
    \caption{Prescribed motion of the end points, as defined in \cref{eq:Ss5_ypos}.}
    \label{fig:Ss5_ypos}
\end{figure}

A structure is held in place by its two ends such that it initially touches 
a line obstacle at a \ang{45} angle,
as sketched in \cref{fig:Ss5a}.
The clamped ends are then translated downwards,
imposing the position
$\vectorsym{\centerlinepos}|_{\arclength=0}
= \vectorsym{\centerlinepos}|_{\arclength=\bladeL}
= \transposed{\left[0, y\ind{0}\of{t}, 0\right]}$ at both ends
while maintaining the initial inclination.
\begin{figure}[bt]%
    \begin{minipage}[c]{40mm}
        \begin{table}[H]
            \begin{tabular}{l r @{\,} l}
                \toprule
                &    \multicolumn{2}{c}{\textbf{Ss5}}    \\ \midrule
                $\alpha$              & $\ang{45}$         &                   \\
                $\bladeL$             & $\num{250}$        & $\si{mm}$         \\
                $\bladeW$             & $\num{15}$         & $\si{mm}$         \\
                $\bladeT$             & $\num{0.063}$      & $\si{mm}$         \\
                $\bladeE$             & $\num{250}$        & $\si{MPa}$        \\
                $\bdensity$           & $\num{920}$        & $\si{kg.m^{-3}}$  \\
                $\bladePoissonsratio$ & $\num{0.5}$        &                   \\
                $\dampingratio$       & $\num{0.01}$       &                   \\ \midrule
                $\bladeks{}$          & $\num{0.83}$       &                   \\
                $\bladekt$            & $\num{7e-5}$       &                   \\
                $\bladeNe$            & $\num{107}$        &                   \\
                $\dt$                 & $\num{1e-4}$       & $\si{s}$          \\ \bottomrule
            \end{tabular}
            \caption{Parameters of configuration Ss5.}
            \label{tab:Ss5}
        \end{table}
    \end{minipage}
    \hfil
    \begin{minipage}[c]{70mm}
        \begin{figure}[H]
  \begingroup%
  \newcommand{\datapath}{figures/}%
  \includestandalone[]{\datapath/Ss5}%
  \endgroup
            {\phantomsubcaption\label{fig:Ss5a}}
            {\phantomsubcaption\label{fig:Ss5b}}
            \caption{Sketch of configuration Ss5
                involving a structure that collides laterally
                with a rigid line obstacle (not to scale).
                When the \usuk{center}{centre}-most cross-section is rotated fully flat
                the entire line A--B below this section is in contact.
                Such a situation is depicted with broken lines, where the angle $\alpha$
                vanishes and the $y$-coordinate of the upper structure is $y=0$, throughout.}
            \label{fig:Ss5}
        \end{figure}
    \end{minipage}
\end{figure}
The vertical position $y\ind{c}$ was prescribed according to
\begin{gather}\label{eq:Ss5_ypos}
    y\ind{c}\of{t} = \begin{cases}
        a/6 t^3 + y\ind{c0}                                        & \text{if}\quad\hphantom{0.1\,T}\mathllap{0} < t \leq 0.1\,T \\
        a/20 T t^2 - a/200 T^2 t + a/6000 T^3 + y\ind{c0}          & \text{if}\quad0.1\,T < t \leq 0.3\,T \\
        -a/6 t^3 + a/5 T t^2 - a/20 T^2t + 7a/1500 T^3 + y\ind{c0} & \text{if}\quad0.3\,T < t \leq 0.4\,T \\
        3a/100 T^2t - 3a/500 T^3 + y\ind{c0}                       & \text{if}\quad0.4\,T < t \leq 0.5\,T \\
        y\ind{ce} - y\ind{c}\of{T - t}                             & \text{if}\quad0.5\,T < t \leq 1.0\,T \\
    \end{cases}
    \\\nonumber
    y\ind{c0} = \bladeW / 2 \sin\of{\pi/4}
    ,\quad
    y\ind{ce} = y\ind{c0} - 2\bladeW/3
    ,\quad
    T = \SI{5}{s}
    ,\quad
    a = 500 (y\ind{ce}-y\ind{c0}) / (9 T^3)
    \eqdot
\end{gather}
This is illustrated in \cref{fig:Ss5_ypos} together with the corresponding derivatives in time.
The purpose of this case is to cover a case where torsion occurs
in addition to the \usuk{centerline}{centreline} being bent.
Parameters defining the colliding structure are given in \cref{tab:Ss5}.
The line obstacle was provided by the edge of a rigid structure plus a relatively large critical distance.
Two sub-configurations were devised which differ only by the element length associated with this  
passive structure.
Corresponding snapshots from simulation with the collision model variant \modelvariant{nb22d}
are shown in \cref{fig:Ss5_visualizations_singlecollision,fig:Ss5_visualizations_multiplecollision}.
Since the collision model 
identifies at most one collision event between two elements,
varying the discretization of the passive structure has the following effect:
When choosing the coarse discretization of \cref{fig:Ss5_visualizations_singlecollision},
at most one collision event is generated at any time.
When choosing the fine discretization of \cref{fig:Ss5_visualizations_multiplecollision},
multiple collision events occur simultaneously underneath the \usuk{center}{centre} element of the displaced structure.

Until $t \approx \SI{2.6}{s}$ the skeleton line remains approximately straight
and the structure gets twisted by the collision force.
During this phase, 
the single shortest minimum distance identifies the correct contact point needed for a physically accurate 
collision impulse.
In the subsequent phase, the structure \usuk{centerline}{centreline} end points move below the obstacle, 
and the collision force needs to increase 
to prevent the mobile structure from penetrating the obstacle.
This goes along with an elongation of the \usuk{centerline}{centreline} 
and generates large contact forces.
At the same time, the amount of rotation, and, hence, the torque applied through the collision force
must remain constant.
A single contact point would, therefore, need to move towards the \usuk{center}{centre} of the structure
to reduce leverage.
Since the contact point is determined simply as the minimum distance between colliding elements,
the collision model is, however, unable to do so, when not permitting multiple collision vectors.
Instead, the contact point remains at the position \usuk{labeled}{labelled} A in \cref{fig:Ss5} and the structure sinks in
with the line connecting A and B remaining parallel to the obstacle.
In contrast, when multiple collision vectors are present along the \usuk{centerline}{centreline}, 
they manage to produce the right amount of force and torque,
preventing penetration.

In summary, the collision model is able to handle the presented case of a collision invoking torsion in addition to bending.
When there is a contact line
the shortest distance does not provide the theoretical position of the equivalent single contact point.
Instead, the collision model needs 
at least two contact points on either side of the \usuk{centerline}{centreline}.
When the colliding structures are resolved with elements much shorter than their width this is automatically the case.

\begin{figure}[h]
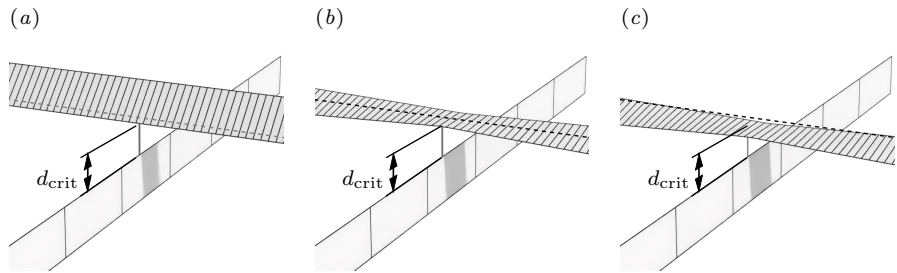
%
  \begingroup%
  \newcommand{\datapath}{figures/}%
  \includestandalone[]{\datapath/Ss5_visualizations_singlecollision}%
  \endgroup
    \caption[]{Snapshots from simulations of test case Ss5
        with a single collision in the contact line,
        using the model \modelvariant{nb22d}.
        Snapshots extracted at
        \subcaptionmarker{a}~$t=\SI{0}{s}$,
        \subcaptionmarker{b}~$t=\SI{2.8}{s}$,
        \subcaptionmarker{c}~$t=\SI{4.0}{s}$.
        Collision distance vectors are drawn as lines.
        The dashed line corresponds to the straight skeleton line in \cref{fig:Ss5}b, \usuk{i.e.,}{i.e.} at $y=0$ and $t=t_1$.
    }\label{fig:Ss5_visualizations_singlecollision}
\end{figure}

\begin{figure}[h]
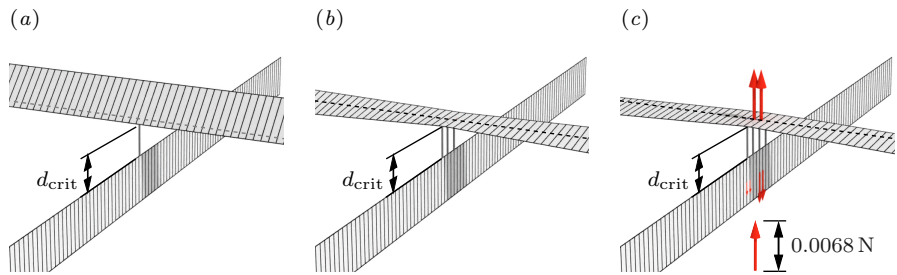

  \begingroup%
  \newcommand{\datapath}{figures/}%
  \includestandalone[]{\datapath/Ss5_visualizations_multiplecollision}%
  \endgroup
    \caption[]{Snapshots from a simulation of test case Ss5 
        with multiple collisions coexist along the contact line, 
        using the model \modelvariant{nb22d}.
        Snapshots extracted at
        \subcaptionmarker{a}~$t=\SI{0}{s}$,
        \subcaptionmarker{b}~$t=\SI{2.8}{s}$,
        \subcaptionmarker{c}~$t=\SI{4.0}{s}$.
        Collision distance vectors are drawn as lines.
        Red arrows indicate the nodal forces due to collision impulses,
        all scaled by the magnitudes of $\externalforcedensity$. 
        Since the same scaling is applied in all figures these arrows are visible only in 
        \subcaptionmarker{c} where collision forces are the largest.
        The dashed line corresponds to the straight skeleton line in \cref{fig:Ss5}b, \usuk{i.e.,}{i.e.} at $y=0$ and $t=t_1$.
    }\label{fig:Ss5_visualizations_multiplecollision}
\end{figure}

\FloatBarrier
\levelstay{Beams touching statically at multiple contact points}
\label{sec:Sm1}

In the test case Sm1 two beams are positioned side by side as sketched in \cref{fig:Sm1},
without any gravitational force present.
To obtain this placement the blades were initially positions in parallel to each other, 
their surfaces aligned with the $x$-$y$-plane 
and separated by a distance slightly less than the critical distance $\dcrit$.
They were then rotated around the $y$-axis in opposite directions by $\alpha_y/2=\SI{5}{\degree}$, each,
followed by rotations around the $x$-axis by 
$\alpha_x/2 = \pm\SI{0.5}{\degree}$, each.
The intention behind this arrangement was to produce a large number of \usuk{neighboring}{neighbouring} distance vectors that are neither parallel to one another nor parallel to any of the axes of the beams.
To achieve this situation a deliberatively large value of $\dcrit$, exceeding the one of \cref{eq:dcrit}, was chosen.
The structures are clamped at one end each, but this boundary condition is not really needed, 
since they are initialized in their static shape, with zero velocity, and no gravity.
The parametrization of the beams is given in \cref{tab:Sm1}.

\begin{figure}[htb]%
    \begin{minipage}[c]{45mm}
        \begin{table}[H]
            \begin{tabular}{l r @{\,} l}
                \toprule
                &    \multicolumn{2}{c}{\textbf{Sm1}}     \\ \midrule
                $\bladeL$             & $\num{250}$         & $\si{mm}$         \\
                $\bladeW$             & $\num{15}$          & $\si{mm}$         \\
                $\bladeT$             & $\num{0.063}$       & $\si{mm}$         \\
                $\bladeE$             & $\num{250}$         & $\si{MPa}$        \\
                $\bdensity$           & $\num{920}$         & $\si{kg.m^{-3}}$  \\
                $\bladePoissonsratio$ & $\num{0.5}$         &                   \\
                $\dampingratio$       & $\num{0.01}$        &                   \\ \midrule
                $\bladeks{}$          & $\num{0.83}$        &                   \\
                $\bladekt$            & $\num{7e-5}$        &                   \\
                $\bladeNe$            & $\num{108}$         &                   \\
                $\dt$                 & $\num{1e-4}$        & $\si{s}$          \\
                $\dcrit$              & $\num{2.5}$         & $\si{mm}$         \\ \bottomrule
            \end{tabular}
            \caption{Parameters of configuration Sm1 involving structures that touch along half their edges.
                No fluid is present and no gravity.}
            \label{tab:Sm1}
        \end{table}
    \end{minipage}
    \hfil
    \begin{minipage}[c]{72mm}
        \begin{figure}[H]
  \begingroup%
  \newcommand{\datapath}{figures/}%
  \includestandalone[]{\datapath/Sm1}%
  \endgroup
            \caption{Sketch of configuration Sm1 involving structures that touch along half their edges.}
            \label{fig:Sm1}
        \end{figure}
    \end{minipage}
\end{figure}

\begin{figure}[htb]
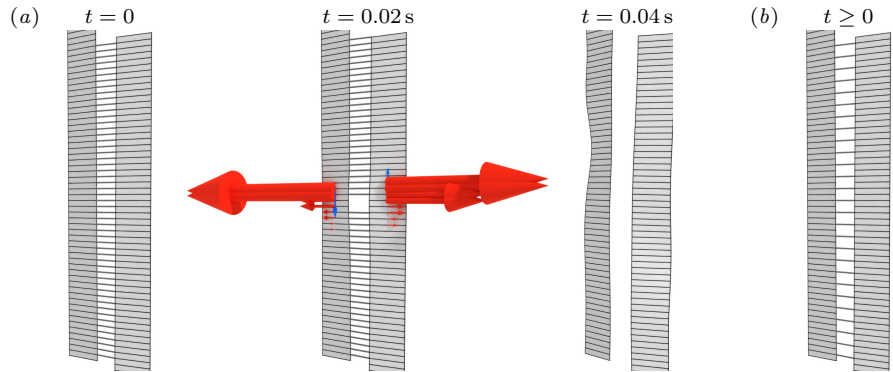

  \begingroup%
  \newcommand{\datapath}{figures/}%
  \includestandalone[]{\datapath/Sm1_vis}%
  \endgroup
    {\phantomsubcaption\label{fig:vis_Sm1_nobend2d}}
    {\phantomsubcaption\label{fig:vis_Sm1_nobend2g}}
    \caption[]{Visualizations from simulation runs of Sm1.
        \subref{fig:vis_Sm1_nobend2d}~Result with model variant \modelvariant{nb22d} at different times;
        \subref{fig:vis_Sm1_nobend2g}~result with model \modelvariant{final}.
        Collision distance vectors are drawn as lines.
        Red arrows in \subref{fig:vis_Sm1_nobend2d} indicate the nodal forces due to (spurious) collision impulses,
        while relatively small, vertically oriented, blue arrows represent the generated torque.
        Both are scaled by the magnitudes of $\externalforcedensity$ and $\externaltorquedensity/(0.5 \bladeW)$, respectively, 
        with the same scaling applied.}
    \label{fig:vis_Sm1}
\end{figure}

\begin{figure}[h]
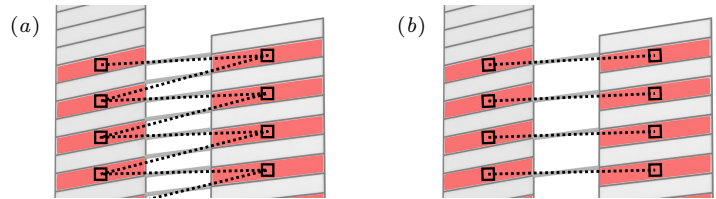

  \begingroup%
  \newcommand{\datapath}{figures/}%
  \includestandalone[]{\datapath/issue_crossover_distances}%
  \endgroup
    {\phantomsubcaption\label{fig:issue_crossover_distances_issue}}
    {\phantomsubcaption\label{fig:issue_crossover_distances_resolved}}
    \caption[]{Detail from simulation runs of Sm1 near the free end of one of the two structures
        with colliding elements highlighted in red and 
        with the collision network indicated by dashed lines.
        \subref{fig:issue_crossover_distances_issue}~For model variant \modelvariant{nb22d};
        \subref{fig:issue_crossover_distances_resolved}~for variant \modelvariant{final}.}
\end{figure}

With no relative velocity of the blades and no deformation, the 
analytical solution of the problem is a trivial zero-velocity steady state with the beams resting in their initial shape 
at zero velocity.
Instead, without further measures,
and using the model variant \modelvariant{nb22d} or any of the other variants considered,
an instability develops, 
which is illustrated by the snapshots in \cref{fig:vis_Sm1_nobend2d},
showing the \usuk{center}{centre} region $-\bladeL / 2 \lessapprox y \lessapprox \bladeL / 2$.
It can be observed that spurious collision impulses build up at an increasingly fast rate,
quickly pushing the structures apart.
The detected collision distance vectors are separated by the length of a collision element,
despite only every other element being involved in a collision, as discussed in 
\cref{sec:fix_removal_competing_collisions}
with $\dnLc = 2$.
The topology of the underlying collision network is drawn in \cref{fig:issue_crossover_distances_issue}
with the structure elements associated with the collision events marked using a separate \usuk{color}{colour}.
The network is in fact one long chain of \enquote{collisions}.
To break this chain a stricter definition of \enquote{competing} collisions was implemented.
For the present situation of \cref{fig:issue_crossover_distances_issue}
this could be accomplished by further filtering 
$\CollPairsNC$ from \cref{eq:collisionpairs_noncompeting} via%
\begin{equation}
    \CollPairs\ind{nc2} 
    = \Big\lbrace
    \left\lbrace\masterelement_1, \masterelement_2\right\rbrace 
    \in \CollPairs\ind{nc2}
    \Bigm|
    \forall
    \left\lbrace\othermasterelement_1,\othermasterelement_2\right\rbrace\in\CollPairsNC
    :
    \begin{aligned}[t]
        &     (\text{$\masterelement_1, \masterelement_2, \othermasterelement_1, \othermasterelement_2$ on 4 different blades}) \\
        &\vee \sum_\text{$\forall\lbrace\masterelement_i, \othermasterelement_j\rbrace$ \rlap{on same structure}} \! \snorm{\masterelement_i - \othermasterelement_j} = 0   \\
        &\vee \sum_\text{$\forall\lbrace\masterelement_i, \othermasterelement_j\rbrace$ \rlap{on same structure}} \! \snorm{\masterelement_i - \othermasterelement_j} \ge 3 
        \,\Big\rbrace
        \eqdot
    \end{aligned}
\end{equation}%
The corresponding model variant \modelvariant{final} reproduces the stable steady state solution 
pictured in \cref{fig:vis_Sm1_nobend2g}
with the underlying collision distance network of \cref{fig:issue_crossover_distances_resolved}.
That is, all collisions produce the expected zero collision force and the structures remain motionless.

%% file: chapters/collision-model/chapters/tests-complex.tex
%
\FloatBarrier
\leveldown{Beam dropping on a perpendicular one}

Two structures are initially stacked forming a cross as shown in \cref{fig:CS1_initial}, configured according to \cref{tab:CS1}.
The lower structure is held in place by its two fixed ends, 
while the upper structure can move freely 
as it is accelerated downwards by gravity.
When it then collides with the lower structure, 
both ribbons become entangled for a while, until the upper one slips off to one side,
since no friction is present in the contact.

\begin{figure}[h]%
    \begin{minipage}[c]{45mm}
        \begin{table}[H]%
            \begin{tabular}[c]{l r @{\,} l}
                \toprule
                &           \multicolumn{2}{c}{\textbf{CS1}}           \\ \midrule
                $\bladeL$             & $\num{250}$       & $\si{mm}$                        \\
                $\bladeW$             & $\num{15}$        & $\si{mm}$                        \\
                $\bladeT$             & $\num{0.063}$     & $\si{mm}$                        \\
                $\bladeE$             & $\num{250}$       & $\si{MPa}$                       \\
                $\bdensity$           & $\num{920}$       & $\si{kg.m^{-3}}$                 \\
                $\bladePoissonsratio$ & $\num{0.5}$       &                                  \\
                $\dampingratio$       & $\num{0.01}$      &                                  \\
                $\grav_y$             & $\num{-9.81}$     & $\si{\meter\per\second\squared}$ \\ \midrule
                $\bladeks{}$          & $\num{0.83}$      &                                  \\
                $\bladekt$            & $\num{7e-5}$      &                                  \\
                $\bladeNe$            & $\num{108}$       &                                  \\
                $\dt$                 & $\num{1e-5}$      & $\si{s}$                         \\
                $\dcrit$              & $\num{1}$         & $\si{mm}$                        \\ \bottomrule
            \end{tabular}
            \caption{Parameters defining the test configuration CS1.}
            \label{tab:CS1}
        \end{table}
    \end{minipage}
    \hfil
    \begin{minipage}[c]{70mm}
        \begin{figure}[H]%
  \begingroup%
  \newcommand{\datapath}{figures/}%
  \includestandalone[]{\datapath/CS1_initial}%
  \endgroup
            \caption[]{The initial geometry ($t=\SI{0}{s}$) of test configuration CS1 involving two structures that collide at their center
                as the upper structure falls on the lower one which is fixed at its two ends.}
            \label{fig:CS1}
            \label{fig:CS1_initial}
        \end{figure}
    \end{minipage}
\end{figure}

The goal of this test is to verify that the collision successfully 
inhibits the intersection of both structures
without introducing unnatural-looking collision impulses that would 
result in a twitching of the structures or jittering force and torque vectors.
The model configuration \modelvariant{nb2g} was found to perform well.

\begin{figure}[htb]
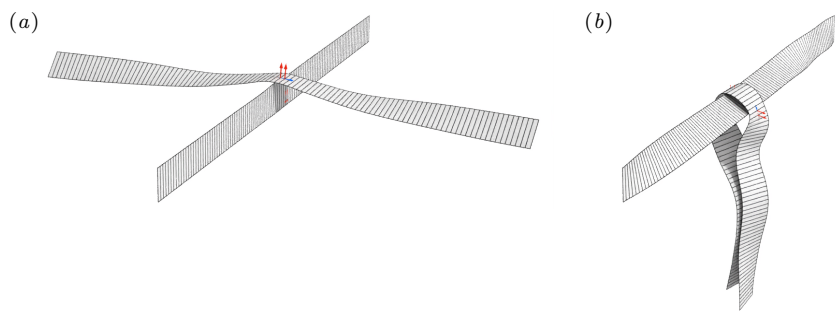
\hspace*{0pt}
  \begingroup%
  \newcommand{\datapath}{figures/}%
  \includestandalone[]{\datapath/CS1_visualizations}%
  \endgroup
    \caption[]{Visualizations of the test CS1 involving two structures that collide at their center
        as the upper structure falls on the lower one which is fixed at its two ends.
        \subcaptionmarker{a}~result at $t=\SI{0.05}{s}$, 
        \subcaptionmarker{b}~result at $t=\SI{0.2}{s}$. 
        Collision distance vectors are drawn as lines.
        Red arrows indicate the nodal forces due to collision impulses,
        all scaled by the magnitudes of $\externalforcedensity$, with the same scaling applied for all times.}
    \label{fig:CS1_visualizations}
\end{figure}

\FloatBarrier
\levelstay{Rings dropping on plate}

The beams \usuk{modeled}{modelled} so far exhibit two ends, free or clamped.
But the method also allows to identify both ends with the same position,
so as to obtain a ring-type structure.
This is done in the case CR1
with R standing for \enquote{ring},
featuring multiple ring-shaped rods falling on an inclined solid wall at the bottom of a box
without the presence of a fluid.
Here, 27 randomly oriented rings are placed at a distance from the wall such that their \usuk{centers}{centres} form a $3\times3\times3$ matrix,
as depicted in \cref{fig:CR1_initial}.

They are accelerated downwards due to gravitation 
and eventually collide with the bottom plate.
During the impact phase the structures are highly compressed and a variety of different collision scenarios occur including chained collisions between elements of stacked rings.
The rings then gather in the lower corner of the domain where they settle in their static arrangement.
This problem was originally presented in \cite[§5.2]{TschisgaleConstraintbased2019}.
The essential parameters defining the problem are given in \cref{tab:CR1}.
\begin{figure}[htb]%
    \begin{minipage}[c]{45mm}%
        \begin{table}[H]
            \begin{tabular}[c]{l r @{\,} l}
                \toprule
                &             \multicolumn{2}{c}{\textbf{CR1}}               \\ \midrule
                $\nBlades$       & $\num{27}$           &                                     \\
                $\grav_y$        & $\num{-10}$          & $\si{\meter\per\second\squared}$    \\
                $\bladeL$        & $\num{0.5}$          & $\si{m}$                            \\
                $\bladeW$        & $\num{0.04}$         & $\si{m}$                            \\
                $\bladeT$        & $\num{0.01}$         & $\si{m}$                            \\
                $\bladeE$        & $\num{ 1}$           & $\si{MPa}$                          \\
                $\bladeG$        & $\num{ 1}$           & $\si{MPa}$                          \\
                $\bdensity$      & $\num{1000}$         & $\si{\kilogram\per\meter\cubed}$    \\
                $\bladeCg{i}$    & $\num{1e-2}$         & $\si{\newton\second}$               \\ 
                $\bladeCk{i}$    & $\num{2e-5}$         & $\si{\newton\second\meter\squared}$ \\ \midrule 
                $\bladeNe$       & $\num{64}$           &                                     \\
                $\dt$            & $\num{1e-3}$         & $\si{s}$                            \\
                $\dcrit$         & $\num{0.01}$         & $\si{m}$                            \\ \bottomrule
            \end{tabular}
            \caption{Parameters defining the test configuration CR1 involving 27 rings dropping on a wall.}
            \label{tab:CR1}
        \end{table}
    \end{minipage}
    \hfil
    \begin{minipage}[c]{67mm}
        \begin{figure}[H]
  \begingroup%
  \newcommand{\datapath}{figures/}%
  \includestandalone[]{\datapath/CR1_initial}%
  \endgroup
            \caption{Initial arrangement of the rings in test CR1 at $t=\SI{0.3}{s}$ as they approach the tilted floor.
                The rings are placed with their \usuk{centers}{centres} 
                on a $3\times3\times3$ lattice with random orientation.}
            \label{fig:CR1_initial}
        \end{figure}
    \end{minipage}
\end{figure}

\begin{figure}[htb]
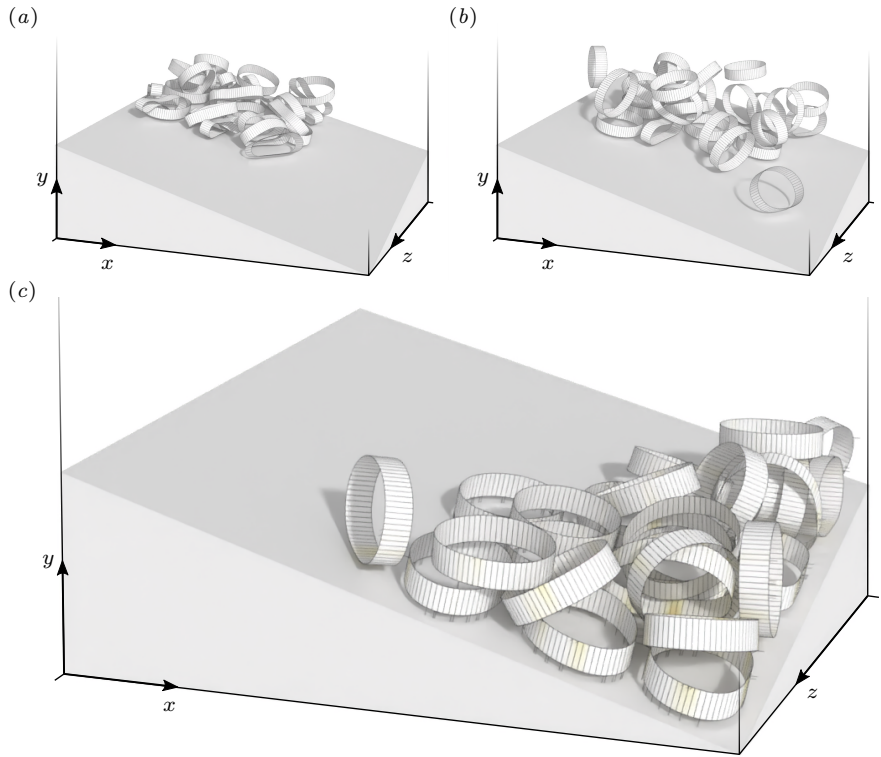

  \begingroup%
  \newcommand{\datapath}{figures/}%
  \includestandalone[]{\datapath/CR1_vis}%
  \endgroup
    \caption{Snapshots from the simulation of CR1 involving ring-shaped rods dropping on a surface, for 
        \subcaptionmarker{a}~$t=\SI{0.5}{s}$,
        \subcaptionmarker{b}~$t=\SI{0.7}{s}$,
        \subcaptionmarker{c}~$t=\SI{5.0}{s}$.
        The visualizations show the ribbons without their nominal thickness, and 
        with the individual structure elements marked.
        The ribbons are painted white when not in contact and with shades of orange indicating the amount of surface-specific loads due to collisions.
        Small lines of length $\distance\approx\ddry=\bladeT$ between different surfaces represent the collision distance vectors.}
    \label{fig:CR1_results}
\end{figure}

The performance of the model was evaluated qualitatively,
verifying that rings are prevented from penetration during the entire simulated time.
The deformation of the rings looks plausible, without any unnatural-looking twitching motion of the rings.
Some snapshots are shown in \cref{fig:CR1_results}.

\FloatBarrier
\levelstay{Beam floating against a perpendicular one, with fluid--structure interaction}

The motivation behind this test is to verify 
the \usuk{behavior}{behaviour} of the collision model 
combined with fluid-structure coupling.
The configuration consists of a periodic box in which two beams are placed
as shown in \cref{fig:FSI2_initial}.
One relatively long, free-floating beam is placed
at $x=\SI{15}{mm}$
near the domain face ($x=0$), aligned with the $y$-axis.
The second, rigid beam is aligned with the $z$-axis and
spans the entire width of the domain.
Its position is fixed at half the domain height,
\usuk{i.e.,}{i.e.} $y=L_y/2$,
slightly downstream the longer blade,
at $x=\SI{30}{mm}$.
The fluid flow and the velocity of the long beam are initialized with a streamwise velocity $\ubulk$\!,
and this value of the spatially averaged flow velocity is kept constant throughout the simulation
by applying a controlled, spatially constant volume force.
The parameters defining the case are given in \cref{tab:CFSI2}.

\begin{figure}[htb!]
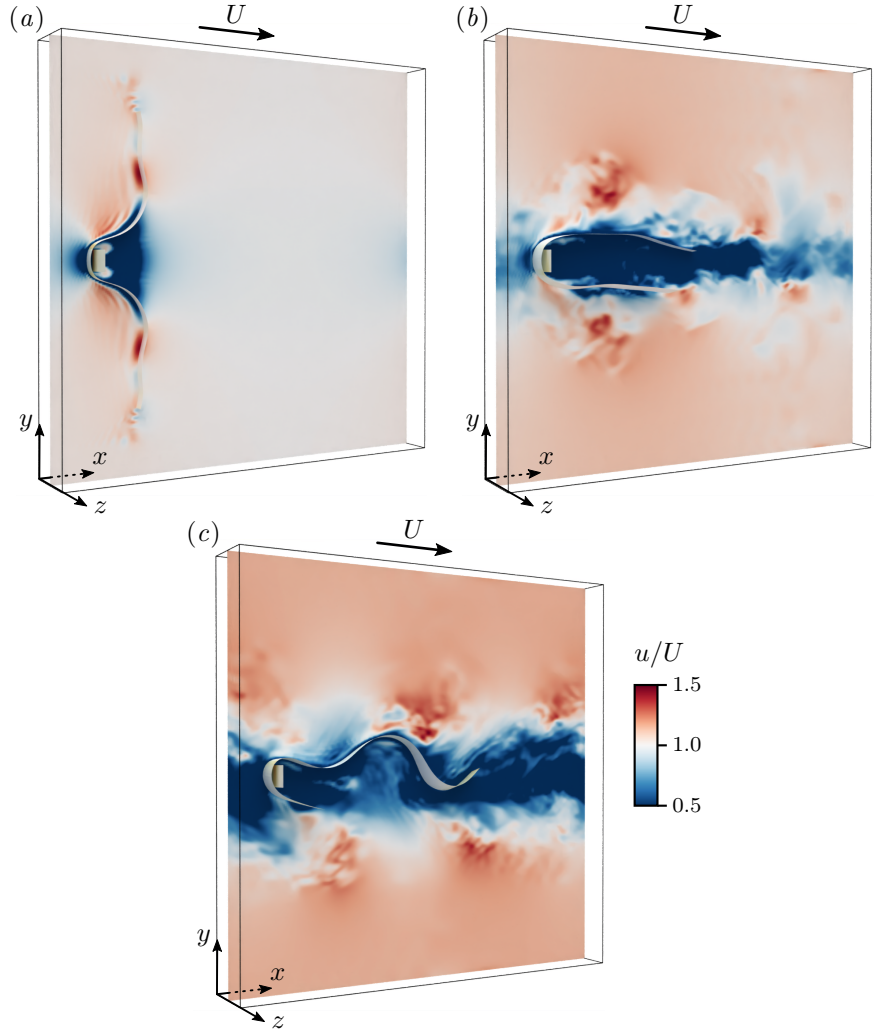
%
  \begingroup%
  \newcommand{\datapath}{figures/}%
  \includestandalone[]{\datapath/CFSI2_vis}%
  \endgroup
    \caption{Visualizations of snapshots from CFSI2,
        at different times.
        \subcaptionmarker{a}~$t=\SI{0.47}{s}$
        \subcaptionmarker{b}~$t=\SI{5.22}{s}$
        \subcaptionmarker{c}~$t=\SI{9.97}{s}$.
        The slice at $z=L_z/2$ is \usuk{colored}{coloured} according to the instantaneous streamwise velocity component.
        White corresponds to $u=\ubulk$, blue is slower, red is faster.}
    \label{fig:FSI2_visualization}
\end{figure}

Consequently, the long beam floats against and wraps around the shorter beam
as depicted in \cref{fig:FSI2_visualization}.
As described in \cref{sec:collisionelementfsi,sec:collisionelementinertia} above,
the inertia associated with a collision element
is the sum of a part that is related to the inertia of the associated structure segment
and a second part that resembles the inertia of an element-attached fluid layer.
The latter results from numerical aspects of the fluid-structure coupling.
Its significance was checked by simulating the same configuration with this term omitted, \usuk{i.e.,}{i.e.}
using \cref{eq:collisionelementinertia_dry} instead of \cref{eq:CMEOMIntegrated_wet_inertia}.
With the term omitted the collision model was unable to produce proper collision reactions,
and instead instabilities occurred.

\begin{figure}[htb]%
    \begin{minipage}[c]{45mm}%
        \begin{table}[H]%
            \begin{tabular}{l r @{\,} l}
                \toprule
                &                   \multicolumn{2}{c}{\textbf{CFSI2}}                   \\ \midrule
                $\nBlades$            & $\lbrace\num{1}, \num{1}\rbrace$    &                                  \\
                $\bladeL$             & $\lbrace\num{250}, \num{45}\rbrace$ & $\si{mm}$                        \\
                $\bladeW$             & $\num{15}$                          & $\si{mm}$                        \\
                $\bladeT$             & $\num{0.063}$                       & $\si{mm}$                        \\
                $\bladeE$             & $\num{250}$                         & $\si{MPa}$                       \\
                $\bladePoissonsratio$ & $\num{0.5}$                         &                                  \\
                $\bdensity$           & $\num{920}$                         & $\si{\kilogram\per\meter\cubed}$ \\ \midrule
                $\ubulk$              & $\num{0.1}$                         & $\si{\meter\per\second}$         \\
                $L_x$                 & $\num{0.3}$                         & $\si{m}$                         \\
                $L_y$                 & $\num{0.3}$                         & $\si{m}$                         \\
                $L_z$                 & $\num{0.045}$                       & $\si{m}$                         \\
                $\fdensity$           & $\num{1000}$                        & $\si{\kilogram\per\meter\cubed}$ \\
                $\kinvis$             & $\num{1e-6}$                        & $\si{\meter\squared\per\second}$ \\ \midrule
                $\bladeNe$            & $\lbrace\num{108}, \num{19}\rbrace$ &                                  \\
                $\nCells_x$           & $\num{240}$                         &                                  \\
                $\nCells_y$           & $\num{240}$                         &                                  \\
                $\nCells_z$           & $\num{36}$                          &                                  \\
                $\bladeks{}$          & $\num{0.83}$                        &                                  \\
                $\bladekt$            & $\num{7e-5}$                        &                                  \\
                $\dt$                 & $\num{6.25e-4}$                     & $\si{s}$                         \\
                $\dcrit$              & $\num{1.25}$                        & $\si{mm}$                        \\ \bottomrule
            \end{tabular}
            \caption{Parameters defining the test configuration CFSI2.
                Where two values are given, the first value describes the floating structure, and the second one the rigid structure.}
            \label{tab:CFSI2}
        \end{table}
    \end{minipage}
    \hfil
    \begin{minipage}[c]{60mm}%
        \begin{figure}[H]%
  \begingroup%
  \newcommand{\datapath}{figures/}%
  \includestandalone[]{\datapath/CFSI2}%
  \endgroup
            \caption{Perspective view of initial situation in test CFSI2. A free-floating beam approaches a rigid, perpendicular beam
                with its skeleton axis in $z$-direction.
                Both are embedded in a fluid which is driven by a spatially constant volume force,
                adjusted in time to maintain a mean velocity $\ubulk$ in the $x$-direction.
                The motion of the fluid is made visible with streamlines in the \usuk{center}{centre} $x$-$y$ plane.}
            \label{fig:FSI2_initial}
        \end{figure}
    \end{minipage}
\end{figure}

\FloatBarrier
\levelstay{Canopies, with and without FSI}

Finally, the model is applied in the simulation of canopies
made up of a highly flexible blades.
Aiming to employ the collision model eventually in the configuration of \cref{fig:SampleFlowHighCauchy},
a small cut-out of this full domain is considered here, measuring 
$L_x \times L_y \times L_z = \num{2.3}\channelH \times \channelH \times \num{0.32}\channelH$\!,
as reported in \cref{tab:CC1_CC2}.
No-slip boundary conditions were applied at the bottom wall,
and the rigid-lid slip boundary condition was used at the top wall.
Two variants were defined, CC1 and CC2.

\begin{figure}[htb]
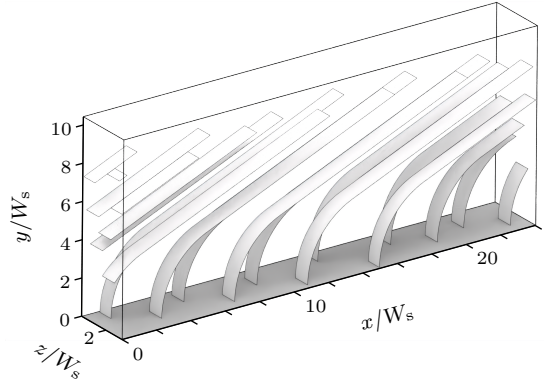
%
  \begingroup%
  \newcommand{\datapath}{figures/}%
  \includestandalone[]{\datapath/CC1_initial}%
  \endgroup
    \caption{Initial situation of the test cases CC1 and CC2.}
    \label{fig:CC1_CC2_initial}
\end{figure}

In CC1, periodic boundary conditions were applied in $x$-direction,
and free-slip boundaries in $z$-direction,
whereas in CC2 periodic boundaries were employed both in $x$ and in $z$.
The initial situation is shown in \cref{fig:CC1_CC2_initial}.
Since their lengths exceeds the height of the simulation domain, 
the blades were bent from their relaxed straight state into a reconfigured shape.
In CC1 gravitation was set in a $\SI{45}{\degree}$-angle such that the blades drop on the floor plate.
In CC2 gravitation was set oriented upwards. Due to their buoyancy this has the ribbons float downwards where they gather on the floor.
The flow itself was not driven such that the ideal converged solution is zero fluid velocity and motionless blades.
In the intended application shown in \cref{fig:SampleFlowHighCauchy}
the canopy is instead subjected to a turbulent flow driven by a horizontal volume force.
The present configuration CC2 was designed differently specifically to verify that the static contact of the gathered blades does not produce instabilities,
and hence spurious fluid velocities that could be mistaken for turbulence.
The ability of the model to handle dynamic situations was verified in the other tests.

\begin{table}[htb]%
    \vspace{-4mm}%
    \newenvironment{subtableenv}{\begin{adjustbox}{varwidth=\linewidth, trim=0 0 0 -4mm}}{\end{adjustbox}}
    \begin{subtableenv}\begin{tabular}[t]{l r @{\,} l r @{\,} l}
            \multicolumn{5}{c}{\subcaptionmarker{a}~Structure properties}                      \\[1mm] \toprule
            & \multicolumn{2}{c}{\textbf{CC1 (dry)}} & \multicolumn{2}{c}{\textbf{CC2 (wet)}} \\ \midrule
            $\bladeL$             & $\num{250}$   & $\si{mm}$              &       \multicolumn{2}{c}{(same)}       \\
            $\bladeW$             & $\num{15}$    & $\si{mm}$              &       \multicolumn{2}{c}{(same)}       \\
            $\bladeT$             & $\num{0.063}$ & $\si{mm}$              &       \multicolumn{2}{c}{(same)}       \\
            $\bladeE$             & $\num{250}$   & $\si{MPa}$             &       \multicolumn{2}{c}{(same)}       \\
            $\bdensity$           & $\num{920}$   & $\si{kg.m^{-3}}$       &       \multicolumn{2}{c}{(same)}       \\
            $\bladePoissonsratio$ & $\num{0.5}$   &                        &       \multicolumn{2}{c}{(same)}       \\
            $\dampingratio$       & $\num{0.01}$  &                        & $\num{0}$ &                            \\ \bottomrule
    \end{tabular}\end{subtableenv}
    \hfil
    \begin{subtableenv}\begin{tabular}[t]{l r @{\,} l r @{\,} l}
            \multicolumn{5}{c}{\subcaptionmarker{b}~Global and fluid properties}                                 \\[1mm] \toprule
            &         \multicolumn{2}{c}{\textbf{CC1 (dry)}}          &         \multicolumn{2}{c}{\textbf{CC2 (wet)}}         \\ \midrule
            $L_x$            & $\num{360}$          & $\si{mm}$                        &               \multicolumn{2}{c}{(same)}               \\
            $L_y$            & $\num{156.25}$       & $\si{mm}$                        &               \multicolumn{2}{c}{(same)}               \\
            $L_z$            & $\num{50}$           & $\si{mm}$                        & $\num{45}$          & $\si{mm}$                        \\
            $\fdensity$      & ---                  &                                  & $\num{998}$         & $\si{\kilogram\per\meter\cubed}$ \\
            $\kinvis$        & ---                  &                                  & $\num{1e-6}$        & $\si{\meter\squared\per\second}$ \\
            $\grav_x$        & $\num{ 6.94}$        & $\si{\meter\per\second\squared}$ & $\num{0}$           & $\si{\meter\per\second\squared}$ \\
            $\grav_y$        & $\num{-6.94}$        & $\si{\meter\per\second\squared}$ & $\num{9.81}$        & $\si{\meter\per\second\squared}$ \\ \bottomrule
    \end{tabular}\end{subtableenv}
    \hfil
    \begin{subtableenv}\begin{tabular}[t]{l r @{\,} l r @{\,} l}
            \multicolumn{5}{c}{\subcaptionmarker{c}~Numerical settings}                        \\[1mm] \toprule
            &  \multicolumn{2}{c}{\textbf{CC1 (dry)}}  &   \multicolumn{2}{c}{\textbf{CC2 (wet)}}    \\ \midrule
            $\nCells_x$      & ---                  &                   & $\num{256}$          &                      \\
            $\nCells_y$      & ---                  &                   & $\num{125}$          &                      \\
            $\nCells_z$      & ---                  &                   & $\num{32}$           &                      \\
            $\bladeNe$       & $\num{216}$          &                   & $\num{108}$          &                      \\
            $\bladeks{}$     & $\num{0.83}$         &                   &         \multicolumn{2}{c}{(same)}          \\
            $\bladekt$       & $\num{7e-5}$         &                   &         \multicolumn{2}{c}{(same)}          \\
            $\dt$            & $\num{1e-5}$         & $\si{s}$          & $\num{0.625}$        & $\si{\milli\second}$ \\
            $\dcrit$         & $\num{1.25}$         & $\si{mm}$         & $\num{2.81}$         & $\si{mm}$            \\ \bottomrule
    \end{tabular}\end{subtableenv}
    \caption{Parameters of tests involving canopies composed of highly flexible structures.}
    \label{tab:CC1_CC2}
\end{table}

\begin{figure}[htb]
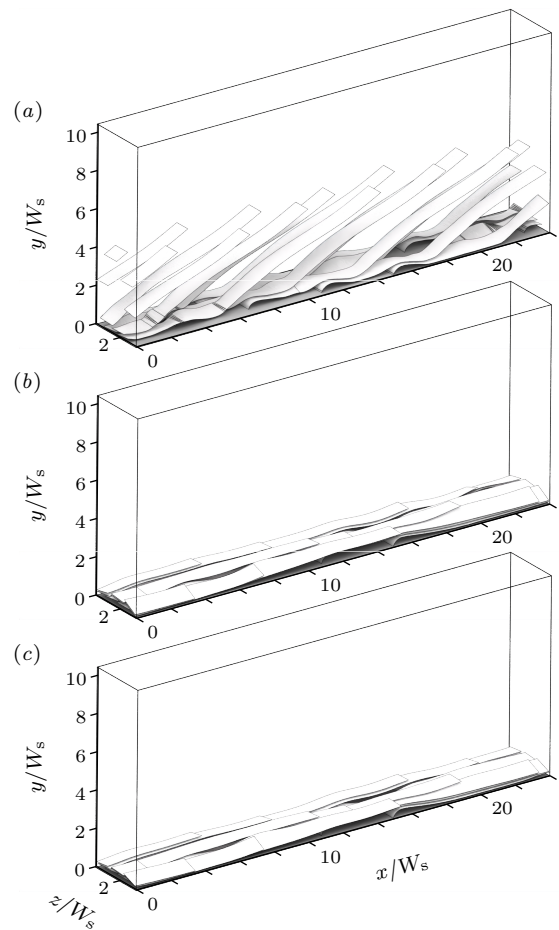
\sidecaption%
  \begingroup%
  \newcommand{\datapath}{figures/}%
  \includestandalone[]{\datapath/CC1_vis}%
  \endgroup
    \caption[]{Snapshots from the simulations CC1 (dry) where blades form a canopy and sink down due to gravity;
        at 
        \subcaptionmarker{a}~$t=\SI{0.12}{s}$, 
        \subcaptionmarker{b}~$t=\SI{0.56}{s}$, 
        \subcaptionmarker{c}~$t=\SI{9.94}{s}$.}
    \label{fig:CC1_visualization}
\end{figure}

\begin{figure}[htb]
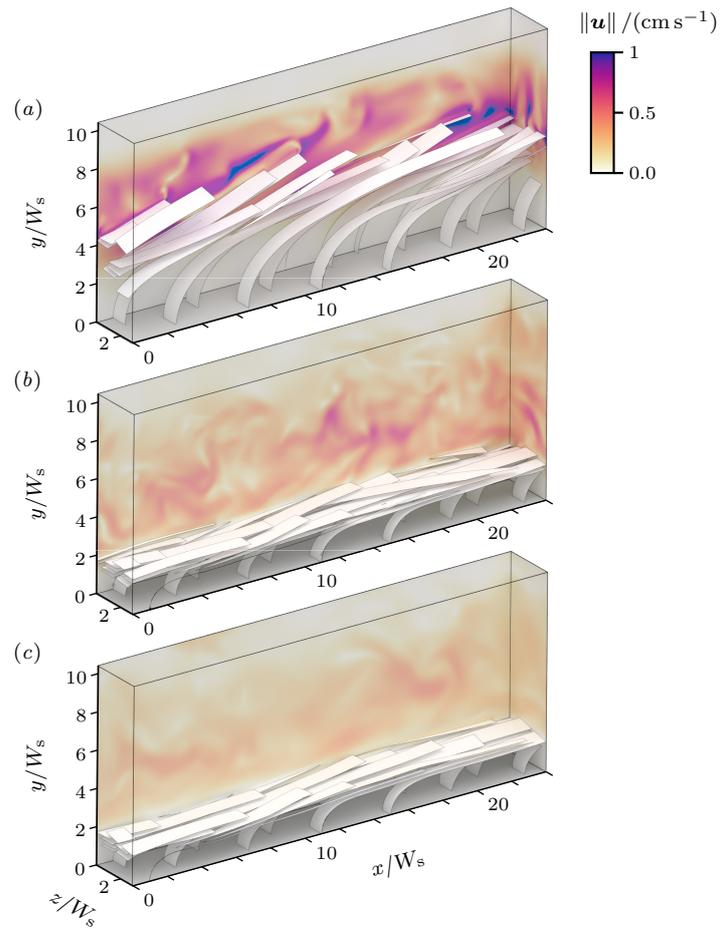
\sidecaption%
  \begingroup%
  \newcommand{\datapath}{figures/}%
  \includestandalone[]{\datapath/CC2_vis}%
  \endgroup
    \caption[]{Snapshots from the simulations CC2 (wet) where blades form a canopy and sink down due to gravity;
        at
        \subcaptionmarker{a}~$t=\SI{10}{s}$,
        \subcaptionmarker{b}~$t=\SI{30}{s}$,
        \subcaptionmarker{c}~$t=\SI{50}{s}$.
        Two periodic boundaries are \usuk{colored}{coloured} according to the fluid velocity magnitude (white is zero velocity).}
    \label{fig:CC2_visualization}
\end{figure}

\Cref{fig:CC1_visualization,fig:CC2_visualization} give a qualitative impression of the solution by visualizing two solution snapshots towards the beginning of the respective simulation,
and one snapshot at the end of the simulated period.
The collision model succeeds at maintaining the desired collision distance, the collisions look plausible, and jittering is minimal.

%% file: chapters/collision-model/conclusions.tex

The present paper proposes a collision model for Cosserat rods.
It considers the collisions between individual elements of the discretized rods
with each element being treated as a rigid body.
The required collision impulses are determined from a constraint-based procedure,
involving the iterative solution of \iac{LCP} to account for simultaneous collisions.
The collision reaction depends on the prescribed coefficient of restitution,
on frictional effects, the inertia of the elements, 
as well as external and internal loads.
The internal loads relate the isolated structure segment to the longer, flexible structure of which it is a part.

The basic formulation of the model was adopted from Tschisgale \textit{et al.} \cite{TschisgaleConstraintbased2019}
and thoroughly revised.
An extensive set of test cases was simulated to demonstrate the functionality of the model.
Among others, an instability was identified, and shown to result from the staggered discretization of the Cosserat rod formulation employed.
Choosing collision elements of length $\bladeLc = 2\bladeLe$, remedied the issue.
The model was further extended by considering fluid--structure coupling realized with an \ac{ibm}
exempt of iterations between the fluid solver and the structure solver.
This introduces coupling loads to be addressed as external loads acting on the colliding elements.
However, part of the these loads depends on the updated solution of the rod movement
which is not available at the time of solving the collision problem.
It was shown that these loads act as inertia and can be included 
by appropriately increasing the inertia of the collision elements.

The resulting method is well suited for cases involving highly flexible structures immersed in a flow,
as simulated by the present authors in
\cite{LohrerFirst2020,LohrerLargeeddysubmitted}
where abstracted aquatic canopies were simulated.
Also, more general applications can benefit from the method proposed,
such as the simulation of hairy surfaces, fur, etc.
Further physical processes can be added,
such as sediment transported inside an aquatic canopy.
With the colliding structure segments represented as rigid discrete elements
in the context of the collision model,
modelling collisions with ordinary rigid discrete elements (sediment particles, walls, etc.) 
is straight forward.
Since collision impulses are determined prior to updating the solution of the structures
the method requires no modification of the structure solver itself,
which simplifies the implementation of the collision model 
and maintains the efficiency of the structure solver.

%% file: chapters/funding.tex
%
This work was funded by \emph{Deutsche Forschungsgemeinschaft} (DFG, grant 316798177).

%% file: chapters/acknowledgements.tex
%
The authors gratefully acknowledge the computing time made available to them on
the high-performance computers and at the NHR Centers at TU Dresden (project name p\_fsi\_canopy) and
RWTH Aachen (project id p0020399). These centres are jointly supported by the Federal Ministry of
Education and Research and the state governments participating in the 
\emph{Verbund für Nationales Hochleistungsrechnen} (NHR, https://www.nhr-verein.
de/unsere-partner).

The authors thank Silvio Tschisgale for helpful discussions on the topic of the manuscript.

%% file: chapters/appendix.tex

\section{Deriving quaternion-based equations of motion}
\label{sec:deriving-quaternion-based-equations}
\subimport{./structure-solver/chapters/}{deriving-quaternion-based-equations}

\clearpage
\section{Structure EOM for IBM, origin of fluid layer mass}
\label{sec:fluid-structure-coupling_appendix}
\setcounter{currentlevel}{5}
\subimport{./fluid-structure-coupling/}{appendix}

\clearpage
\section{Algorithms}
\subimport{./collision-model/}{appendix}

%% file: chapters/structure-solver/chapters/deriving-quaternion-based-equations.tex
%
In this appendix
the quaternion-based equations of motion \eqref{eq:str_EOM_quaternion}
are derived from the \ac{EOM} \eqref{eq:str_EOM} involving angular velocities.
This is based on \cites[12-13]{LangLagrangian2009}{LangMultibody2011,LinnDerivation2013},
and is included here to provide the equations in the same notation as above.
It is also useful to have both formulations, in angular velocities and quaternion-based, at hand
since the quaternion-based formulation is solved when updating the dynamics of the Cosserat rods,
whereas the angular-velocity-based variant is used in order to determine collision and fluid loads.
After all, the two differ substantially,
as can be seen by comparing \cref{eq:str_EOM_rot} with \eqref{eq:str_EOM_quaternion_rot_appendix}
such that it is not trivial to link terms from one representation to the other.

    \begin{enumerate}
        
        \item Starting from \cref{eq:str_EOM}, insert rotational velocity, 
              interior forces and 
              interior torques in the reference configuration
              by forward-rotating them into the instantaneous configuration, 
            $\vectorsym{\angvel} = \rotationmatrix\bcdot\vectorsym{\relref{\angvel}}$,
            $\vectorsym{\internalforce} = \rotationmatrix\bcdot\vectorsym{\relref{\internalforce}}$, and
            $\vectorsym{\internaltorque} = \rotationmatrix\bcdot\vectorsym{\relref{\internaltorque}}$,
            \cite[eqs.~38--39]{LangLagrangian2009}
        \begin{subequations}
            \begin{gather}
                \bdensity\bladeA\ttderiv{\vectorsym{\centerlinepos}} = \frac{\partial\rotationmatrix\bcdot\vectorsym{\relref{\internalforce}}}{\partial\arclength} + \externalforcedensity \label{eq:str_EOM_step2a}\\
                \bdensity \rotationmatrix \bcdot \left( \relref{\crosssectiontensorofinertia}\bcdot\relref{\tderiv{\vectorsym{\angvel}}} + \vectorsym{\relref{\angvel}}\times\relref{\crosssectiontensorofinertia}\bcdot\vectorsym{\relref{\angvel}} \right)
                = \frac{\partial\rotationmatrix\bcdot\vectorsym{\relref{\internaltorque}}}{\partial\arclength}
                + \frac{\partial\vectorsym{\centerlinepos}}{\partial\arclength} \times \rotationmatrix\bcdot\vectorsym{\relref{\internalforce}}
                + \externaltorquedensity\label{eq:str_EOM_step2b}
                = \rhsvecomega
                \eqdot
            \end{gather}
        \end{subequations}
        
        \item Left-multiplying \eqref{eq:str_EOM_step2b} with $\transposed{\rotationmatrix}$ yields, since $\transposed{\rotationmatrix}\bcdot\rotationmatrix = \identity$,
        \begin{subequations}\label{eq:str_EOM_step3}
        \begin{gather}
            \bdensity \left( \relref{\crosssectiontensorofinertia}\bcdot\relref{\tderiv{\vectorsym{\angvel}}} + \vectorsym{\relref{\angvel}}\times\relref{\crosssectiontensorofinertia}\bcdot\vectorsym{\relref{\angvel}} \right)
                    = \transposed{\rotationmatrix}\bcdot\rhsvecomega
                    = \relref{{\rhsvecomega}}
            \\
            \relref{{\rhsvecomega}}
            = \transposed{\rotationmatrix}\bcdot\frac{\partial\rotationmatrix}{\partial\arclength}\bcdot\vectorsym{\relref{\internaltorque}}
            + \frac{\partial\vectorsym{\relref{\internaltorque}}}{\partial\arclength}
            + \left(\transposed{\rotationmatrix}\bcdot\frac{\partial\vectorsym{\centerlinepos}}{\partial\arclength}\right) \times \vectorsym{\relref{\internalforce}}
            + \transposed{\rotationmatrix}\bcdot\externaltorquedensity
            \eqdot
        \end{gather}
        \end{subequations}
        
        \item Translate \eqref{eq:str_EOM_step2a} and \eqref{eq:str_EOM_step3} into quaternion language.
        The former equation \eqref{eq:str_EOM_step2a} can readily be converted by using the convention
        $\vectorsym{\internalforce}  = \rotationmatrix\bcdot\vectorsym{\relref{\internalforce}}  = \quaternion\qcdot\vectorsym{\relref{\internalforce}} \qcdot\qconj{\quaternion}$.
        The conversion of \eqref{eq:str_EOM_step3} requires more advanced techniques and involves a number of intermediate steps, not printed here.
        The reader is referred to \cite[p.13]{LangLagrangian2009} for details.
        The resulting equations then read
        \begin{subequations}
            \begin{gather}
                \bdensity\bladeA\ttderiv{\vectorsym{\centerlinepos}} = \frac{\partial\quaternion\qcdot\vectorsym{\relref{\internalforce}}\qcdot\qconj{\quaternion}}{\partial\arclength} + \externalforcedensity
                \\
                \label{eq:str_EOM_step4b}
                \bdensity ( \relref{\crosssectiontensorofinertia}\bcdot\relref{\tderiv{\vectorsym{\angvel}}}
                          - \vectorsym{\relref{\angvel}}\times\relref{\crosssectiontensorofinertia}\bcdot\vectorsym{\relref{\angvel}})
                   = \relref{{\rhsvecomega}}
            \eqcomma
            \intertext{with}
            \begin{align}
                \relref{\crosssectiontensorofinertia}\bcdot\relref{\tderiv{\vectorsym{\angvel}}}
                    &= \tfrac{1}{2}\qconj{\quaternion}\qcdot\quaternionmatrixofinertia\bcdot\ttderiv{\quaternion}
                \\
                \vectorsym{\relref{\angvel}}\times\relref{\crosssectiontensorofinertia}\bcdot\vectorsym{\relref{\angvel}}
                    &= 4\qconj{\quaternion}\qcdot\tderiv{\quaternion}\qcdot\quaternionmatrixofinertia\bcdot(\tderiv{\qconj{\quaternion}}\qcdot\quaternion) + \vectorsym{\eta}\ind{\ell}
                \\
                \relref{{\rhsvecomega}}
                    = \qconj{\quaternion}\qcdot\rhsvecomega\qcdot\quaternion
                    &= 2\qconj{\quaternion}\qcdot\frac{\partial\quaternion}{\partial\arclength}\qcdot\vectorsym{\relref{\internaltorque}}
                    + \frac{\partial\vectorsym{\relref{\internaltorque}}}{\partial\arclength}
                    + \qconj{\quaternion}\qcdot\frac{\partial\vectorsym{\centerlinepos}}{\partial\arclength}\qcdot\quaternion\qcdot\vectorsym{\relref{\internalforce}}
                    + \qconj{\quaternion}\qcdot\externaltorquedensity\qcdot\quaternion
                    + \vectorsym{\eta}\ind{r}
                    \eqcomma
            \end{align}
            \end{gather}
        \end{subequations}
        where $\vectorsym{\eta}\ind{\ell}$ and $\vectorsym{\eta}\ind{r}$ collect purely real summands
        and $\quaternionmatrixofinertia = \quaternionmatrixofinertia\of{\quaternion}$ is the quaternion matrix of inertia
        (that actually also depends on the constant tensor $\relref{\crosssectiontensorofinertia}$).
        The latter is defined such that 
        \begin{equation}
            \frac{\bdensity}{2} \transposed{\tderiv{\quaternion}}\bcdot\quaternionmatrixofinertia\bcdot\tderiv{\quaternion}
            = \frac{\bdensity}{2} \transposed{\vectorsym{\relref{\angvel}}}\bcdot\relref{\crosssectiontensorofinertia}\bcdot\vectorsym{\relref{\angvel}}
        \end{equation}
        provides the rotary kinetic energy density.
        The rather lengthy expression for 
        $\quaternionmatrixofinertia$
        can be found in \cite[p.8~\enquote{$\mu\of{p}$}]{LangLagrangian2009}.
        
        \item Left-multiply \eqref{eq:str_EOM_step4b} with $2\quaternion$.
        \begin{multline}
            \bdensity( \quaternionmatrixofinertia\bcdot\ttderiv{\quaternion} - 8\tderiv{\quaternion}\qcdot\quaternionmatrixofinertia\bcdot(\tderiv{\qconj{\quaternion}}\qcdot\quaternion) + 2\quaternion\qcdot\vectorsym{\eta}\ind{\ell} )
            \\\begin{aligned}
            &=    4\frac{\partial\quaternion}{\partial\arclength}\qcdot\vectorsym{\relref{\internaltorque}}
                + 2\quaternion\qcdot\frac{\partial\vectorsym{\relref{\internaltorque}}}{\partial\arclength}
                + 2\frac{\partial\vectorsym{\centerlinepos}}{\partial\arclength}\qcdot\quaternion\qcdot\vectorsym{\relref{\internalforce}}
                + 2\externaltorquedensity\qcdot\quaternion
                + 2\quaternion\qcdot\vectorsym{\eta}\ind{r}
            \\
            &=    2\frac{\partial\quaternion}{\partial\arclength}\qcdot\vectorsym{\relref{\internaltorque}}
                + 2\frac{\partial\quaternion\qcdot\vectorsym{\relref{\internaltorque}}}{\partial\arclength}
                + 2\frac{\partial\vectorsym{\centerlinepos}}{\partial\arclength}\qcdot\quaternion\qcdot\vectorsym{\relref{\internalforce}}
                + 2\externaltorquedensity\qcdot\quaternion
                + 2\quaternion\qcdot\vectorsym{\eta}\ind{r}
            \\
            & 
               = 2\quaternion\qcdot\relref{{\rhsvecomega}}
            \end{aligned}
        \end{multline}
        
        \item Defining the \enquote{\enquote{tangential inverse} quaternion mass}~\cite[p.9]{LangLagrangian2009} $\inversequaternionmatrixofinertia = \inversequaternionmatrixofinertia\of{\quaternion}$, 
        with $\quaternionmatrixofinertia\bcdot\inversequaternionmatrixofinertia = \identity - \quaternion\dyaddot\quaternion$,
        multiplication with $\inversequaternionmatrixofinertia$ yields
        \begin{align}
            ( \identity - \quaternion\dyaddot\quaternion )\qcdot\ttderiv{\quaternion}
            &=  \frac{2}{\bdensity}\inversequaternionmatrixofinertia\bcdot
                \begin{multlined}[t]
                \Big[
                    \underbrace{
                          4\bdensity\tderiv{\quaternion}\qcdot\quaternionmatrixofinertia\bcdot(\tderiv{\qconj{\quaternion}}\qcdot\quaternion)
                        - \quaternion\qcdot\vectorsym{\eta}\ind{\ell}
                    \vphantom{\bigg(}}_{=-\quaternion\qcdot(\vectorsym{\relref{\angvel}}\times\relref{\crosssectiontensorofinertia}\bcdot\vectorsym{\relref{\angvel}})}
                    \\
                    + \underbrace{
                      \frac{\partial\quaternion}{\partial\arclength}\qcdot\vectorsym{\relref{\internaltorque}}
                    + \frac{\partial\quaternion\qcdot\vectorsym{\relref{\internaltorque}}}{\partial\arclength}
                    + \frac{\partial\vectorsym{\centerlinepos}}{\partial\arclength}\qcdot\quaternion\qcdot\vectorsym{\relref{\internalforce}}
                    + \externaltorquedensity\qcdot\quaternion
                    + \quaternion\qcdot\vectorsym{\eta}\ind{r}
                    \vphantom{\bigg(}}_{
                        = \quaternion\qcdot\relref{{\rhsvecomega}}
                    }
                \Big]\eqdot
                \end{multlined}
        \end{align}

        \item The terms with $\vectorsym{\eta}\ind{\ell}$ and $\vectorsym{\eta}\ind{r}$ disappear, because $\qimag{\vectorsym{\eta}\ind{\ell}} = \qimag{\vectorsym{\eta}\ind{r}} = 0$
        and $\inversequaternionmatrixofinertia\bcdot\quaternion=0$~\cite[p.9]{LangLagrangian2009}:
        \begin{equation}
            \inversequaternionmatrixofinertia\bcdot(\quaternion\qcdot\vectorsym{\eta})
            = \inversequaternionmatrixofinertia\bcdot\quaternion\,\qreal{\vectorsym{\eta}}
            = 0
            \eqcomma
        \end{equation}
        and it follows that
        \begin{equation}\label{eq:str_EOM_quaternion_rot_appendix}
            \begin{multlined}[b]
               \ttderiv{\quaternion} = 
               \frac{2}{\bdensity}\inversequaternionmatrixofinertia\bcdot\Big[
                \underbrace{
                      4\bdensity\tderiv{\quaternion}\qcdot\quaternionmatrixofinertia\bcdot(\tderiv{\qconj{\quaternion}}\qcdot\quaternion)
                \vphantom{\bigg(}}_{\mathclap{=-\quaternion\qcdot(\vectorsym{\relref{\angvel}}\times\relref{\crosssectiontensorofinertia}\bcdot\vectorsym{\relref{\angvel}})+\quaternion\qcdot\vectorsym{\eta}\ind{\ell}}}
                + \underbrace{
                  \frac{\partial\quaternion}{\partial\arclength}\qcdot\vectorsym{\relref{\internaltorque}}
                + \frac{\partial\quaternion\qcdot\vectorsym{\relref{\internaltorque}}}{\partial\arclength}
                + \frac{\partial\vectorsym{\centerlinepos}}{\partial\arclength}\qcdot\quaternion\qcdot\vectorsym{\relref{\internalforce}}
                + \externaltorquedensity\qcdot\quaternion
                \vphantom{\bigg(}}_{
                    = \quaternion\qcdot\relref{{\rhsvecomega}} - \quaternion\qcdot\vectorsym{\eta}\ind{r}
                }
            \Big]
            \\+ (\quaternion\dyaddot\quaternion)\qcdot\ttderiv{\quaternion}
            \eqdot
            \end{multlined}
        \end{equation}

        \item Finally, with \cite[p.12--13]{LangLagrangian2009}\cite[p.293]{LangMultibody2011}
        \begin{gather}
            (\quaternion\dyaddot\quaternion)\qcdot\ttderiv{\quaternion}
            = ( \quaternion\qcdot\ttderiv{\quaternion} ) \qcdot \quaternion
            = -\vnorm{\tderiv{\quaternion}}^2 \quaternion
            \\
            \vnorm{\quaternion} = 1
        \end{gather}
        the quaternion-based equations of motion are obtained.
    \end{enumerate}

%% file: chapters/fluid-structure-coupling/appendix.tex

This appendix serves to determine 
the fluid-structure coupling loads employed in the \ac{EOM} of the structures,
\eqref{eq:str_EOM_withfluidloads} and \eqref{eq:structure_eom_coupled2}.
The strategy is inspired by the one of
Tschisgale \cite[§2.4.2--2.4.3]{TschisgaleNumerical2018},
but is different with regard to the control geometry chosen.
Here, it is based on a control plane aligned with a cross-section of the structure, 
depicted in \cref{fig:controlplane}.
It was chosen this way because it links very well to the \ac{EOM} of the Cosserat rod.
This somewhat simplified approach is different from the one in \cite{TschisgaleNumerical2018}
that is based on a small control volume.
In addition, this appendix aims to provide all intermediate steps needed to follow the derivation,
thus supplementing the compact description in \cite{TschisgaleNumerical2018}.

\leveldown{Dynamic coupling via surface-specific force}
\label{sec:dynamiccoupling_surfacespecific}
\subimport{chapters/}{dynamic-via-volume-force}

\levelstay{Kinematic coupling}
\label{sec:kinematicoupling}
\subimport{chapters/}{kinematic-coupling}

\levelstay{Modification of the coupling terms for continuous time integration}
\label{sec:coupling_for_continuous_integration}
\subimport{chapters/}{mod-for-continuous-time-integration}

\levelstay{Separating fluid layer inertia from coupling loads}
\label{sec:effective_parameters}
\subimport{chapters/}{separating-fluid-layer-inertia}

%% file: chapters/fluid-structure-coupling/chapters/dynamic-via-volume-force.tex

Integration of \cref{eq:NSE_} over the control plane depicted in \cref{fig:controlplane} yields
\begin{subequations}
\begin{align}
    \int\limits_{\planespace\setminus\crosssectionspace}\!\left(\fdensity\frac{\Tdiff{\vectorsym{\vel}}}{\tdiff{t}}-\fdensity\massspecificforce\right)\,\dSurface
    &= \int\limits_{\planespace\setminus\crosssectionspace}\!\nabla\bcdot\tensorsym{\hydrodynamicstress}\,\dSurface
    \\
    &= \int\limits_{\planespace}\!\nabla\bcdot\tensorsym{\hydrodynamicstress}\,\dSurface
     -\int\limits_{\crosssectionspace}\!\nabla\bcdot\tensorsym{\hydrodynamicstress}\,\dSurface
    \eqdot
\end{align}
\end{subequations}
Applying Gauss' theorem, the rightmost term can be converted into a line integral along the fluid structure interface 
\begin{equation}
    \int\limits_{\planespace\setminus\crosssectionspace}\!\left(\fdensity\frac{\Tdiff{\vectorsym{\vel}}}{\tdiff{t}}-\fdensity\massspecificforce\right)\,\dSurface
    = \int\limits_{\planespace}\!\nabla\bcdot\tensorsym{\hydrodynamicstress}\,\dSurface
     +\int\limits_{\partial\crosssectionspace}\!\tensorsym{\hydrodynamicstress}\bcdot\normalvec\,\dCurve
\end{equation}

By reducing the width of the blade to zero,
$\bladeT\to0$,
the situation in \cref{fig:controlplane} is converted into that of \cref{fig:controlplanezerowidth}, 
\usuk{i.e.,}{i.e.} the blade is a ribbon with
    zero thickness $\bladeT\to0$, 
    zero cross-sectional area $\crosssectionspace\to\emptyset$,
    and its cross-sectional circumference reducing to a single line $\partial\crosssectionspace\to\fsinterfacesectionspace$
    normal to which is the unit vector $\normalvec\coloneqq\normalvec\Ind{+}=-\normalvec\Ind{-}$.
    With
      $\planespace\setminus\crosssectionspace \to \planespace$
    it follows
\begin{subequations}
\begin{gather}
    \int\limits_{\planespace}\!\left(
        \fdensity\frac{\Tdiff{\vectorsym{\vel}}}{\tdiff{t}}
        -\nabla\bcdot\tensorsym{\hydrodynamicstress}
        -\fdensity\massspecificforce
    \right)\,\dSurface
    = \int\limits_{\fsinterfacesectionspace}\!\surfacecouplingforce\,\dCurve
    ,\quad
    \surfacecouplingforce \coloneqq (\tensorsym{\relpside{\hydrodynamicstress}}-\tensorsym{\relmside{\hydrodynamicstress}})\bcdot\normalvec
    \quad \forall\xvec\in\fsinterface
    \\
    \label{eq:coupled2_dynamiccoupling}
    \externalforcedensityrelfsinterface = -\int\limits_\fsinterfacesectionspace\!\surfacecouplingforce\,\dCurve,\qquad
    \externaltorquedensityrelfsinterface = -\int\limits_\fsinterfacesectionspace\!\vectorsym{\relpos}\times\surfacecouplingforce\,\dCurve
    \\
    \label{eq:coupled2_kinematiccoupling}
    \vectorsym{\vel}\relfsinterface 
    = \left.\vectorsym{\vel}\right|_{\xvec=\vectorsym{\centerlinepos} + \vectorsym{\relpos}} 
    = \tderiv{\vectorsym{\centerlinepos}} + \vectorsym{\angvel}\times\vectorsym{\relpos}
    \eqdot
\end{gather}
\end{subequations}

The surface-specific coupling force $\surfacecouplingforce$ establishes the dynamic coupling between the structure and the fluid,
but it is only properly defined on the coupling surface.
With the aid of an appropriate Dirac delta distribution $\diracfsinterface\of{\xvec}$ with $\diracfsinterface\of{\xvec\notin\fsinterface} = 0$,
the line integral over $\fsinterfacesectionspace$ can be converted into an area integral over $\planespace$
\begin{equation}
    \int\limits_{\planespace}\!\left(
        \fdensity\frac{\Tdiff{\vectorsym{\vel}}}{\tdiff{t}}
        -\nabla\bcdot\tensorsym{\hydrodynamicstress}
        -\fdensity\massspecificforce
    \right)\,\dSurface
    = \int\limits_{\planespace}\diracfsinterface\surfacecouplingforce\,\dSurface
\end{equation}
such that the momentum equation of the fluid reads
\begin{equation}\label{eq:NSE_2}
    \fdensity\frac{\Tdiff{\vectorsym{\vel}}}{\tdiff{t}}
    = \nabla\bcdot\tensorsym{\hydrodynamicstress}
    + \fdensity\massspecificforce
    + \diracfsinterface\surfacecouplingforce
    \eqdot
\end{equation}

%% file: chapters/fluid-structure-coupling/chapters/kinematic-coupling.tex
%
Integrating \cref{eq:NSE_2} over $t\in]t^{(\rkstep-1)},t^{(\rkstep)}]$ gives
\begin{equation}
      \int\limits_{\mathclap{t^{(\rkstep-1)}}}^{\mathclap{t^{(\rkstep)}}}\!\frac{\partial{\vectorsym{\vel}}}{\partial{t}}\,\tdiff{t}
    = \vectorsym{\vel}^{(\rkstep)} - \vectorsym{\vel}^{(\rkstep-1)}
    = \int\limits_{\mathclap{t^{(\rkstep-1)}}}^{\mathclap{t^{(\rkstep)}}}\!\left(- \nabla\bcdot(\vectorsym{\vel}\dyaddot\vectorsym{\vel}) + \frac{1}{\fdensity}\nabla\bcdot\tensorsym{\hydrodynamicstress} + \massspecificforce\right)\,\tdiff{t} 
    + \int\limits_{\mathclap{t^{(\rkstep-1)}}}^{\mathclap{t^{(\rkstep)}}}\!\frac{\diracfsinterface\surfacecouplingforce}{\fdensity}\,\tdiff{t}
    \eqdot
\end{equation}
This equation can be written as 
\begin{equation}
    \vectorsym{\vel}^{(\rkstep)} - \uprelimvec
    = \int\limits_{\mathclap{t^{(\rkstep-1)}}}^{\mathclap{t^{(\rkstep)}}}\!\frac{\diracfsinterface\surfacecouplingforce}{\fdensity}\,\tdiff{t}
    ,\qquad
    \uprelimvec
    \coloneqq
    \vectorsym{\vel}^{(\rkstep-1)} + \int\limits_{\mathclap{t^{(\rkstep-1)}}}^{\mathclap{t^{(\rkstep)}}}\!\left(- \nabla\bcdot(\vectorsym{\vel}\dyaddot\vectorsym{\vel}) + \frac{1}{\fdensity}\nabla\bcdot\tensorsym{\hydrodynamicstress} + \massspecificforce\right)\,\tdiff{t} 
    \eqcomma
\end{equation}
where $\uprelimvec$ is a preliminary updated fluid velocity which does not account for the fluid structure coupling.

By considering $\diracfsinterface\surfacecouplingforce$ as the effective average coupling force during the integration step,
\usuk{i.e.,}{i.e.}
$\diracfsinterface\surfacecouplingforce = \mathrm{const.}\ \forall\ t\in]t^{(\rkstep-1)},t^{(\rkstep)}]$,
the remaining integral can be resolved,
\begin{equation}
    \frac{\diracfsinterface\surfacecouplingforce}{\fdensity}
    = \frac{\vectorsym{\vel}^{(\rkstep)} - \uprelimvec}{t^{(\rkstep)}-t^{(\rkstep-1)}}
    \ \text{.}
\end{equation}

The surface-specific force $\surfacecouplingforce$ is only defined at the fluid structure interface.
In the framework of the immersed boundary method, spatial integrals over $\diracfsinterface\surfacecouplingforce$
are transformed into integrals over a mass-specific force $\vectorsym{\volumecouplingforce}$, 
as proposed by~\cite{TschisgaleGeneral2018}.
This is accomplished with common techniques of distribution theory~\cite{OnuralImpulse2006,FarassatIntroduction1994},
resulting in
\begin{equation}
    \label{eq:volumecouplingforce}
    \frac{\diracfsinterface\surfacecouplingforce}{\fdensity}
    \approx
    \vectorsym{\volumecouplingforce}
    \coloneqq
    \Spr\braceof{
        \frac{\vectorsym{\vel}\relfsinterface^{(\rkstep)} - \uprelimvec\relfsinterface}{t^{(\rkstep)}-t^{(\rkstep-1)}}
    }
    ,\quad
    \uprelimvec\relfsinterface
    \coloneqq
    \IntOp\braceof{\uprelimvec}
    \eqcomma
\end{equation}
where the spreading operator $\Spr$ spreads in the compact proximity of the interface,
such that $\vectorsym{\volumecouplingforce} = 0\ \forall\ \xvec\notin\forcingvolumespace$,
while the operation $\IntOp$ interpolates field values at the interface.
This gives
\begin{subequations}
\begin{gather}
    \label{eq:NSE_3}
    \fdensity\frac{\Tdiff{\vectorsym{\vel}}}{\tdiff{t}}
    = \nabla\bcdot\tensorsym{\hydrodynamicstress}
    + \fdensity\massspecificforce
    + \fdensity\vectorsym{\volumecouplingforce}
    \\
    \label{eq:coupled3_dynamiccoupling_lin}
    \externalforcedensityrelfsinterface 
        = -\int\limits_\planespace\!\diracfsinterface\surfacecouplingforce\,\dSurface
        \approx -\int\limits_\planespace\!\fdensity\vectorsym{\volumecouplingforce}\,\dSurface
        = -\int\limits_\forcingcrosssectionspace\!\fdensity\vectorsym{\volumecouplingforce}\,\dSurface
        = -\int\limits_\forcingcrosssectionspace\!\fdensity\,
            \Spr\braceof{ \frac{\vectorsym{\vel}\relfsinterface^{(\rkstep)} - \uprelimvec\relfsinterface}{t^{(\rkstep)}-t^{(\rkstep-1)}} }
          \,\dSurface
    \\
    \label{eq:coupled3_dynamiccoupling_rot}
    \begin{multlined}
    \externaltorquedensityrelfsinterface
        = -\int\limits_\planespace\!\vectorsym{\relpos}\times\diracfsinterface\surfacecouplingforce\,\dSurface
        \approx -\int\limits_\planespace\!\vectorsym{\relpos}\times\fdensity\vectorsym{\volumecouplingforce}\,\dSurface
        = -\int\limits_\forcingcrosssectionspace\!\vectorsym{\relpos}\times\fdensity\vectorsym{\volumecouplingforce}\,\dSurface
        \\
        = -\int\limits_\forcingcrosssectionspace\!\vectorsym{\relpos}\times\fdensity\,
             \Spr\braceof{ \frac{\vectorsym{\vel}\relfsinterface^{(\rkstep)} - \uprelimvec\relfsinterface}{t^{(\rkstep)}-t^{(\rkstep-1)}} }
          \,\dSurface
    \eqdot
    \end{multlined}
\end{gather}
\end{subequations}

Inserting the kinematic coupling relation of \cref{eq:coupled2_kinematiccoupling},
\begin{subequations}
\begin{gather}
    \externalforcedensityrelfsinterface 
        = -\int\limits_\forcingcrosssectionspace\!\fdensity\,
            \Spr\braceof{ \frac{ \left[\tderiv{\vectorsym{\centerlinepos}} + \vectorsym{\angvel}\times\vectorsym{\relpos}\right]^{(\rkstep)}
                                 - \uprelimvec\relfsinterface}{t^{(\rkstep)}-t^{(\rkstep-1)}}
                               }
          \,\dSurface
    \\
    \externaltorquedensityrelfsinterface
        = -\int\limits_\forcingcrosssectionspace\!\vectorsym{\relpos}\times\fdensity\,
             \Spr\braceof{\frac{ \left[\tderiv{\vectorsym{\centerlinepos}} + \vectorsym{\angvel}\times\vectorsym{\relpos}\right]^{(\rkstep)}
                                 - \uprelimvec\relfsinterface}{t^{(\rkstep)}-t^{(\rkstep-1)}}
                               }
          \,\dSurface
    \eqcomma
\end{gather}
\end{subequations}
and taking advantage of the linearity of $\Spr$,
of its conservative nature
    $\int_{\forcingcrosssectionspace}\!\Spr\braceof{\uprelimvec\relfsinterface}\,\dSurface 
    = \int_{\fsinterfacesectionspace}\!\uprelimvec\relfsinterface\,\dCurve$,
of $\fdensity=\mathrm{const}$,
and that $\vectorsym{\centerlinepos}=\vectorsym{\centerlinepos}\of{\arclength}$ and $\vectorsym{\angvel}=\vectorsym{\angvel}\of{\arclength}$ 
are constants for a given integral over $\forcingcrosssectionspace$,
\cref{eq:coupled3_dynamiccoupling_lin,eq:coupled3_dynamiccoupling_rot} can be written as
\begin{subequations}\label{eq:coupled3b}
\begin{gather}
    \externalforcedensityrelfsinterface 
    = -(t^{(\rkstep)}-t^{(\rkstep-1)})^{-1}
       \left(
           \tderiv{\vectorsym{\centerlinepos}}^{(\rkstep)} \int_{\forcingcrosssectionspace}\!\fdensity\,\dSurface
         + \vectorsym{\angvel}^{(\rkstep)} \times    \int_{\forcingcrosssectionspace}\!\vectorsym{\relpos}^{(\rkstep)}\fdensity\,\dSurface
         -                              \int_{\forcingcrosssectionspace}\!\uprelimvec\relfsinterface\fdensity\,\dSurface
       \right)
    \\
    \externaltorquedensityrelfsinterface
    = -(t^{(\rkstep)}-t^{(\rkstep-1)})^{-1}
       \left(
          \int_{\forcingcrosssectionspace}\!\vectorsym{\relpos}\fdensity\,\dSurface\times\tderiv{\vectorsym{\centerlinepos}}
        + \int_{\forcingcrosssectionspace}\!\transposed{\skewmatrix{\vectorsym{\relpos}}}\bcdot\skewmatrix{\vectorsym{\relpos}^{(\rkstep)}}\fdensity\,\dSurface\bcdot\vectorsym{\angvel}^{(\rkstep)}
        - \int_{\forcingcrosssectionspace}\!\vectorsym{\relpos}\times\uprelimvec\relfsinterface\fdensity\,\dSurface
       \right)
    \eqdot
\end{gather}
\end{subequations}
Rearranging the terms gives
\begin{subequations}\label{eq:coupled3}
\begin{alignat}{3}
    -\externalforcedensityrelfsinterface 
    &
   = \frac{{\sderiv{\vectorsym{\linmom}}}^{(\rkstep)} - \sderiv{\tilde{\vectorsym{\linmom}}}}{t^{(\rkstep)}-t^{(\rkstep-1)}}
   &&\quad\text{with}\quad
   {\sderiv{\vectorsym{\linmom}}}^{(\rkstep)} = \tderiv{\vectorsym{\centerlinepos}}^{(\rkstep)} \sderiv{\tilde{m}} + \zero{ \vectorsym{\angvel}^{(\rkstep)} \times    {\sderiv{\vectorsym{s}}}^{(\rkstep)}}
   ,\quad
   &&\sderiv{\tilde{\vectorsym{\linmom}}}   = - \int_{\forcingcrosssectionspace}\!\uprelimvec\relfsinterface\fdensity\,\dSurface
   \\
    -\externaltorquedensityrelfsinterface
    &
    = \frac{{\sderiv{\vectorsym{\angmom}}}^{(\rkstep)} - \sderiv{\tilde{\vectorsym{\angmom}}}}{t^{(\rkstep)}-t^{(\rkstep-1)}}
    &&\quad\text{with}\quad
    {\sderiv{\vectorsym{\angmom}}}^{(\rkstep)} = \zero{\sderiv{\vectorsym{s}}\times\tderiv{\vectorsym{\centerlinepos}}^{(\rkstep)}} + \forcingcrosssectiontensorofinertia^{(\rkstep)}\bcdot\vectorsym{\angvel}^{(\rkstep)}
    ,\quad
    &&\sderiv{\tilde{\vectorsym{\angmom}}} = \int_{\forcingcrosssectionspace}\!\vectorsym{\relpos}\times\uprelimvec\relfsinterface\fdensity\,\dSurface
\end{alignat}
\begin{equation}
    \sderiv{\tilde{m}}          = \int\limits_\forcingcrosssectionspace\!\fdensity\,\dSurface
    ,\qquad
    \zero{ \sderiv{\vectorsym{s}}     = \int\limits_\forcingcrosssectionspace\!\vectorsym{\relpos}\fdensity\,\dSurface = 0 }
    ,\qquad
    \forcingcrosssectiontensorofinertia^{(\rkstep)} = \int\limits_{\forcingcrosssectionspace}\!\transposed{\skewmatrix{\vectorsym{\relpos}}}\bcdot\skewmatrix{\vectorsym{\relpos}^{(\rkstep)}}\fdensity\,\dSurface
    \eqdot
\end{equation}
\end{subequations}

%% file: chapters/fluid-structure-coupling/chapters/mod-for-continuous-time-integration.tex
%
For the efficient numerical solution of the equations of motion of the structures,
established black-boxed solvers are desirable.
These solvers require a function for the right hand side that can be evaluated at any intermediate time, 
between the discrete time steps, 
to enable an implicit solution of the problem.
With the formulations of \eqref{eq:coupled3} this is not the case, because 
$\externalforcedensityrelfsinterface$ and $\externaltorquedensityrelfsinterface$
depend on the solution at $t^{(\rkstep)}$.
To address this issue, 
the momentum densities ${\sderiv{\vectorsym{\linmom}}}^{(\rkstep)}$, ${\sderiv{\vectorsym{\angmom}}}^{(\rkstep)}$
are extrapolated linearly from ${\sderiv{\vectorsym{\linmom}}}\of{t}$, ${\sderiv{\vectorsym{\angmom}}}\of{t}$, \usuk{i.e.,}{i.e.}
\begin{subequations}
\begin{align}\label{eq:constantaccelerationapproximation}
    {\sderiv{\vectorsym{\linmom}}}^{(\rkstep)} &\overset{!}{=} {\sderiv{\vectorsym{\linmom}}}^{(\rkstep-1)} + \frac{t^{(\rkstep)}-t^{(\rkstep-1)}}{t-t^{(\rkstep-1)}}\left( {\sderiv{\vectorsym{\linmom}}}\of{t}-{\sderiv{\vectorsym{\linmom}}}^{(\rkstep-1)} \right)\ \\
    {\sderiv{\vectorsym{\angmom}}}^{(\rkstep)} &\overset{!}{=} {\sderiv{\vectorsym{\angmom}}}^{(\rkstep-1)} + \frac{t^{(\rkstep)}-t^{(\rkstep-1)}}{t-t^{(\rkstep-1)}}\left( {\sderiv{\vectorsym{\angmom}}}\of{t}-{\sderiv{\vectorsym{\angmom}}}^{(\rkstep-1)} \right) 
    \eqcomma
\end{align}
\end{subequations}
and the approximations
\begin{equation}\label{eq:relposvecapproximation}
    \vectorsym{\relpos}\of{t}     \overset{!}{=} \vectorsym{\relpos}^{(\rkstep-1)},\quad
    \vectorsym{\relpos}^{(\rkstep)} \overset{!}{=} \vectorsym{\relpos}^{(\rkstep-1)}
\end{equation}
are made wherever needed for the computations of
$\sderiv{\tilde{\vectorsym{\angmom}}}$, $\forcingcrosssectiontensorofinertia$, and $\sderiv{\vectorsym{s}}$,
\begin{subequations}\label{eq:coupled4}
    \input{equations/coupling-load-densities4}
\end{subequations}

%% file: chapters/fluid-structure-coupling/chapters/equations/coupling-load-densities4.tex
\begin{alignat}{3}
-\externalforcedensityrelfsinterface
&=   \frac{ {\sderiv{\vectorsym{\linmom}}}\of{t}       - {\sderiv{\vectorsym{\linmom}}}^{(\rkstep-1)} }{t        -t^{(\rkstep-1)}}
+ \frac{ {\sderiv{\vectorsym{\linmom}}}^{(\rkstep-1)} - \sderiv{\tilde{\vectorsym{\linmom}}}       }{t^{(\rkstep)}-t^{(\rkstep-1)}}
&&\quad\text{with}\quad
\begin{alignedat}[t]{2}
    &{\sderiv{\vectorsym{\linmom}}}\of{t}       &&= \sderiv{\tilde{m}}\tderiv{\vectorsym{\centerlinepos}}\,           + \zero{ \vectorsym{\angvel}             \times \sderiv{\vectorsym{s}} },\\
    &{\sderiv{\vectorsym{\linmom}}}^{(\rkstep-1)} &&= \sderiv{\tilde{m}}\tderiv{\vectorsym{\centerlinepos}}^{(\rkstep-1)} + \zero{ \vectorsym{\angvel}^{(\rkstep-1)} \times \sderiv{\vectorsym{s}} },\\
    & \sderiv{\tilde{\vectorsym{\linmom}}}      &&= -\int_{\forcingcrosssectionspace}\!\uprelimvec\relfsinterface\fdensity\,\dSurface
\end{alignedat}
\\
-\externaltorquedensityrelfsinterface
&=   \frac{ {\sderiv{\vectorsym{\angmom}}}\of{t}       - {\sderiv{\vectorsym{\angmom}}}^{(\rkstep-1)} }{t        -t^{(\rkstep-1)}}
+ \frac{ {\sderiv{\vectorsym{\angmom}}}^{(\rkstep-1)} - \sderiv{\tilde{\vectorsym{\angmom}}}       }{t^{(\rkstep)}-t^{(\rkstep-1)}}
&&\quad\text{with}\quad
\begin{alignedat}[t]{2}
    &{\sderiv{\vectorsym{\angmom}}}\of{t}       &&= \zero{\sderiv{\vectorsym{s}}\times\tderiv{\vectorsym{\centerlinepos}}            } + \forcingcrosssectiontensorofinertia\bcdot\vectorsym{\angvel},\\
    &{\sderiv{\vectorsym{\angmom}}}^{(\rkstep-1)} &&= \zero{\sderiv{\vectorsym{s}}\times\tderiv{\vectorsym{\centerlinepos}}^{(\rkstep-1)}} + \forcingcrosssectiontensorofinertia\bcdot\vectorsym{\angvel}^{(\rkstep-1)},\\
    &\sderiv{\tilde{\vectorsym{\angmom}}}       &&= \int_{\forcingcrosssectionspace}\!\vectorsym{\relpos}^{(\rkstep-1)}\times\uprelimvec\relfsinterface\fdensity\,\dSurface
\end{alignedat}
\end{alignat}
\begin{equation}
\sderiv{\tilde{m}}                          = \int\limits_{\forcingcrosssectionspace}\!\fdensity\,\dSurface
,\qquad
\zero{ \sderiv{\vectorsym{s}}               = \int\limits_{\forcingcrosssectionspace}\!\vectorsym{\relpos}^{(\rkstep-1)}\fdensity\,\dSurface = 0 }
,\qquad
\forcingcrosssectiontensorofinertia = \int\limits_{\forcingcrosssectionspace}\!\transposed{\skewmatrix{\vectorsym{\relpos}^{(\rkstep-1)}}}\bcdot\skewmatrix{\vectorsym{\relpos}^{(\rkstep-1)}}\fdensity\,\dSurface
\eqdot
\end{equation}

%% file: chapters/fluid-structure-coupling/chapters/separating-fluid-layer-inertia.tex
%
\Cref{eq:coupled3_dynamiccoupling_lin,eq:coupled3_dynamiccoupling_rot} can be written
\begin{subequations}
\begin{gather}
    \externalforcedensityrelfsinterface 
        = -\int\limits_\forcingcrosssectionspace\!\fdensity\,
            \Spr\braceof{
                  \tderiv{\vectorsym{\vel}}\relfsinterface
                + \frac{\vectorsym{\vel}\relfsinterface^{(\rkstep-1)} - \uprelimvec\relfsinterface}{t^{(\rkstep)}-t^{(\rkstep-1)}}
            }\,\dSurface
    \\
    \externaltorquedensityrelfsinterface
        = -\int\limits_\forcingcrosssectionspace\!\vectorsym{\relpos}\times\fdensity\,
             \Spr\braceof{ 
                  \tderiv{\vectorsym{\vel}}\relfsinterface
                + \frac{\vectorsym{\vel}\relfsinterface^{(\rkstep-1)} - \uprelimvec\relfsinterface}{t^{(\rkstep)}-t^{(\rkstep-1)}} 
             }\,\dSurface
    \eqdot
\end{gather}
\end{subequations}
where $\tderiv{\vectorsym{\vel}}\relfsinterface$ is the effective acceleration of the interface during $t\in]t^{(\rkstep-1)},t^{(\rkstep)}]$
that can be substituted with the time derivative of the kinematic coupling condition, \usuk{i.e.,}{i.e.}
\begin{equation}
    \tderiv{\vectorsym{\vel}}\relfsinterface
        = \frac{\vectorsym{\vel}\relfsinterface^{(\rkstep)} - \vectorsym{\vel}\relfsinterface^{(\rkstep-1)}}{t^{(\rkstep)}-t^{(\rkstep-1)}}
        = \frac{ \pdiff{ ( \tderiv{\vectorsym{\centerlinepos}} + \vectorsym{\angvel}\times\vectorsym{\relpos} ) } }{ \pdiff{t} }
        = \ttderiv{\vectorsym{\centerlinepos}} + \tderiv{\vectorsym{\angvel}}\times\vectorsym{\relpos} + \vectorsym{\angvel}\times(\vectorsym{\angvel}\times\vectorsym{\relpos})
    \eqdot
\end{equation}
where the assumptions 
$\ttderiv{\vectorsym{\centerlinepos}}\overset{!}{=}\mathrm{const}$,
$\tderiv{\vectorsym{\angvel}}\overset{!}{=}\mathrm{const}$,
are employed, both of which were implied by \cref{eq:constantaccelerationapproximation} above.
\begin{subequations}
\begin{gather}
    \externalforcedensityrelfsinterface 
        = -\int\limits_\forcingcrosssectionspace\!\fdensity\,
            \Spr\braceof{
                  \ttderiv{\vectorsym{\centerlinepos}} + \tderiv{\vectorsym{\angvel}}\times\vectorsym{\relpos} + \vectorsym{\angvel}\times(\vectorsym{\angvel}\times\vectorsym{\relpos})
                + \frac{\vectorsym{\vel}\relfsinterface^{(\rkstep-1)} - \uprelimvec\relfsinterface}{t^{(\rkstep)}-t^{(\rkstep-1)}}
            }\,\dSurface
    \\
    \externaltorquedensityrelfsinterface
        = -\int\limits_\forcingcrosssectionspace\!\vectorsym{\relpos}\times\fdensity\,
             \Spr\braceof{ 
                  \ttderiv{\vectorsym{\centerlinepos}} + \tderiv{\vectorsym{\angvel}}\times\vectorsym{\relpos} + \vectorsym{\angvel}\times(\vectorsym{\angvel}\times\vectorsym{\relpos})
                + \frac{\vectorsym{\vel}\relfsinterface^{(\rkstep-1)} - \uprelimvec\relfsinterface}{t^{(\rkstep)}-t^{(\rkstep-1)}} 
             }\,\dSurface
    \eqdot
\end{gather}
\end{subequations}
Splitting these integrals as done for \cref{eq:coupled3b}, and with 
some of these summands vanishing for geometrical reasons,
\begin{subequations}\label{eq:simplifications1}
\begin{gather}
    \int\limits_{\forcingcrosssectionspace}\!\tderiv{\vectorsym{\angvel}}\times\vectorsym{\relpos}\,\dSurface = 0           \\
    \int\limits_{\forcingcrosssectionspace}\!\vectorsym{\angvel}\times(\vectorsym{\angvel}\times\vectorsym{\relpos})\,\dSurface
    = \int\limits_{\forcingcrosssectionspace}\!\vectorsym{\angvel}\bcdot\vectorsym{\relpos}\,\dSurface\vectorsym{\angvel}
    - \vectorsym{\angvel}\bcdot\vectorsym{\angvel}\int\limits_{\forcingcrosssectionspace}\!\vectorsym{\relpos}\,\dSurface = 0      \\
    \int\limits_{\forcingcrosssectionspace}\!\vectorsym{\relpos}\times\ttderiv{\vectorsym{\centerlinepos}}\,\dSurface = 0
    \eqcomma
\end{gather}
\end{subequations}
the following expressions are obtained,
\begin{subequations}
\begin{align}
    -\externalforcedensityrelfsinterface 
    &=   \fdensity\,\ttderiv{\vectorsym{\centerlinepos}}\int\limits_{\forcingcrosssectionspace}\!\dSurface
       + \int\limits_{\forcingcrosssectionspace}\!\fdensity \frac{\vectorsym{\vel}\relfsinterface^{(\rkstep-1)} - \uprelimvec\relfsinterface}{t^{(\rkstep)}-t^{(\rkstep-1)}} \,\dSurface \\
    -\externaltorquedensityrelfsinterface
    &= \fdensity\int\limits_{\forcingcrosssectionspace}\!\vectorsym{\relpos}\times(\tderiv{\vectorsym{\angvel}}\times\vectorsym{\relpos})\,\dSurface
     + \fdensity\int\limits_{\forcingcrosssectionspace}\!\vectorsym{\relpos}\times(\vectorsym{\angvel}\times(\vectorsym{\angvel}\times\vectorsym{\relpos}))\,\dSurface
     + \int\limits_{\forcingcrosssectionspace}\!\vectorsym{\relpos}\times\fdensity\frac{\vectorsym{\vel}\relfsinterface^{(\rkstep-1)} - \uprelimvec\relfsinterface}{t^{(\rkstep)}-t^{(\rkstep-1)}}\,\dSurface\\
    &= \fdensity\int\limits_{\forcingcrosssectionspace}\!\transposed{\skewmatrix{\vectorsym{\relpos}}}\bcdot\skewmatrix{\vectorsym{\relpos}}\,\dSurface\bcdot\tderiv{\vectorsym{\angvel}}
     + \fdensity\,\vectorsym{\angvel}\times\int\limits_{\forcingcrosssectionspace}\!\transposed{\skewmatrix{\vectorsym{\relpos}}}\bcdot\skewmatrix{\vectorsym{\relpos}}\,\dSurface\bcdot\vectorsym{\angvel}
     + \int\limits_{\forcingcrosssectionspace}\!\vectorsym{\relpos}\times\fdensity\frac{\vectorsym{\vel}\relfsinterface^{(\rkstep-1)} - \uprelimvec\relfsinterface}{t^{(\rkstep)}-t^{(\rkstep-1)}}\,\dSurface
    \eqdot
\end{align}
\end{subequations}
By utilizing the approximations of \cref{eq:relposvecapproximation}, the \ac{EOM} can be written as
\begin{subequations}\label{eq:structure_eom_coupled}
    \input{equations/structure_eom_coupled}
\end{subequations}

Comparison with \cref{eq:coupled4} shows that the terms 
$ ({\sderiv{\vectorsym{\linmom}}}\of{t}-{\sderiv{\vectorsym{\linmom}}}^{(\rkstep-1)})/(t-t^{(\rkstep-1)})$ and
$({\sderiv{\vectorsym{\angmom}}}\of{t}-{\sderiv{\vectorsym{\angmom}}}^{(\rkstep-1)})/(t-t^{(\rkstep-1)})$,
included in
$\externalforcedensityrelfsinterface$ and
$\externaltorquedensityrelfsinterface$, respectively,
actually correspond to the inertia of the fluid contained in $\forcingcrosssectionspace$.
It is possible to define effective inertial parameters with
    $\releff{{\bdensity}}\releff{\bladeA} = \bdensity\bladeA + \fdensity\forcingcrosssectionarea$
    and
    $\releff{{\bdensity}}\releff{\crosssectiontensorofinertia} = \bdensity\crosssectiontensorofinertia + \fdensity\forcingcrosssectiontensorofinertia$
to account for this.
\begin{subequations}\label{eq:structure_eom_coupled2}
    \input{equations/structure_eom_coupled2}
\end{subequations}

%% file: chapters/fluid-structure-coupling/chapters/equations/structure_eom_coupled.tex
		\begin{alignat}{2}
			(\bdensity\bladeA + \fdensity\forcingcrosssectionarea ) \ttderiv{\vectorsym{\centerlinepos}} 
			&= \rhsvecforce + \modexternalforcedensityrelfsinterface,
			\qquad
			&\modexternalforcedensityrelfsinterface &= - \frac{{\sderiv{\vectorsym{\linmom}}}^{(\rkstep-1)}-\sderiv{\vectorsym{{\tilde{\linmom}}}}}{t^{(\rkstep)}-t^{(\rkstep-1)}}
			\\
			(\bdensity\crosssectiontensorofinertia + \fdensity\forcingcrosssectiontensorofinertia)\bcdot\tderiv{\vectorsym{\angvel}}
			+ \vectorsym{\angvel}\times(\bdensity\crosssectiontensorofinertia+\fdensity\forcingcrosssectiontensorofinertia)\bcdot\vectorsym{\angvel}
			&= \rhsvecomega + \modexternaltorquedensityrelfsinterface,
			\qquad
			&\modexternaltorquedensityrelfsinterface &= - \frac{{\sderiv{\vectorsym{\angmom}}}^{(\rkstep-1)}-\sderiv{\vectorsym{{\tilde{\angmom}}}}}{t^{(\rkstep)}-t^{(\rkstep-1)}}
		\end{alignat}\label{eq:structure_eom_w_coupling_final}
		\begin{alignat}{2}
			{\sderiv{\vectorsym{\linmom}}}^{(\rkstep-1)} &= \sderiv{\tilde{m}}\tderiv{\vectorsym{\centerlinepos}}^{(\rkstep-1)} \zero{\ + \ \vectorsym{\angvel}^{(\rkstep-1)}\times\sderiv{\vectorsym{s}} }, \qquad
			&\sderiv{\vectorsym{{\tilde{\linmom}}}} &= \int\limits_\forcingcrosssectionspace\!\uprelimvec\relfsinterface\fdensity\,\dSurface,\\
			{\sderiv{\vectorsym{\angmom}}}^{(\rkstep-1)} &= \zero{\sderiv{\vectorsym{s}}\times\tderiv{\vectorsym{\centerlinepos}}^{(\rkstep-1)}} + \fdensity\forcingcrosssectiontensorofinertia\bcdot\vectorsym{\angvel}^{(\rkstep-1)}, \qquad
			&\sderiv{\vectorsym{{\tilde{\angmom}}}} &= \int\limits_\forcingcrosssectionspace\!\vectorsym{\relpos}^{(\rkstep-1)}\times\uprelimvec\relfsinterface\fdensity\,\dSurface
		\end{alignat}
		\begin{equation}\label{eq:structure_eom_w_coupling_final_forcingcrosssectioninertia}
			\forcingcrosssectionarea = \int\limits_{\forcingcrosssectionspace}\!\dSurface,                                                                              \qquad
			\forcingcrosssectiontensorofinertia = \int\limits_{\forcingcrosssectionspace}\!\transposed{\skewmatrix{\vectorsym{\relpos}^{(\rkstep-1)}}}\bcdot\skewmatrix{\vectorsym{\relpos}^{(\rkstep-1)}}\,\dSurface
        \end{equation}
        \begin{equation}
			\sderiv{\tilde{m}}          = \int\limits_\forcingcrosssectionspace\!\fdensity\,\dSurface,                                                                          \qquad
			\zero{ \sderiv{\vectorsym{s}}     = \int\limits_\forcingcrosssectionspace\!\vectorsym{\relpos}^{(\rkstep-1)}\fdensity\,\dSurface = 0 }
		\end{equation}

%% file: chapters/fluid-structure-coupling/chapters/equations/structure_eom_coupled2.tex
\begin{gather}
\begin{alignat}{2}
    \releff{{\bdensity}}\releff{\bladeA} \ttderiv{\vectorsym{\centerlinepos}} 
    &= \rhsvecforce + \modexternalforcedensityrelfsinterface,
    \qquad
    &\modexternalforcedensityrelfsinterface &= - \frac{{\sderiv{\vectorsym{\linmom}}}^{(\rkstep-1)}-\sderiv{\vectorsym{{\tilde{\linmom}}}}}{t^{(\rkstep)}-t^{(\rkstep-1)}}
    \\
    \label{eq:structure_eom_w_coupling_final2}
    \releff{{\bdensity}}\releff{\crosssectiontensorofinertia}\bcdot\tderiv{\vectorsym{\angvel}}
    + \vectorsym{\angvel}\times\releff{{\bdensity}}\releff{\crosssectiontensorofinertia}\bcdot\vectorsym{\angvel}
    &= \rhsvecomega + \modexternaltorquedensityrelfsinterface,
    \qquad
    &\modexternaltorquedensityrelfsinterface &= - \frac{{\sderiv{\vectorsym{\angmom}}}^{(\rkstep-1)}-\sderiv{\vectorsym{{\tilde{\angmom}}}}}{t^{(\rkstep)}-t^{(\rkstep-1)}}
\end{alignat}
\\
\begin{alignat}{2}
    {\sderiv{\vectorsym{\linmom}}}^{(\rkstep-1)} &= \sderiv{\tilde{m}}\tderiv{\vectorsym{\centerlinepos}}^{(\rkstep-1)} \zero{\ + \ \vectorsym{\angvel}^{(\rkstep-1)}\times\sderiv{\vectorsym{s}} }, \qquad
    &\sderiv{\vectorsym{{\tilde{\linmom}}}} &= \int\limits_\forcingcrosssectionspace\!\uprelimvec\relfsinterface\fdensity\,\dSurface,\\
    {\sderiv{\vectorsym{\angmom}}}^{(\rkstep-1)} &= \zero{\sderiv{\vectorsym{s}}\times\tderiv{\vectorsym{\centerlinepos}}^{(\rkstep-1)}} + \fdensity\forcingcrosssectiontensorofinertia\bcdot\vectorsym{\angvel}^{(\rkstep-1)}, \qquad
    &\sderiv{\vectorsym{{\tilde{\angmom}}}} &= \int\limits_\forcingcrosssectionspace\!\vectorsym{\relpos}^{(\rkstep-1)}\times\uprelimvec\relfsinterface\fdensity\,\dSurface
\end{alignat}
\\
    \releff{{\bdensity}} \coloneqq \bdensity + \fdensity
    ,\qquad
    \releff{\bladeA} \coloneqq \frac{\bdensity\bladeA+\fdensity\forcingcrosssectionarea}{\releff{{\bdensity}}}
    ,\quad
    \releff{\crosssectiontensorofinertia} \coloneqq \frac{\bdensity\crosssectiontensorofinertia + \fdensity\forcingcrosssectiontensorofinertia}{\releff{{\bdensity}}}
\\
    \forcingcrosssectionarea = \int\limits_{\forcingcrosssectionspace}\!\dSurface,                                                                              \qquad
    \forcingcrosssectiontensorofinertia = \int\limits_{\forcingcrosssectionspace}\!\transposed{\skewmatrix{\vectorsym{\relpos}^{(\rkstep-1)}}}\bcdot\skewmatrix{\vectorsym{\relpos}^{(\rkstep-1)}}\,\dSurface \\
    \sderiv{\tilde{m}}          = \int\limits_\forcingcrosssectionspace\!\fdensity\,\dSurface,                                                                          \qquad
    \zero{ \sderiv{\vectorsym{s}}     = \int\limits_\forcingcrosssectionspace\!\vectorsym{\relpos}^{(\rkstep-1)}\fdensity\,\dSurface = 0 }
\end{gather}

%% file: chapters/collision-model/appendix.tex
%
This appendix provides an algorithmic representation of the collision model.
\Cref{alg:singlecollision} determines the collision impulse for a single collision, while
\cref{alg:multiplecollisions} handles multiple simultaneous contacts.
These algorithms are essentially equivalent to the ones in \cite[Algs.~1--2]{TschisgaleConstraintbased2019},
but are formulated a bit differently, closer to the actual implementation,
and matching the present nomenclature.

\begin{algorithm}[h]\smaller
  \DontPrintSemicolon
  \setstretch{1.15}
  \subimport{./algorithms/}{singlecollision}
  \caption{Computation of the collision impulse $\dptot = \dpw + \dpv$ 
           in case of a single collision
           in response to the relative motion of two colliding elements and/or
           the force transferred through collision contact in case of static contact.
           The relative velocity is expressed as the velocity ${\vrelvecv}$ of the two contact points,
           while the loads acting in both elements are accounted for via a projected velocity ${\vrelvecw}$ of the contact points due to the resulting acceleration.
           \\
           \begin{tabular}{ p{12mm} @{} l }
               ${\vrelvecv}$      \dotfill & current relative velocity of the contact points                                                   \\
               ${\vrelvecw}$      \dotfill & projected additional relative velocity due to loads acting on both elements                       \\
               $\distancedir$            \dotfill & direction of collision                                                                            \\
           \end{tabular}
           }
  \label{alg:singlecollision}
\end{algorithm}

\begin{algorithm}[h]\smaller
  \DontPrintSemicolon
  \setstretch{1.15}
  \subimport{./algorithms/}{multiplecollisions}
  \caption{Computation of all collision impulses in an iterative manner, permitting elements to be involved in multiple collisions at once.
  The routine \texttt{Collision} is described in algorithm~\ref{alg:singlecollision}.
  }
  \label{alg:multiplecollisions}
\end{algorithm}

%% file: chapters/collision-model/algorithms/singlecollision.tex
%
\newcommand\mycommfont[1]{\textcolor{gray}{#1}}
\SetCommentSty{mycommfont}
\SetKwComment{textcomment}{}{}
\SetKw{KwTo}{\dots}
\SetKwBlock{PlainBlock}{\hspace*{-0.25em}}{}
\SetKwProg{Routine}{routine}{}{}
\SetKwInOut{Input}{input}
\SetKwFunction{Collision}{Collision}

\Routine{$\Collision\of{ \vrelvecv, \vrelvecw, \distancedir, \systemmatrix, \restitutioncoefficient, \frictioncoeffstatic, \frictioncoeffdynamic, \vectorsym{\linmom}_0=\transposed{[0,0,0]} }$}{
	
	
	\PlainBlock(impulses for non-sliding reaction){
	$\dpv \assign -\systemmatrix^{-1}\cdot( \vrelvecv + \restitutioncoefficient (\vrelvecv\cdot\distancedir)\distancedir )$ \textcomment*[r]{\primecoderef{\texttt{dpv}}} 
	$\dpw \assign -\systemmatrix^{-1}\cdot\vrelvecw $                                                                       \textcomment*[r]{\primecoderef{\texttt{dpw}}} 
	}
	
	
	\PlainBlock(correct impulses to achieve sliding reaction){
		\If{ $(\dpw\cdot\distancedir>0) \lor (\dpv\cdot\distancedir>0)$ }{  
		
			$\vectorsym{\linmom} \assign \vectorsym{\linmom}_0 + \dpv + \dpw$\; 
			$\linmom\ind{n} \assign \vectorsym{\linmom}\cdot\distancedir$                       \textcomment*[r]{\primecoderef{\texttt{abs\_pn}}}
			$\linmom\ind{t} \assign \vnorm{ \vectorsym{\linmom} - \linmom\ind{n}\distancedir }$ \textcomment*[r]{\primecoderef{\texttt{abs\_pt}}}
		
			\If{$\linmom\ind{t} > \frictioncoeffstatic \linmom\ind{n}$}{ 
			
				$\vectorsym{v}\ind{t} \assign \vrelvecv - (\vrelvecv\cdot\distancedir)\distancedir$ \textcomment*[r]{\primecoderef{\texttt{vrt}}} 
				\leIf{$\vnorm{\vectorsym{v}\ind{t}}=0$}
				     {$\vectorsym{t} \assign \transposed{[0,0,0]}$}
				     {$\vectorsym{t} \assign \vectorsym{v}\ind{t}/\vnorm{\vectorsym{v}\ind{t}}$}
				$\dpw \assign \frac{(\systemmatrix\cdot\dpw)\cdot\distancedir}{(\systemmatrix\cdot(\distancedir-\frictioncoeffdynamic\vectorsym{t}))\cdot\distancedir}(\distancedir-\frictioncoeffdynamic\vectorsym{t})$ \textcomment*[r]{\primecoderef{\texttt{dpw}}}
				$\dpv \assign \frac{(\systemmatrix\cdot\dpv)\cdot\distancedir}{(\systemmatrix\cdot(\distancedir-\frictioncoeffdynamic\vectorsym{t}))\cdot\distancedir}(\distancedir-\frictioncoeffdynamic\vectorsym{t})$ \textcomment*[r]{\primecoderef{\texttt{dpv}}}
			
			}
		}
	}
	
	\PlainBlock(prevent attraction){
		\lIf{$\dpw\cdot\distancedir\leq0$}{ $\dpw = \transposed{[0,0,0]}$ }
		\lIf{$\dpv\cdot\distancedir\leq0$}{ $\dpv = \transposed{[0,0,0]}$ }
	}
	
%
%
%
%
%
	
	\Return{$\dpw$, $\dpv$}
}




\let\mycommfont\undefined

%% file: chapters/collision-model/algorithms/multiplecollisions.tex
%
  \newcommand\mycommfont[1]{\textcolor{gray}{#1}}
  \SetCommentSty{mycommfont}
  \SetKwComment{textcomment}{}{}
  \SetKw{KwTo}{\dots}
  \SetKwBlock{PlainBlock}{\hspace*{-0.25em}}{}

\newcommand{\lcpit}{\ensuremath{k}}
\newcommand{\globalrun}{\ensuremath{\mathit{run}}}
\newcommand{\ncoll}{\ensuremath{m}}

\ForEach{collision pair $\ncoll$}{
	$\vectorsym{\linmom}_\ncoll \assign \transposed{\left[0,0,0\right]}$ \textcomment*[r]{\primecoderef{{\texttt{pair\%p} in \texttt{pair\_variables}}}}
	
}

\ForEach{collision element $\masterelement$}{
	${\relvelconstr{\vectorsym{\linmom}}}_\masterelement, 
	 {\relextloads{\vectorsym{\linmom}}}_\masterelement, 
	 {\relvelconstr{\vectorsym{\angmom}}}_\masterelement, 
	 {\relextloads{\vectorsym{\angmom}}}_\masterelement \assign \transposed{[0,0,0]}$\;
}

\While{collision impulses ${\relextloads{\vectorsym{\linmom}}}_\masterelement$, ${\relvelconstr{\vectorsym{\linmom}}}_\masterelement$ are not converged}{

	\ForEach{collision $\ncoll$ involving the elements $\masterelement_1$, $\masterelement_2$}{

		${\vrelvecv}_\ncoll \assign {\vrelvecv} + \Big[ \mass^{-1} {\relvelconstr{\vectorsym{\linmom}}} + (\tensorofinertia^{-1}\cdot{\relvelconstr{\vectorsym{\angmom}}})\times\contactpointvec \Big]_{\masterelement_2}  
			                                   -\Big[ \mass^{-1} {\relvelconstr{\vectorsym{\linmom}}} + (\tensorofinertia^{-1}\cdot{\relvelconstr{\vectorsym{\angmom}}})\times\contactpointvec \Big]_{\masterelement_1}  
			             $ \textcomment*[r]{\primecoderef{\texttt{vr} in \texttt{collision\_dry}}}
		
        ${\vrelvecw}_\ncoll \assign {\vrelvecw} + \Big[ \mass^{-1} {\relextloads{\vectorsym{\linmom}}} + (\tensorofinertia^{-1}\cdot{\relextloads{\vectorsym{\angmom}}})\times\contactpointvec \Big]_{\masterelement_2} 
		                                      - \Big[ \mass^{-1} {\relextloads{\vectorsym{\linmom}}} + (\tensorofinertia^{-1}\cdot{\relextloads{\vectorsym{\angmom}}})\times\contactpointvec \Big]_{\masterelement_1} 
			              $ \textcomment*[r]{\primecoderef{\texttt{vex} in \texttt{collision\_dry}}}  
		
        	$ \dpv, \dpw \assign $ \Collision{$ {\vrelvecv}_\ncoll, {\vrelvecw}_\ncoll, \distancedir, \systemmatrix, \restitutioncoefficient, \frictioncoeffstatic, \frictioncoeffdynamic, \vectorsym{\linmom}_0=\vectorsym{\linmom}_\ncoll $ }\;
			
			$ {\relextloads{\vectorsym{\linmom}}}_{\masterelement_2} \plusassign  \dpw $, \qquad $ {\relextloads{\vectorsym{\angmom}}}_{\masterelement_2} \plusassign  \contactpointvec_{\masterelement_2}\times\dpw $   \textcomment*[r]{\primecoderef{\texttt{element\_2\%pw\_c}, \texttt{element\_2\%lw\_c} in \texttt{collision\_dry}}}
			$ {\relextloads{\vectorsym{\linmom}}}_{\masterelement_1} \minusassign \dpw $, \qquad $ {\relextloads{\vectorsym{\angmom}}}_{\masterelement_1} \minusassign \contactpointvec_{\masterelement_1}\times\dpw $   \textcomment*[r]{\primecoderef{\texttt{element\_1\%pw\_c}, \texttt{element\_1\%lw\_c} in \texttt{collision\_dry}}}
			$ {\relvelconstr{\vectorsym{\linmom}}}_{\masterelement_2} \plusassign  \dpv $, \qquad $ {\relvelconstr{\vectorsym{\angmom}}}_{\masterelement_2} \plusassign  \contactpointvec_{\masterelement_2}\times\dpv $ \textcomment*[r]{\primecoderef{\texttt{element\_2\%pv\_c}, \texttt{element\_2\%lv\_c} in \texttt{collision\_dry}}}
			$ {\relvelconstr{\vectorsym{\linmom}}}_{\masterelement_1} \minusassign \dpv $, \qquad $ {\relvelconstr{\vectorsym{\angmom}}}_{\masterelement_1} \minusassign \contactpointvec_{\masterelement_1}\times\dpv $ \textcomment*[r]{\primecoderef{\texttt{element\_1\%pv\_c}, \texttt{element\_1\%lv\_c} in \texttt{collision\_dry}}}

		
        $\vectorsym{\linmom}_\ncoll \assign \vectorsym{\linmom}_\ncoll + \dpw$\; 
		
	}
	
	
}

\ForEach(){collision element $\masterelement$}{
	$\vectorsym{\linmom}_\masterelement \assign {\relextloads{\vectorsym{\linmom}}}_\masterelement + {\relvelconstr{\vectorsym{\linmom}}}_\masterelement$ \textcomment*[r]{\primecoderef{\texttt{element\_mas(e)\%p\_c} in \texttt{element\_collision}}} 
	$\vectorsym{\angmom}_\masterelement \assign {\relextloads{\vectorsym{\angmom}}}_\masterelement + {\relvelconstr{\vectorsym{\angmom}}}_\masterelement$ \textcomment*[r]{\primecoderef{\texttt{element\_mas(e)\%l\_c} in \texttt{element\_collision}}} 
}

\let\mycommfont\undefined